\documentclass[fleqn,usenatbib]{mnras}

\usepackage{newtxtext,newtxmath}

\usepackage[T1]{fontenc}

\DeclareRobustCommand{\VAN}[3]{#2}
\let\VANthebibliography\thebibliography
\def\thebibliography{\DeclareRobustCommand{\VAN}[3]{##3}\VANthebibliography}

\usepackage{graphicx}	
\usepackage{amsmath}	
\usepackage{booktabs,siunitx} 
\usepackage{xcolor}           
\usepackage{multirow} 
\usepackage[normalem]{ulem}

\newcommand{\shb}{South Haze Bubble}	
\newcommand{\nhb}{North Haze Bubble}	

\title[The Haze as seen by QUIJOTE]{QUIJOTE scientific results -- VI. The Haze as seen by QUIJOTE}

\author[F. Guidi et al.]{F. Guidi,$^{1,2,3}$\thanks{E-mail: federica.guidi@iap.fr}
R.~T. G\'enova-Santos,$^{1,2}$\thanks{E-mail: rgs@iac.es}
J.~A. Rubi{\~n}o-Mart\'in,$^{1,2}$
M.~W. Peel,$^{1,2}$ 
\newauthor
M. Fern\'andez-Torreiro,$^{1,2}$
C.~H. L\'opez-Caraballo,$^{1,2}$
R. Vignaga,$^{1,2}$
E. de la Hoz,$^{4,5}$
\newauthor
P. Vielva,$^{4}$
R.~A. {Watson},$^{6}$ 
M. Ashdown,$^{7,8}$
C. Dickinson,$^{6}$
E. Artal,$^{9}$
R.~B. Barreiro,$^{4}$
\newauthor
F.~J. Casas,$^{4}$
D. Herranz,$^{4}$
R.~J. Hoyland,$^{1,2}$
A.~N. Lasenby,$^{7,8}$
E. Martinez-Gonzalez,$^{4}$
\newauthor
L. Piccirillo,$^{6}$
F. Poidevin,$^{1,2}$
R. Rebolo,$^{1,2,10}$
B. Ruiz-Granados,$^{1,2,11}$
D. Tramonte,$^{12,13,1,2}$
\newauthor
F. Vansyngel$^{1,2}$
\\
$^{1}$Instituto de Astrof\'{\i}sica de Canarias, E-38200 La Laguna, Tenerife, Spain\\
$^{2}$Departamento de Astrof\'{\i}sica, Universidad de La Laguna,
E-38206 La Laguna, Tenerife, Spain\\
$^{3}$Institut d'Astrophysique de Paris, UMR 7095, CNRS \& Sorbonne Universit\'e, 98 bis boulevard Arago, 75014 Paris, France\\
$^{4}$Instituto de Fisica de Cantabria (IFCA), CSIC-Univ. de Cantabria, Avda. los
Castros, s/n, E-39005 Santander, Spain\\
$^{5}$Dpto. de Física Moderna, Universidad de Cantabria, Avda. los Castros s/n, E-39005 Santander, Spain \\
$^{6}$Jodrell Bank Centre for Astrophysics, Alan Turing Building, Department of Physics and Astronomy, School of Nature Sciences, 
University of Manchester,\\ Oxford Road, Manchester M13 9PL, U.K\\
$^{7}$Astrophysics Group, Cavendish Laboratory, University of Cambridge, J J Thomson Avenue, Cambridge CB3 0HE, UK\\
$^{8}$Kavli Institute for Cosmology, University of Cambridge, Madingley Road, Cambridge CB3 0HA \\
$^{9}$Departamento de Ingenieria de COMunicaciones (DICOM), Edificio Ingenieria de Telecomunicacion, Plaza de la Ciencia s/n, E-39005 Santander, Spain\\ 
$^{10}$Consejo Superior de Investigaciones Cientificas, E-28006 Madrid, Spain\\
$^{11}$Departamento de F\'{\i}sica. Facultad de Ciencias. Universidad de C\'ordoba.  Campus de Rabanales, 
Edif. C2. Planta Baja.  E-14071 C\'ordoba, Spain\\
$^{12}$Purple Mountain Observatory, CAS, No.10 Yuanhua Road, Qixia District, Nanjing 210034, China \\
$^{13}$NAOC-UKZN Computational Astrophysics Center (NUCAC), University of Kwazulu-Natal, Durban 4000, South Africa
}

\date{Accepted XXX. Received YYY; in original form ZZZ}

\pubyear{2023}

\begin{document}
\label{firstpage}
\pagerange{\pageref{firstpage}--\pageref{lastpage}}
\maketitle

\begin{abstract}
The Haze is an excess of microwave intensity emission surrounding the Galactic centre. It is spatially correlated with
the $\gamma$-ray \textit{Fermi} bubbles, and with the S-PASS radio polarization plumes, suggesting a possible common provenance. 
The models proposed to explain the origin of the Haze, including energetic events at the Galactic centre and dark matter decay in the Galactic halo, do not yet provide a clear physical interpretation.
In this paper we present a re-analysis of the Haze including new observations from the Multi-Frequency Instrument (MFI) of the Q-U-I Joint TEnerife (QUIJOTE) experiment, at 11 and 13\,GHz.
We analyze the Haze in intensity and polarization, characterizing its spectrum. 
We detect an excess of diffuse intensity signal ascribed to the Haze. The spectrum at frequencies 11$\,\leq\nu\leq\,$70\,GHz is a power-law with spectral index $\beta^{\rm H}=-2.79\pm0.08$, which is flatter than the Galactic synchrotron in the same region ($\beta^{\rm S}=-2.98\pm0.04$), but steeper than that obtained from previous works ($\beta^{\rm H}\sim-2.5$ at 23$\,\leq\,\nu\leq\,$70\,GHz).
We also observe an excess of polarized signal in the QUIJOTE-MFI maps in the Haze area. This is a first hint detection of polarized Haze, or a consequence of curvature of the synchrotron spectrum in that area. 
Finally, we show that the spectrum of polarized structures associated with Galactic centre activity is steep at low frequencies ($\beta \sim -3.2$  at 2.3\,$\leq\nu\leq$\,23\,GHz), and becomes flatter above 11\,GHz. 

\end{abstract}

\begin{keywords}
diffuse radiation -- Galaxy: centre -- ISM: bubbles -- cosmology: observations
\end{keywords}



\section{Introduction}

During the last decades multiple high sensitivity surveys have been carried out in order to provide an accurate characterization of the Galactic foregrounds at radio and microwave wavelengths, with the final goal of doing cosmology with the Cosmic Microwave Background  (CMB; see e.g., \citealp{Pla2018VI}). The main target of the two satellite missions Wilkinson Microwave Anisotropy Probe (\textit{WMAP}; e.g., \citealp{Bennett2003}) and \textit{Planck} (e.g., \citealp{Pla2018I}) was the CMB, which implied the generation of full sky images of the Galactic emission at multiple frequencies, between 23 and 857\,GHz. These data provided an accurate picture of the emission of our own Galaxy, and enabled a number of discoveries, among them that of the microwave Haze. 

The Haze was discovered in the process of disentangling the Galactic emission from the cosmological CMB signal using \textit{WMAP} data between 23 and 60\,GHz by \citet{Fink2004}, and it was confirmed by further studies  (\citealp{WMAPHaze,2012ApJ...750...17D,Pietrobon2012,PlanckHaze}). During this process, a diffuse and extended signal became evident in the residuals after removing all the already known emission mechanisms from the \textit{WMAP} frequency maps. The microwave (or sometimes \textit{WMAP}) Haze is indeed an excess of diffuse emission with an elliptically symmetric shape centred on the Galactic centre, extending towards the north and the south of the Galactic plane and reaching high Galactic latitudes $|b| \approx 35^{\circ}$. The Haze has been measured to have a relatively flat spectrum ($\beta \approx -2.5$) at the lowest \textit{WMAP} frequencies compared to that of typical Galactic synchrotron emission at high Galactic latitudes ($\beta \approx -3.0$). 

The \textit{Planck} Collaboration also detected the Haze excess \citep{PlanckHaze} with an independent dataset between 30 and 70\,GHz. They measured the spectrum of the South Haze bubble using \textit{Planck} and \textit{WMAP} data, showing a synchrotron-like power-law with a spectral index $\beta=-2.56\pm0.05$. This spectrum is in agreement with what had been previously observed with \textit{WMAP} data alone.

The microwave Haze has a $\gamma$-ray counterpart, the so-called \textit{Fermi} bubbles, which were discovered in the \textit{Fermi}-LAT data at energies $2$--$50\,$GeV (\citealp{GammaHaze,GammaHaze2}). The \textit{Fermi} bubbles are two extended $\gamma$-ray lobes located at a position coincident to that of the \textit{WMAP} Haze, but with a larger extension in Galactic latitude, reaching $|b|\approx50^{\circ}$, and with a flat spectrum. 
This multi-wavelength correspondence confirmed the interpretation of the microwave Haze as a real sky component and it was ascribed to synchrotron emission of a young population of cosmic-ray electrons \citep{GammaHaze}.
Cosmic-ray electrons with energies $10$--$100\,$GeV produce microwave synchrotron during their interaction with a magnetic field, but also $\gamma$-ray photons through Inverse Compton  scattering (IC) with the Interstellar Radiation Field (ISRF). 

In addition, recent observations of the eRosita X-ray space telescope \citep{Merloni2012} detected a distinct but possibly related structure: two circular and symmetric soft-X-ray (0.3--2.3\,keV) bubbles, which extended up to high Galactic latitude $|b|\approx 85^{\circ}$ \citep{Predehl2020}. The eRosita bubbles enclose the \textit{Fermi} bubbles, and the northern one partially overlaps with the North Polar Spur (NPS), a large and polarized filament that emerges from the Galactic centre and goes toward the north. This spatial correlation points towards a possible connection between the NPS and the Haze, which could be generated by an explosive event in the Galactic centre \citep{Sofue1977,Sofue1994}. Several works support this hypothesis by locating the NPS at a distance of $\sim$10\,kpc, which is comparable to the distance to the Galactic centre (e.g., \citealp{Sofue2015, Predehl2020, Kataoka2021}). However this aspect is still controversial. According with different works (e.g., \citealp{pla2015XXV, Panopoulou2021}) the distance to the NPS is smaller than to the Galactic centre, being of the order of $\sim$ 100--200\,pc, identifying therefore the NPS and the Haze as two different components, with respectively a local and a Galactic centre origin. 

The Haze, moreover, is not peculiar to our own Galaxy. \cite{Li2019} reported the first detection of a Haze-like structure in an external galaxy, using radio (C-band) and X-ray (0.8--8\,keV) data. The spectral index of this extra-galactic Haze is $\beta \approx -3.1$ at radio wavelengths, which is typical for synchrotron emission, and takes the slightly flatter value $\beta \approx -2.8$ in the joint fit of radio and X-ray data. 

It is well known that the synchrotron emission is polarized, and to confirm that the Haze has a synchrotron origin it should be possible to observe an associated polarized component. Such a component was identified for the first time by the S-PASS southern sky survey at 2.3\,GHz, which detected two giant radio polarized plumes extending from the centre of our Galaxy (\citealp{SpassHaze}). The plumes spatially correlate with the \textit{Fermi} bubbles, with the microwave Haze, and with X-ray structures observed by ROSAT (\citealp{XrayROSAT,SpassHaze}) that connect the plumes with the centre of the Galaxy. Interestingly, the radio polarized plumes appear to be more extended than the \textit{Fermi} bubbles, reaching $|b|\approx60^{\circ}$.   

The radio polarized plumes can also be roughly identified in the low frequency maps of \textit{Planck} and \textit{WMAP}, although the signal-to-noise is not as good as in S-PASS. The combination of S-PASS and \textit{WMAP} data allowed the measurement of the spectral index of the polarized emission between 2.3\,GHz and $23$\,GHz, which is $\beta=-3.2$ (\citealp{SpassHaze}). It should be noted that the spectral index is significantly flatter in intensity than in polarization, making the interpretation of the Haze/bubbles to be very puzzling. The difference in the spectral index might suggest that the cosmic-ray electrons that generate the intensity of the Haze and the polarization of the plumes belong to two different electron populations. Alternatively, the superposition of different components along the line-of-sight could explain the different spectral index in polarization.  

A variety of scenarios have been proposed in order to explain the possible origin of the Haze signal. One intriguing proposal is that it is generated by secondary emission of dark matter particles (\citealp{Hooper2007,Cholis2009,GammaHaze,Delahaye2012,DMreview,DMhaze}).  However the existence of  $\gamma$-ray bubbles with sharp edges (\citealp{GammaHaze2}) and radio polarized sharp filaments and plumes (\citealp{Biermann2010,Jones2012,Crocker2011,SpassHaze,pla2015XXV}) contradict the dark matter hypothesis as a complete explanation of this phenomenon, while energetic events in the Galactic centre provide a much more likely scenario. Still, it cannot be excluded that a small fraction of the Haze emission could have a dark matter origin (\citealp{DMhaze}). 

Other proposed progenitors for the Haze emission demand energetic events in the Galactic centre. AGN activity of the super-massive black hole in the centre of the Milky Way (SgrA*) (\citealp{Zubovas2012,Guo2012I,Guo2012II,Ackermann2014,Fox2015,Zhang2020,Zhang2021b,Pillepich2021,Yang2022}), nuclear activity in the central Galactic region such as star-formation, star-bursts, or supernovae explosions, which could power outflows of hot and magnetized plasma and accelerate cosmic rays (\citealp{Crocker2011,Crocker2012,Lacki2014,SpassHaze, Zhang2021a}), or more complex scenarios (\citealp{Ashley2022}). 

A study from \cite{Crocker2015} proposed a unified model for the microwave Haze, radio plumes, and \textit{Fermi} bubbles, as generated by outflows powered by nuclear activity. For the first time, this model provided an explanation for the change of the spectral index in the outer and inner part of the bubbles at microwave or radio wavelengths, as suggested by observations \citep{SpassHaze}.

However, even if the scenarios proposed in the literature partially explain some of the Haze characteristics, none of them provide a complete description. New observations are crucial for the understanding of the origin of the Haze, and independent determinations of the spectral index of the emission across the Haze area, as well as polarization measurements, can yield a clearer picture of this complex region. 

In this paper we provide new observational constraints on the Haze microwave emission using data from the Multi-Frequency Instrument of the Q-U-I Joint TEnerife experiment \citep{RubinoSPIE12,MFIstatus12}.  We performed a full reanalysis of the Haze bubbles and filaments first reproducing, in an independent manner, previous results obtained with \textit{WMAP} \citep{WMAPHaze}, \textit{Planck}-LFI \citep{PlanckHaze}, and S-PASS \citep{SpassHaze} data. Afterwards we included in the analysis microwave data from QUIJOTE-MFI at 11 and 13\,GHz. In particular, we performed for the first time a component separation in polarization, searching for a polarized Haze component at the QUIJOTE frequencies. Note that there is a gap of available data between 2.3\,GHz and 23\,GHz, and 2.3\,GHz data are affected by Faraday rotation and depolarization \citep{SPASSmaps}. QUIJOTE effectively extends the frequency coverage of \textit{WMAP} and \textit{Planck} down to 11\,GHz, where the signal is relatively strong and not significantly affected by Faraday effects, providing robust spectral measurements in intensity and polarization.

The paper is organized as follows: we present the new QUIJOTE maps of the Haze and the ancillary data used for the analysis in Sect.\,\ref{sec:data}, we then describe the methodologies applied for this work in Sect.\,\ref{sec:methodology}, consisting of a template fitting component separation described in Sec\,\ref{secch5:templ_fitting}, and a correlation T-T plots analysis in polarization, as described in Sect.\,\ref{secch5:tt-plots}. Afterwards, in Sect.\,\ref{sec:results} we present the results of the template fitting  (Sect.\,\ref{secch5:templatefitting-int} in intensity and  \ref{secch5:templatefitting-pol} in polarization), and of the T-T plots in polarization (Sect.\,\ref{secch5:tt-plots-results}). Finally, we summarize and we conclude in Sect.\,\ref{sec:discussion} and \ref{sec:conclusions}.

\section{Data} \label{sec:data}
\begin{table}
\resizebox{0.5\textwidth}{!}{%
    \centering
    \begin{tabular}{ccccc}
    \hline
         Survey & Freq. & FWHM  & $\sigma_{\rm c}$ & Reference \\
              &  [GHz]    & [deg] &  [\%]    & \\
     \midrule
         S-PASS & \phantom{0}2.3 & 0.15 & 5 & \cite{SPASSmaps} \\
         QUIJOTE & 11.1 & 0.93 & 5 & \cite{mfiwidesurvey} \\ 
         QUIJOTE & 12.9 & 0.92 & 5 & \cite{mfiwidesurvey} \\ 
         \textit{WMAP} K-band & 22.8 & 0.88 & 3 & \cite{WMAPmaps} \\ 
         \textit{Planck}-LFI & 28.4 & 0.54 & 3 & \cite{NPIPE2020} \\ 
         \textit{WMAP} Ka-band & 33.0 & 0.66 & 3 & \cite{WMAPmaps} \\ 
         \textit{WMAP} Q-band & 40.6 & 0.51 & 3 & \cite{WMAPmaps} \\ 
         \textit{Planck}-LFI & 44.1 & 0.45 & 3 & \cite{NPIPE2020} \\ 
         \textit{WMAP} V-band & 60.8 & 0.35 & 3 & \cite{WMAPmaps} \\ 
         \textit{Planck}-LFI & 70.4 & 0.22 & 3 & \cite{NPIPE2020} \\   
    \hline
    \end{tabular}}
 \caption{Summary of the data that are used in this work.  We show central frequencies, beam FWHMs and adopted calibration uncertainties ($\sigma_{\rm c}$) of each survey.}
    \label{tabch5:summary_dsataseta}
\end{table}
We describe here the dataset that is used in this work, which is composed of the QUIJOTE-MFI data at 11 and 13\,GHz (see Sect.~\ref{secch5:QJTdata}), in combination with ancillary data (see Sect.~\ref{secch5:data-overview}) from S-PASS at 2.3\,GHz \citep{SPASSmaps}, \textit{WMAP} at $\sim$23, 33, 41, 61\,GHz \citep{WMAPmaps}, and \textit{Planck}-LFI at $\sim$30, 44, 70\,GHz \citep{NPIPE2020}. A summary  of the dataset can be found in Table.\,\ref{tabch5:summary_dsataseta}.

\subsection{QUIJOTE-MFI data}\label{secch5:QJTdata}

\begin{figure*}
    \centering
    \includegraphics[width=0.32\textwidth]{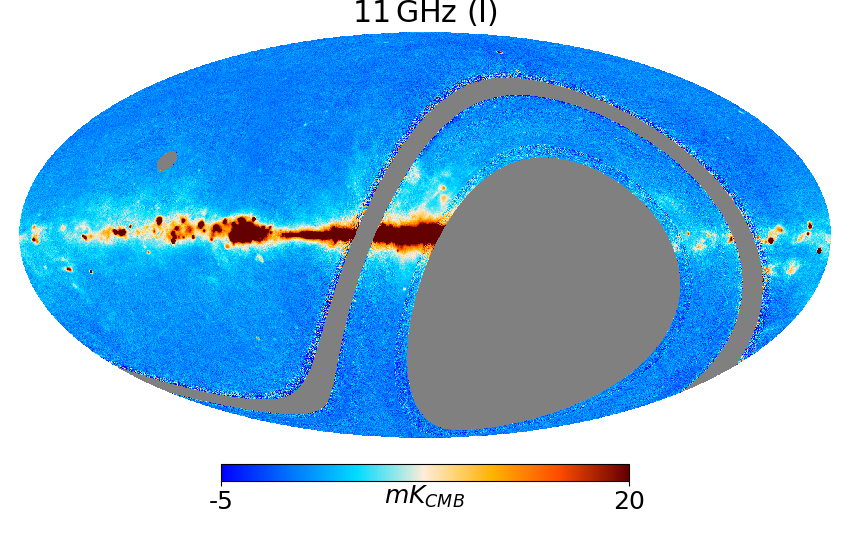}
    \includegraphics[width=0.32\textwidth]{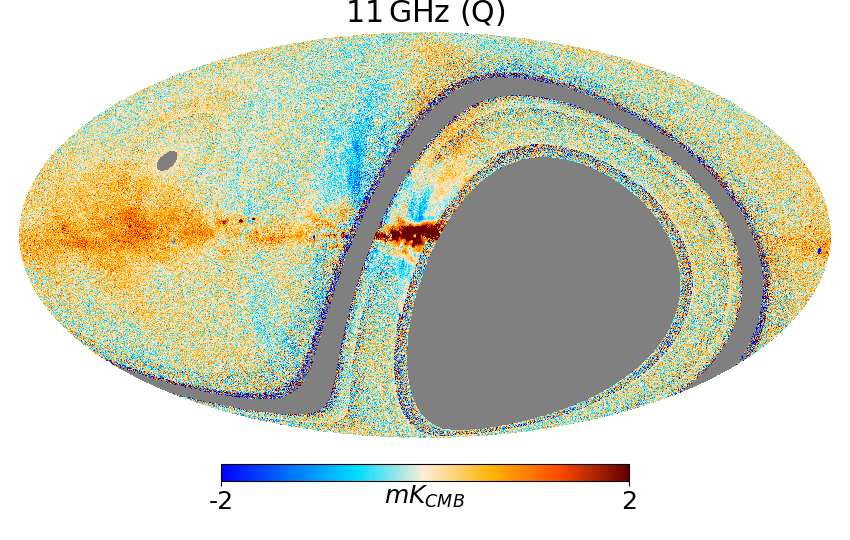}
    \includegraphics[width=0.32\textwidth]{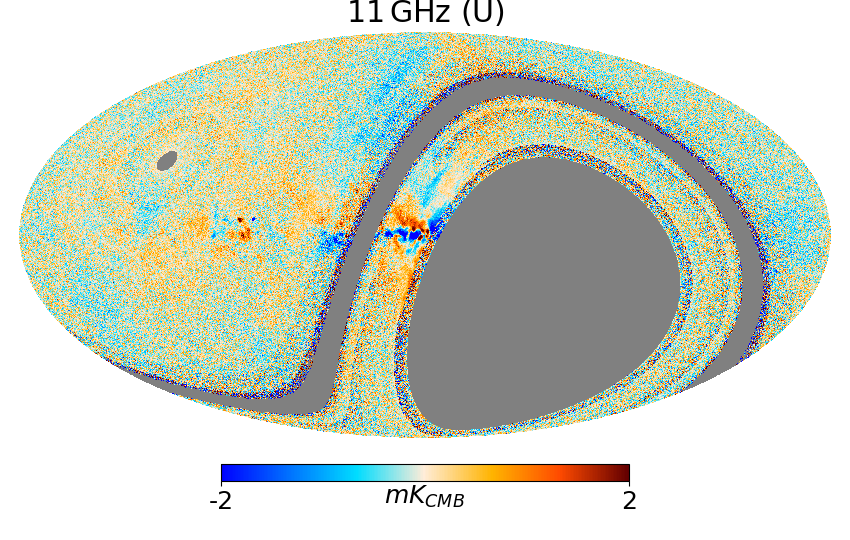}
    \includegraphics[width=0.32\textwidth]{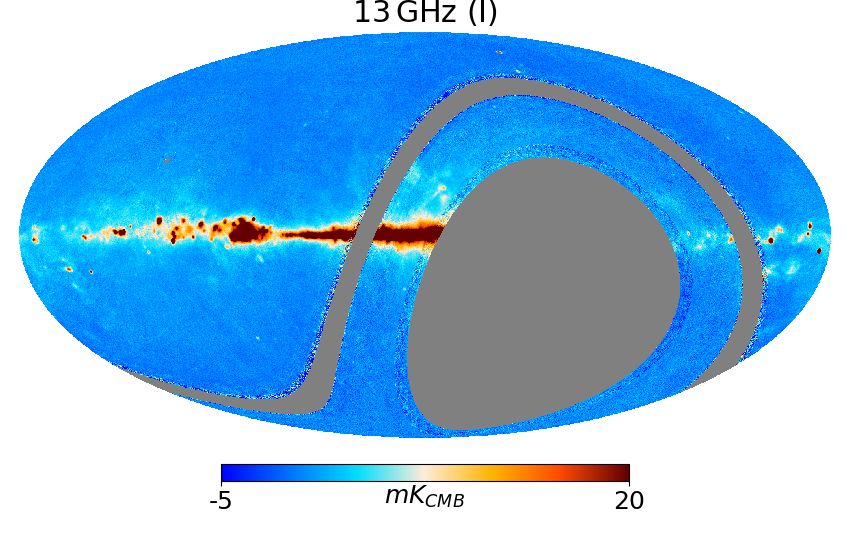}
    \includegraphics[width=0.32\textwidth]{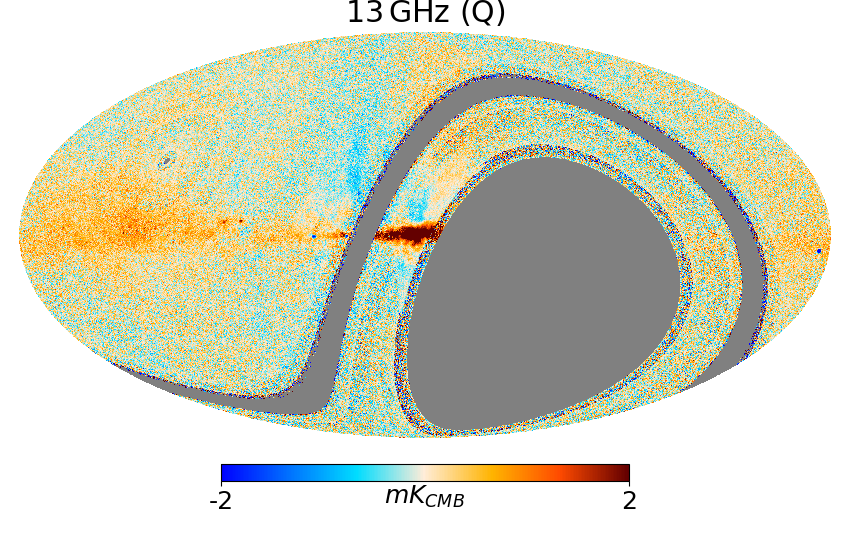}
    \includegraphics[width=0.32\textwidth]{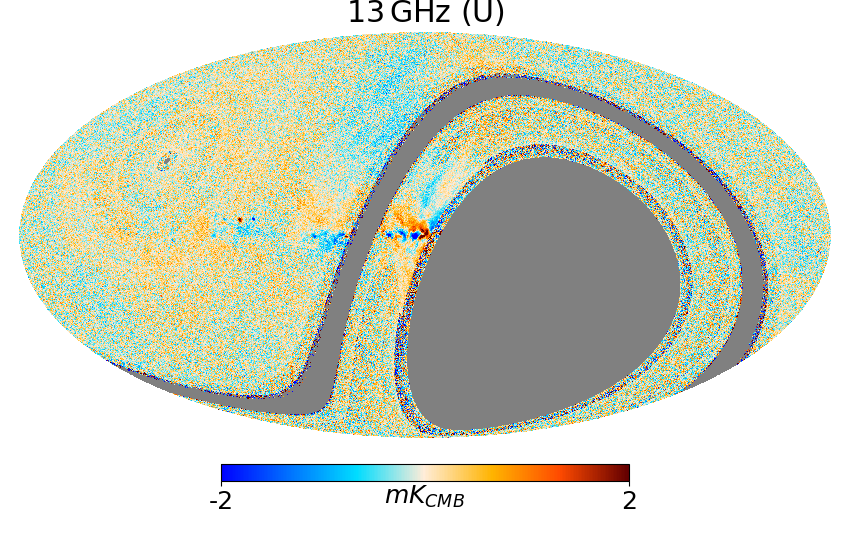}
    \caption{\textit{I}, \textit{Q}, and \textit{U} maps from QUIJOTE-MFI at 11\,GHz (top) and 13\,GHz (bottom) at the original angular resolution and pixel size ($N_{\mathrm{side}}=512$). The maps are mollweide projections and in Galactic coordinates, with the centre of the projection at ($l,b)=(0^{\circ},0^{\circ})$, and with longitude increasing to the left. Colour scales are linear and grey represents missing data or regions contaminated by RFI. The data used to generate these maps combine the wide-survey and the dedicated raster scan observations described in the text. }
    \label{fig:QJT1113}
\end{figure*}

QUIJOTE is a polarimetric ground-based CMB experiment located at the Teide observatory (Tenerife, Spain), at 2400\,m above sea level \citep{RubinoSPIE12}. The MFI instrument of QUIJOTE observes the sky of the Northern hemisphere at four frequency bands in the range 10--20\,GHz, and with an angular resolution of $\approx 1 ^\circ$ \citep{MFIstatus12}. 

The reference dataset for this paper is the survey of the full northern sky performed with QUIJOTE-MFI (hereafter the \textit{wide-survey}). This survey provides an average sensitivity in polarization of $\sim 40$--$55$\,$\mu$K\,deg$^{-1}$ in the four bands centred around 11, 13, 17 and 19\,GHz (see Sect.~4.3 in \citealp{mfiwidesurvey}).
This paper is part of the release that describes the survey and the associated scientific results, concerning principally the characterization of diffuse synchrotron radiation and Anomalous Microwave Emission (AME). 
A complete description of the wide survey can be found in \citet{mfiwidesurvey}.

\begin{figure}
    \centering
    \includegraphics[width=0.5\textwidth]{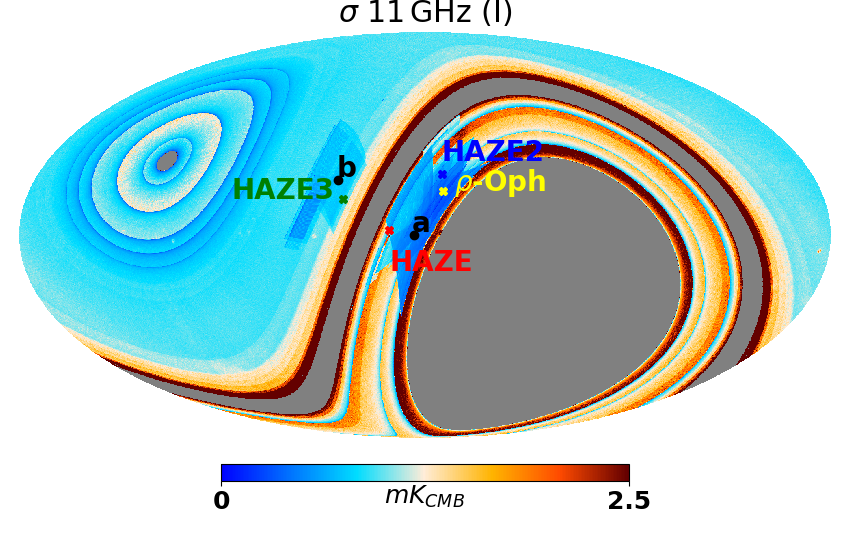}
    \caption{Uncertainty ($\sigma$) of the Intensity map from QUIJOTE at 11\,GHz, at the original angular resolution and pixel size ($N_{\mathrm{side}}=512$; the uncertainty distribution is similar for \textit{Q} and  \textit{U}, and at 13\,GHz). The location of the three fields observed with raster-scans are indicated in the maps (see Table\,\ref{tabch5:hours}), as well as the position of ``a'' and ``b'' where noise estimates are provided (see Table\,\ref{tab:noise_QJT}).}
    \label{fig:figch5:QJT1113_sigma_maps}
\end{figure}

The QUIJOTE-MFI maps used in this work are a combination of this wide-survey data with additional raster-scan observations that were performed specifically around the Haze region in order to improve the signal-to-noise ratio. These raster scan observations consisted of back-and-forth constant elevation scans of the telescope performed with a scanning speed of $1$\,deg/s on the sky in the period  June 2013 -- August 2018. In particular, four sky fields have been observed, which we call the "HAZE", "HAZE2" and "HAZE3" fields, as well as a sky patch enclosing the $\rho$-Ophiuchi cloud complex,\footnote{$\rho$-Ophiuchi observations had a different scientific goal, specifically the study of this specific cloud complex. However, since they lie nearby the Haze fields, we included them in this analysis.} covering, in total, a sky fraction $f_{\rm sky}\sim 5$\%. The approximate central coordinates of each raster scan field are indicated  in Fig.\,\ref{fig:figch5:QJT1113_sigma_maps} and in Table~\ref{tabch5:hours}, where also their total observing time is reported. 

Although we have produced the maps for all the QUIJOTE-MFI channels, here we use only the 11 and 13\,GHz frequency maps from horn 3 (central frequencies 11.1 and 12.9\,GHz), which have sufficiently good signal-to-noise for this analysis. Note that there are some difficulties inherent to the observations, which are: (1) the contamination of Radio Frequency Interference (RFI) from geostationary satellites that requires the flagging of a declination band with $ -10^\circ \lesssim \delta \lesssim -1^\circ$; (2) the fact that elevation of the Haze area from the Teide Observatory (geographical latitude $+28^{\circ}$) is very low ($el \lesssim 35^{\circ}$), so all the observations are taken looking through a large air-mass. Point (2) is the main reason why the two additional QUIJOTE maps at 17 and 19\,GHz are not used here.  

The maps are shown in Fig.\,\ref{fig:QJT1113}, where we present the \textit{I}, \textit{Q}, and \textit{U} maps at the original angular resolution and pixel size ($N_{\mathrm{side}}=512$ in the HEALPix\footnote{ \url{https://sourceforge.net/projects/healpix/}} pixelization scheme; \citealp{Gorski2005}). 
The maps have been generated with the \texttt{PICASSO} map-making code, which was implemented for the construction of maps from the QUIJOTE-MFI data \citep{destriper}. The maps have been obtained with a single run of \texttt{PICASSO}, combining simultaneously the wide-survey data and the additional raster observations with an efficient subtraction of the correlated $1/f$ noise. The parameters adopted for this run (priors on noise properties, baseline length, etc) are identical to those used for the wide survey \citep[see details in Sect.~2.3  of][]{mfiwidesurvey}.

In Fig.\,\ref{fig:figch5:QJT1113_sigma_maps} we show the statistical white noise level ($\sigma$) of the 11\,GHz map in intensity, computed from the propagation of the weights in the TOD through the map-making procedure, and with pixel resolution $N_{\mathrm{side}}=512$.  
The location of the raster observations can be seen as the bluish regions at the center of these maps, corresponding to a decrease of $\sigma$. The raster scan data result in an improvement of the noise level with respect to the wide-survey data alone in two specific areas: in the Galactic centre region (with ``HAZE'' and ``HAZE2'', around the location identified by ``a'' in Fig.\,\ref{fig:figch5:QJT1113_sigma_maps}), and in the proximity of the NPS (with ``HAZE3'', around ``b'' in Fig.\,\ref{fig:figch5:QJT1113_sigma_maps}). We report in Table\,\ref{tab:noise_QJT} the typical noise levels of the new QUIJOTE maps in a $1^{\circ}$ FWHM beam, including wide-survey and raster data, and we compare these values with the noise levels achieved with wide-survey data alone. The numbers have been obtained by computing the median value of the uncertainty maps  within circles with a radius of $5^{\circ}$, centred in two different positions: close to the Galactic centre at $(l,\, b) =(5^{\circ},\,0^{\circ})$ (a), and in the proximity of the NPS at $(l,\, b) =(40^{\circ},\,20^{\circ})$ (b). We observe that the raster scan data improve the noise level, both in intensity and polarization, by a factor $\sim3$ in the Galactic centre, and by $\sim1.2$--$1.5$ in the NPS region. The $\sigma$-maps are scaled-up by a multiplicative factor obtained from the QUIJOTE weight maps ($\sigma=1/\sqrt{w}$) that accounts for the $1/f$ noise contribution, which is characterized by the half-mission wide-survey null-test, as described in Sect.~4.1 of \citet{mfiwidesurvey}. The factors are: $f$ = 5.214 (\textit{I}), 1.333 (\textit{Q}), 1.335 (\textit{U}) at 11\,GHz, and \mbox{$f$ = 4.682 (\textit{I}),} 1.320 (\textit{Q}), 1.321 (\textit{U})  at 13\,GHz. Finally, given the integration time in the same area, we found that the estimated global noise level correspond to an instantaneous sensitivity of $\sim 0.42$--$0.44\,{\rm mK}\sqrt{{\rm s}}$ in intensity, and $\sim 0.13$--$0.14\,{\rm mK}\sqrt{{\rm s}}$ in polarization.

In the analysis presented in this paper, we use the QUIJOTE-MFI maps convolved to $1^{\circ}$ angular resolution with the window function of QUIJOTE-MFI (G\'enova-Santos et al., in preparation), and degraded to $N_{\mathrm{side}}=64$ (pixel resolution $\sim0.9^{\circ}$). In order to obtain uncertainty maps at this resolution, we performed 100 white noise realizations, whose amplitude is given by the $\sigma$ maps presented above. We applied the same smoothing and degradation of the data to the noise simulations, and we computed the standard deviation of the noise realizations to obtain a smoothed and degraded variance map. We also tested different methodologies to determine the variance maps,  accounting for $1/f$ noise correlation at large angular scales, and we obtained no significant differences in the final results.   Finally, the noise of the 11\,GHz and 13\,GHz maps of QUIJOTE is partially correlated between the frequency channels. The correlation of the noise in intensity is $\rho=0.76$ and in polarization it is $\rho=0.35$ (see Sect.~4.3.3 in \citealp{mfiwidesurvey}). We account for this correlation in this work, and for an overall calibration uncertainty of 5\,\% (for more details see Sect.~5 of \citealp{mfiwidesurvey} and G\'enova-Santos et al., in preparation). 

\begin{table}
\centering
\begin{tabular}{c|c|c|c|c} \hline
 & HAZE & HAZE\,2 & HAZE\,3 & $\rho$-Ophiuchi \\\hline
$(l,b)$ & $(16^{\circ}, 2^{\circ})$ & $(352^{\circ}, 22^{\circ})$ & $(37^{\circ}, 13^{\circ})$ & $(352^{\circ}, 16^{\circ})$ \\
$\Delta az$ & $47^{\circ}$ & $33^{\circ}$& $86^{\circ}$& $18^{\circ}$ \\
$el$ & $30^{\circ}$-- $40^{\circ}$ & $32.5^{\circ}, 36^{\circ}, 37^{\circ}$ & $39^{\circ}, 62^{\circ}$ & $32^{\circ}, 33^{\circ}, 37^{\circ}$ \\
Time [h] & 742.5 & 98.8 & 494.4 & 258.7\\
\hline
\end{tabular}
\caption{General characteristics of the raster scan observations for the four fields used in this work. We report the central coordinates of the fields (in Galactic coordinates), the typical length of the azimuth rasters, the approximate elevation at which they were taken, and the total number of hours of the observations.}  \label{tabch5:hours}
\end{table}

\begin{table}
    \centering
    \begin{tabular}{cccccccc}
    \hline
    Map & Area &  \multicolumn{3}{c}{11\,GHz [$\mu K_{\mathrm{CMB}}/1^{\circ}$]} & \multicolumn{3}{c}{13\,GHz [$\mu K_{\mathrm{CMB}}/1^{\circ}$]} \\     \cmidrule(lr){3-5}     
    \cmidrule(lr){6-8}
    & &  $I$ & $Q$ & $U$ & $I$ & $Q$ & $U$ \\  \hline
    \multirow{2}{4.9em}{Rasters +  wide-survey} & a & 47.0 & 19.6 & 19.7 & 37.2 & 18.4 & 18.7 \\
    & b & 82.3 & 28.5 & 28.4 & 57.1 & 24.8 & 24.8 \\ \hline
    \multirow{2}{4.9em}{wide-survey} & a  & 145.5 & 57.7 & 57.9 & 140.6 & 57.4 & 57.6 \\
     & b &  94.7 & 44.0 & 44.4 & 80.7 & 38.5 & 38.7 \\
    \hline
    \end{tabular}
    \caption{Noise level of the QUIJOTE maps of the rasters plus wide-survey data (top two lines), and of the wide-survey data alone (bottom two lines), in a $1^{\circ}$-FWHM beam.
    This is obtained as the median of the uncertainty maps (shown in Fig.\,\ref{fig:figch5:QJT1113_sigma_maps}), computed within a $5^{\circ}$ radius circle centred in two different positions: close to the Galactic centre (a) at $(l,b) =(5^{\circ},0^{\circ})$, and in the proximity of the NPS (b) at $(l, b) =(40^{\circ},20^{\circ})$. }
    \label{tab:noise_QJT}
\end{table}

\subsection{Ancillary data}\label{secch5:data-overview}

We use as ancillary data the \textit{WMAP} 9-year maps \citep{WMAPmaps} in the K, Ka, Q and V bands (central frequencies 22.8, 33.0, 40.7 and 60.7\,GHz) and the \texttt{NPIPE} \textit{Planck}-LFI maps \citep{NPIPE2020}, at 30, 44 and 70\,GHz (central frequencies 28.4, 44.1, and 70.4\,GHz). In addition, for the analysis in polarization, we include the S-PASS data \citep{SPASSmaps} at 2.3\,GHz. As in previous works \citep{PLAerXX,PLAirXV}, in order to take into account the uncertainty due to, for example, beam asymmetries and colour corrections, we adopt a calibration uncertainty of 3\% in \textit{WMAP} and in \textit{Planck}-LFI, and of 5\% in S-PASS \citep{SPASSmaps}.  We summarize the main data parameters in Table\,\ref{tabch5:summary_dsataseta}. 

All the maps are smoothed to the common angular resolution of $1^{\circ}$, and degraded to $N_\mathrm{side}=64$, which corresponds to a pixel size of $\sim0.9^{\circ}$ and prevents noise pixel-to-pixel correlation.  For the computation of spectral indices we apply colour corrections  by using the python code \texttt{fastcc} presented in \cite{fastcc}, which includes colour correction models for different datasets including QUIJOTE-MFI, \textit{Planck}, and \textit{WMAP}. No colour correction for S-PASS data is applied.\footnote{S-PASS uses a spectral back-end which allows to flatten the bandpass and to reduce the necessary colour correction.}

A collection of figures representing the full dataset is shown in appendix \ref{secA:maps} (Fig.\,\ref{figch5:maps_int}, \ref{figch5:maps_q}  and \ref{figch5:maps_u} for, respectively, $I$, $Q$ and $U$). In Fig.\,\ref{figch5:maps_pol_ang_data} we also show the debiased polarization amplitude maps ($P_{\mathrm{MAS}}$, given by Eq.~\ref{eq:pmas}) at some selected frequencies: S-PASS at 2.3\,GHz, QUIJOTE 11\,GHz, \textit{WMAP} K-band and \textit{Planck} 30\,GHz. From a quick visual inspection of the polarization amplitude and polarization angle maps in Fig.\,\ref{figch5:maps_pol_ang_data}, we can see that, while the QUIJOTE, \textit{WMAP} and \textit{Planck} maps show very high similarity in the synchrotron polarized structures and angles, at the S-PASS frequency there is evident depolarization in the Galactic plane, up to $|b| \approx 15$\,deg. We can also see a rotation of the polarization angle up to high Galactic latitudes, which is produced by Faraday rotation along the line-of-sight \citep{SPASSmaps,gradP_SPASS}.  Faraday rotation is important at 2.3\,GHz in some of the regions that are studied in this work. We therefore correct the S-PASS maps for Faraday rotation as described in appendix~\ref{sec:FR}, and shown in Fig.\,\ref{fig:SPASS_FRang}. QUIJOTE, \textit{WMAP} and \textit{Planck} data are not corrected for Faraday rotation, since the effect in the regions we are studying is expected to be negligible at these frequencies, i.e., within the uncertainty of the calibration angle (see e.g., \citealp{Hutschenreuter2022} and \citealp{Vidal2015}, where Faraday rotation is shown to be lower than 1$^{\circ}$ in \textit{WMAP}, everywhere except in the Galactic centre). 

For the subsequent analysis, in order to assign uncertainties to the data,  we use uncertainty maps at the same angular and pixel resolution as for the maps ($N_{\mathrm{side}}=64$ and 1$^\circ$ resolution). 
The variance maps are generated with Monte Carlo realizations, as described in more detail in Peel et al. (in preparation). 

\subsection{Selection of the regions}\label{secch5:regions}

\begin{figure}
    \centering
    \includegraphics[width=0.5\textwidth]{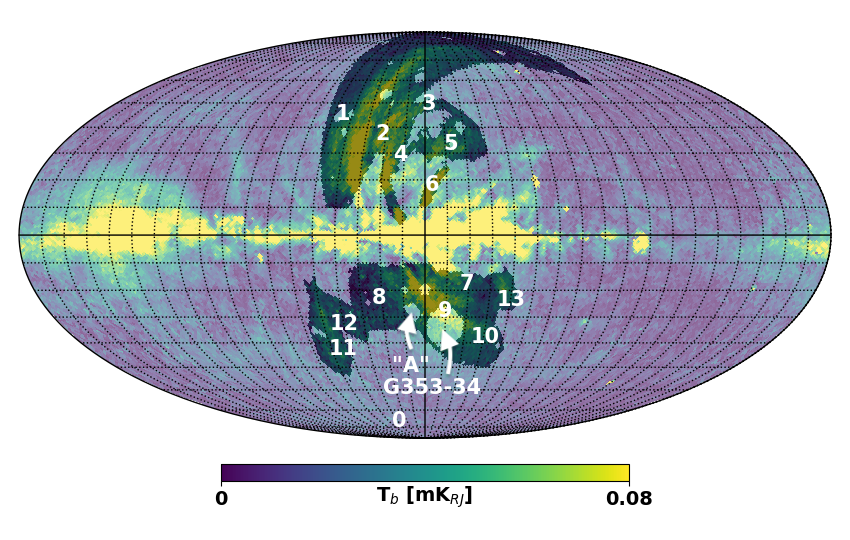}
    \caption{Selected regions for the analysis  overlaid on the \textit{WMAP} K-band polarization amplitude map. The regions are listed and described in Tab.\,\ref{tab:regions}.
    }
    \label{fig:regions_ttplot_data}
\end{figure} 

\begin{table*}
    \centering
    \begin{tabular}{clll}
    \toprule
         Region & Description & Coordinates & Reference\\\midrule
         0  & High-latitudes sky observed by each survey &&  \\
         \specialrule{0.1pt}{1pt}{1pt}
         1  & North Polar Spur (NPS)  & $(l,\,b)\sim(30^{\circ},45^{\circ})$ & \cite{Large1962}\\
        \midrule
        2  & Bright polarized feature between the NPS and the Haze filament  & $(l,\,b)\sim(30^{\circ},35^{\circ})$ & Defined in this work \\
         \specialrule{0.1pt}{1pt}{1pt}
         3  & Filament surrounding the northern \textit{Fermi} bubble in $\gamma$-rays & $(l,\,b)\sim(7^{\circ},47^{\circ})$ & \multirow{2}{15em}{\cite{Vidal2015} (region IX)\\ \cite{pla2015XXV}}\\
           & & & \\
         \specialrule{0.1pt}{1pt}{1pt}
         4  & Polarized structure below the Haze filament & $(l,\,b)\sim(20^{\circ},27^{\circ})$ & Defined in this work \\
         \specialrule{0.1pt}{1pt}{1pt}
         5  & \nhb{} & $(l,\,b)\sim(-4^{\circ},31^{\circ})$ & \cite{SpassHaze}\\
         \specialrule{0.1pt}{1pt}{1pt}
         6  & The Galactic Centre Spur (GCS) & $(l,\,b)\sim(5^{\circ},16^{\circ})$ & (e.g.,) \cite{Vidal2015}\\
         \specialrule{0.1pt}{1pt}{1pt}
         7  & Rectangle enclosing the \shb & 
         \multirow{2}{10em}{ $|l|<35^{\circ}$\\ $-35^{\circ}<b<-10^{\circ}$} & \citealp{PlanckHaze} \\
            &  &  &  \\
         \specialrule{0.1pt}{1pt}{1pt}
         8  & Same as region 7, restricted to the QUIJOTE map area &  \multirow{3}{10em}{ $|l|<35^{\circ}$\\ $-35^{\circ}<b<-10^{\circ}$\\ $-32^{\circ}\lesssim\delta\lesssim-10^{\circ}$} & Defined in this work \\
            & & & \\ 
            & & & \\ 
         \specialrule{0.1pt}{1pt}{1pt}
         9  &\shb{} & $(l,\,b)\sim(-1^{\circ},-30^{\circ})$ & \cite{SpassHaze}\\
         \specialrule{0.1pt}{1pt}{1pt}
         10  & \multirow{2}{20em}{Region 9 excluding Faraday depolarized regions: \\"A" and G353.34} &  $(l,\,b)\sim(-1^{\circ},-30^{\circ})$ & 
         \multirow{2}{12em}{\cite{SpassHaze} \\ \cite{gradP_SPASS}}\\
             & &  \multirow{2}{15em}{"A": $(l,\,b)\sim(20^{\circ},-52^{\circ})$\\ G353.34: $(l,\,b)\sim(30^{\circ},-62^{\circ})$} & \\
             & & & \\
         \specialrule{0.1pt}{1pt}{1pt}
         11  & Western eRosita bubble or South Polar Spur (SPS) & $(l,\,b)\sim(58^{\circ},-45^{\circ})$ & \multirow{3}{15em}{\cite{Predehl2020} \\ \cite{Vidal2015} (region VIIb)\\ \cite{pla2015XXV}} \\
         & & & \\
         & & & \\
         \specialrule{0.1pt}{1pt}{1pt}
         12 & Faint polarized spur with unknown origin  & $(l,\,b)\sim(53^{\circ},-35^{\circ})$ & Defined in this work \\
         \specialrule{0.1pt}{1pt}{1pt}
         13  & Eastern eRosita bubble  & $(l,\,b)\sim(-29^{\circ},-26^{\circ})$ & \cite{Predehl2020} \\
    \bottomrule
    \end{tabular}
    \caption{List and description of the regions selected for this work.  See Fig.\,\ref{fig:regions_ttplot_data} for visualisation.}
    \label{tab:regions}
\end{table*}

We identified thirteen regions of particular interest for the study of the Haze. Six of them are within the footprint of our QUIJOTE map. They are shown in Fig.\,\ref{fig:regions_ttplot_data} and are listed and described in Tab.\,\ref{tab:regions}. 
We use the numbering in the table to identify specific regions throughout this work.

Regions 7 and 8 have been selected in order to reproduce the analysis of the diffuse Haze in intensity presented in \cite{WMAPHaze,PlanckHaze}. We extend the same analysis including QUIJOTE data, using for the first time also polarization data. 

Region 3 has been used in order to reproduce the result by \citet{pla2015XXV}, who measured the polarization spectral index of the filament. We repeated here the analysis also using QUIJOTE data. We also define as regions 2 and 4 two features of diffuse emission extending, respectively, outside and inside the border of region 3.  

Regions 5, 9 and 10 have been defined to identify the polarized radio plumes observed by S-PASS \citep{SpassHaze}, in order to carry out a spectral index analysis using QUIJOTE and ancillary data.

Region 6 is the Galactic Centre Spur (GCS, \citealp{Vidal2015}), which is a very bright polarized feature connected to the Galactic Center. Its relation with the Haze is still unclear.

Regions 11 and 13 have been identified as the borders of the eROSITA bubbles \citep{Predehl2020}, and have been used for a spectral index analysis using QUIJOTE and ancillary data.

Region 12 is defined as an area with some faint diffuse polarized emission with unknown origin. It was identified during the analysis of the polarization template fitting residuals (see Fig.\,\ref{fig:res_p}).

Finally, region 0 is the sky observed by each survey, after excluding the Galactic plane with the mask described in \ref{secch5:templates}, and region 1 corresponds to the North Polar Spur (NPS, \citealp{Large1962}). These two regions are used for comparison purposes.

\section{Methodology}\label{sec:methodology}
In this section, we describe the two methodologies that are applied in this work: the template fitting procedure, both in intensity and in polarization, and the correlation T-T plots analysis in polarization.
\subsection{Template fitting}
\label{secch5:templ_fitting}

In order to isolate the diffuse emission of the Haze from the other Galactic foregrounds, we apply a template fitting technique, following the same formalism as in \cite{Fink2004, WMAPHaze, PlanckHaze}. This methodology relies on the assumption that each frequency map is a linear combination of several templates, which spatially trace the Galactic emission of different mechanisms, such as the synchrotron, free-free, thermal dust and AME. This can be represented analytically as:
\begin{equation}
    d_{\nu}=a_{\nu} \cdot \textbf{P}_{\nu},
\end{equation}
where $d_{\nu}$ is the map at frequency $\nu$,  $\textbf{P}_{\nu}$ is a the template matrix that contains one template map per column, estimated at frequency $\nu$, and $a_{\nu}$ is a vector of coefficients indicating the amplitude of the templates. The template fitting problem consists in determining the amplitudes $a_{\nu}$ that provide the best description of the data with the templates in $\textbf{P}_{\nu}$. We solve the problem with a maximum likelihood approach, by applying an extended formalism to include the correlation between different templates.\footnote{This formalism was developed and applied for the radio/microwaves map-making problem. See for example \cite{Madam}, or \cite{destriper}} The logarithm of the posterior of this problem, including priors for the fitted amplitudes $a_{\nu}$ is given by:
\begin{equation}
    \ln \mathcal{P} \propto (d-a\cdot \textbf{P})^TC_{\rm w}^{-1}(d-a\cdot \textbf{P})+(a-a_{0})^TC_{\rm a}^{-1}(a-a_{0})+c,
    \label{chi2}
\end{equation}
where we neglect the frequency subscript $\nu$ for brevity. Here, $C_{\rm w}$ is the noise covariance matrix of the data, $a_{0}$ is the central value of the amplitude priors, $C_{\rm a}$ is the covariance matrix of the template amplitudes, and $c$ is a global constant.
The solution of equation~\ref{chi2} is:
\begin{equation}
    a=(\textbf{P}^TC_{\rm w}^{-1}\textbf{P}+C_{\rm a}^{-1})^{-1}\cdot(P^TC_{\rm w}^{-1}d+C_{\rm a}^{-1}a_{0}).
    \label{eq:a_priors}
\end{equation}
By means of the term $C_{\rm a}^{-1}a_{0}$, we can apply priors on the fitting of the foreground templates. In particular,  the off-diagonal elements of $C_{\rm a}$ allow us to introduce in the fitting the degree of correlation between the templates, which is measured for synchrotron and dust to be at the order of $20$--$40$\,\%, with some evident spatial variation (\citealp{peel2012,2015JCAP...12..020C,Krachmalnicoff2018}). 

This improved template fitting technique has been tested with simulations based on the foreground templates and frequency scaling used in this work (see Sect.\,\ref{secch5:templates}). We observed that a more precise separation of the foregrounds is achieved by applying priors that account for their spatial correlation. 
In particular we noticed that, at low frequencies, where the dust component is subdominant but spatially correlated with the synchrotron, the separation of synchrotron and dust is significantly improved by applying priors as in Eq.\,\ref{eq:a_priors}. Simulations have also been used to check for biases in the results, when including or excluding priors in the fitting procedure. We observed that the results on the spectrum of the Haze are not significantly different in the two cases, while the spectra of the foregrounds components, especially that of the synchrotron, is affected by significant biases if priors are not adopted. We concluded that the use of priors allows us to have better control on the fitting of the foreground components, while not significantly affecting the results on the spectrum of the Haze.

\subsubsection{Templates}
\label{secch5:templates}

\begin{figure*}
    \centering
    \includegraphics[width=0.32\textwidth]{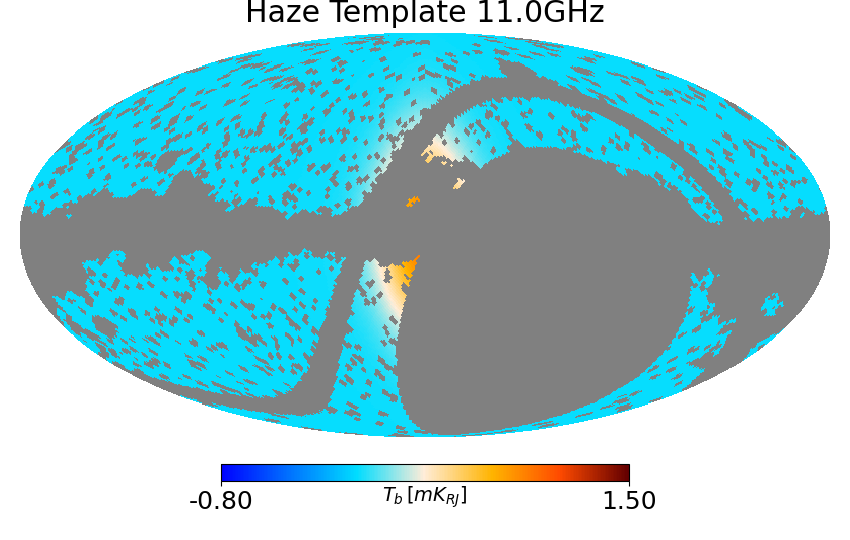}
    \includegraphics[width=0.32\textwidth]{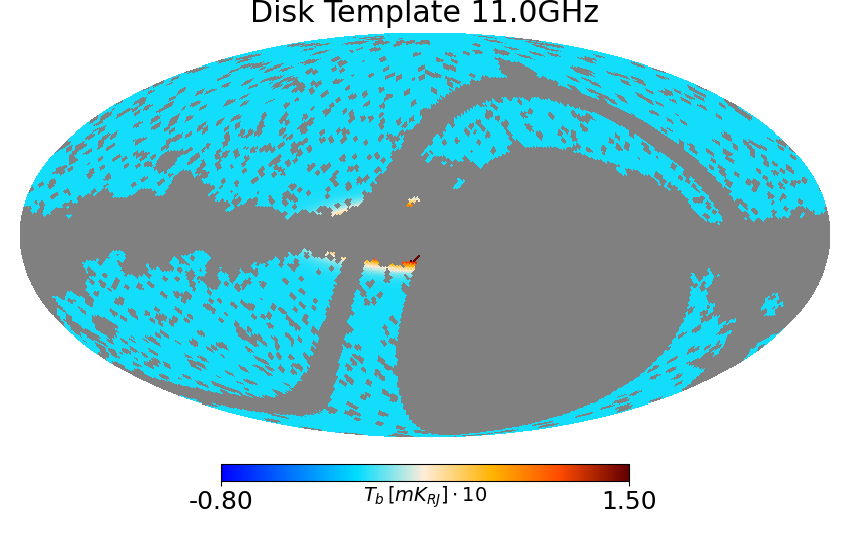}
    \includegraphics[width=0.32\textwidth]{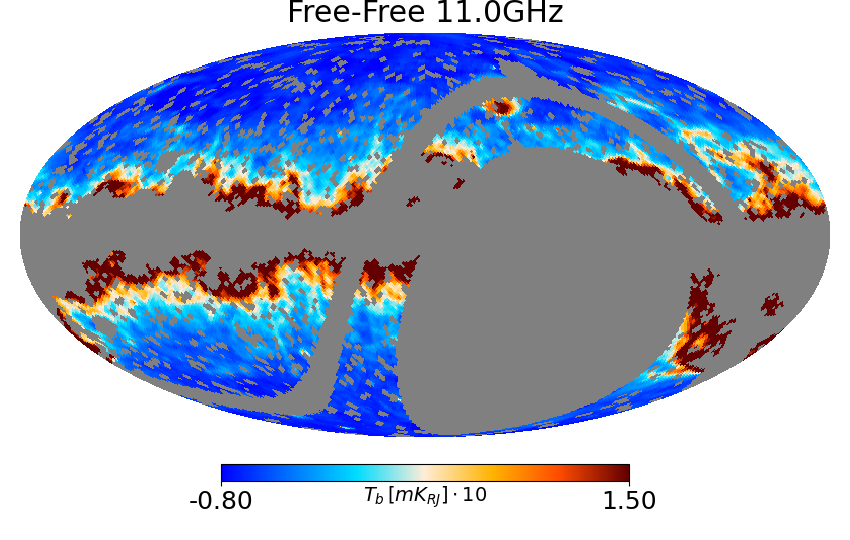}
    \includegraphics[width=0.32\textwidth]{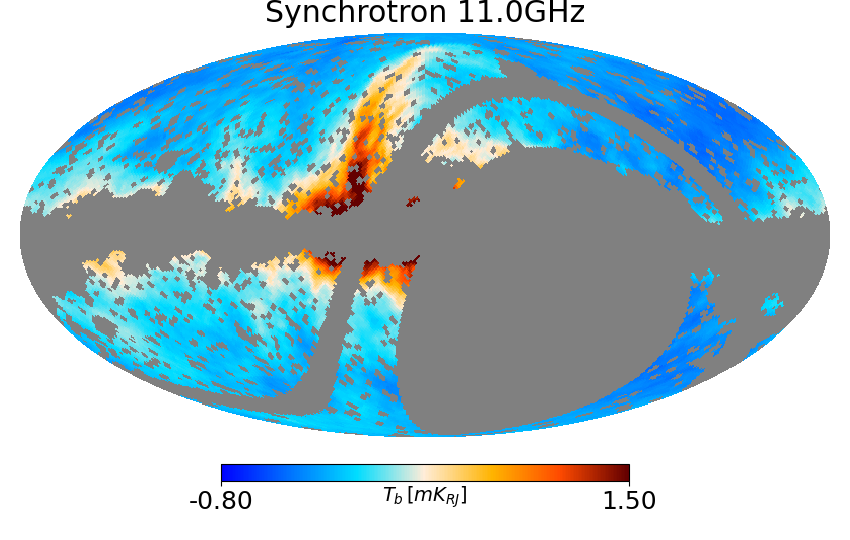}
    \includegraphics[width=0.32\textwidth]{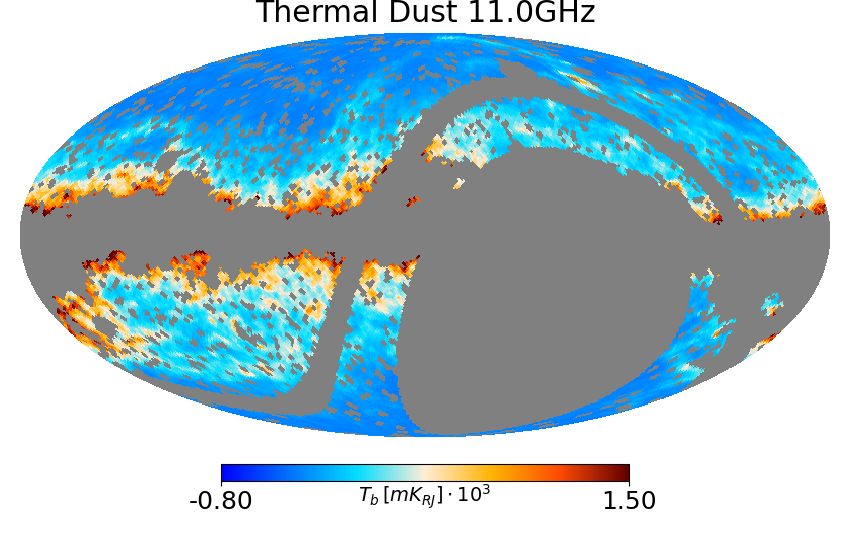}
    \includegraphics[width=0.32\textwidth]{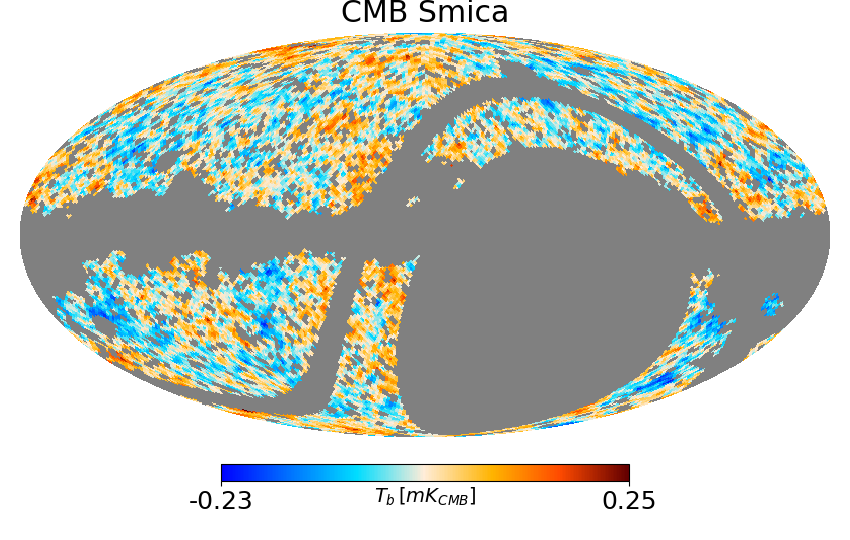}
    \caption{Intensity template maps at 11\,GHz at $N_{\mathrm{side}}=64$. They are: a simple model of the Haze as described in Sect.~\ref{secch5:templates} (top left), a disk template for the Galactic plane diffuse synchrotron emission  (multiplied by 10 for display purposes; top centre), free-free (multiplied by 10 for display purposes; top right), synchrotron (bottom left), dust (multiplied by $10^3$ for display purposes; bottom centre), which is used to fit both thermal dust and AME, and the CMB anisotropies (bottom right). The maps are in units of mK Rayleigh-Jeans, and have had the mean subtracted. The grey area represents the mask that is used for the analysis, which is a combination of the QUIJOTE sky coverage with the free-free and CMB mask, as described in Sect.~\ref{secch5:templates}.}
    \label{figch5:templates_int}
\end{figure*}

\begin{figure*}
    \centering
    \includegraphics[width=0.32\textwidth]{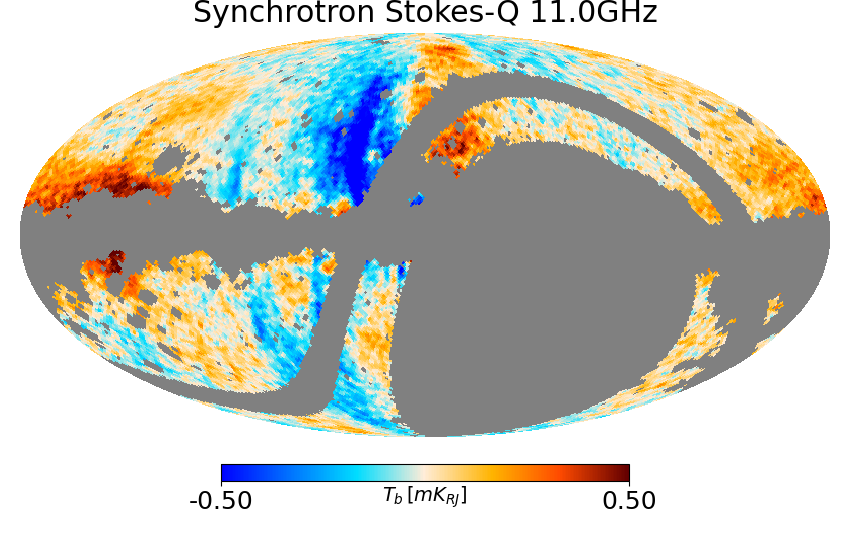}
    \includegraphics[width=0.32\textwidth]{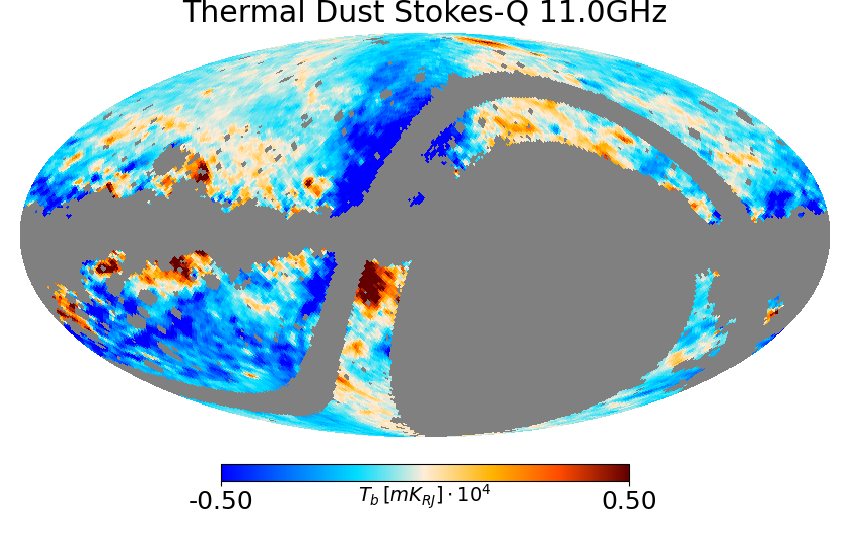}
    \includegraphics[width=0.32\textwidth]{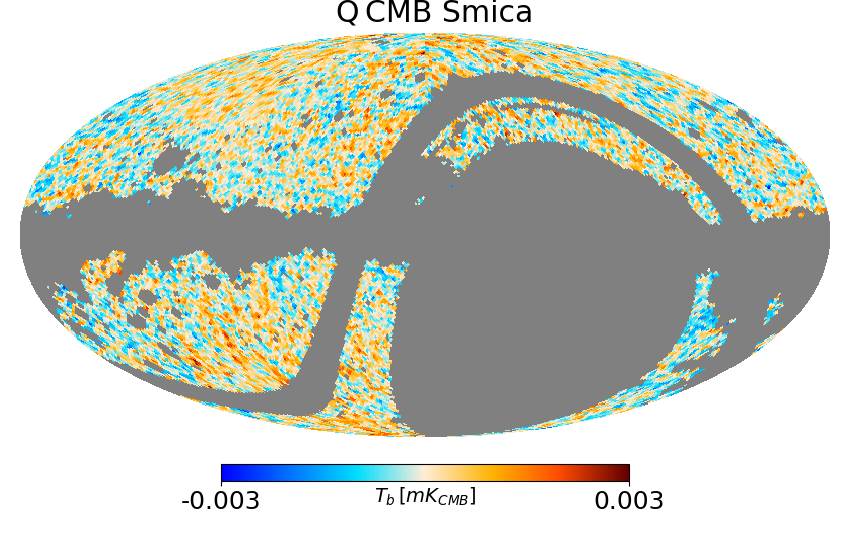}
    \includegraphics[width=0.32\textwidth]{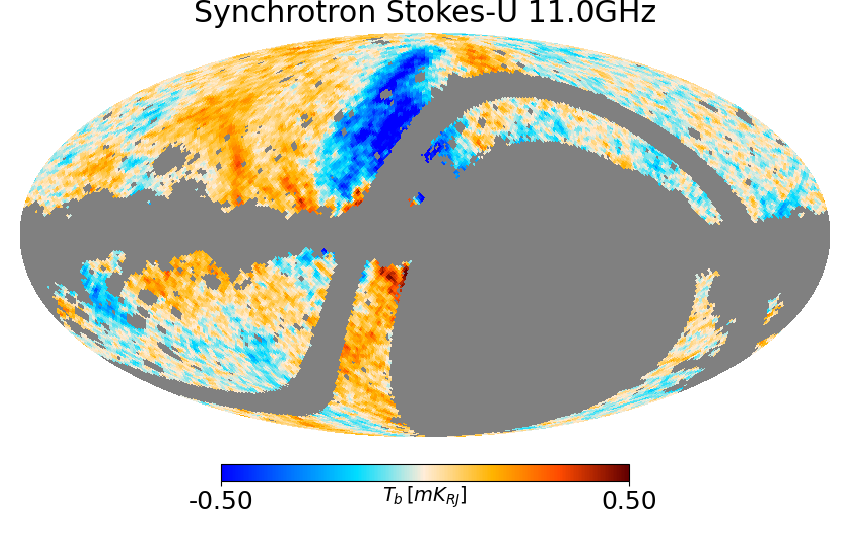}
    \includegraphics[width=0.32\textwidth]{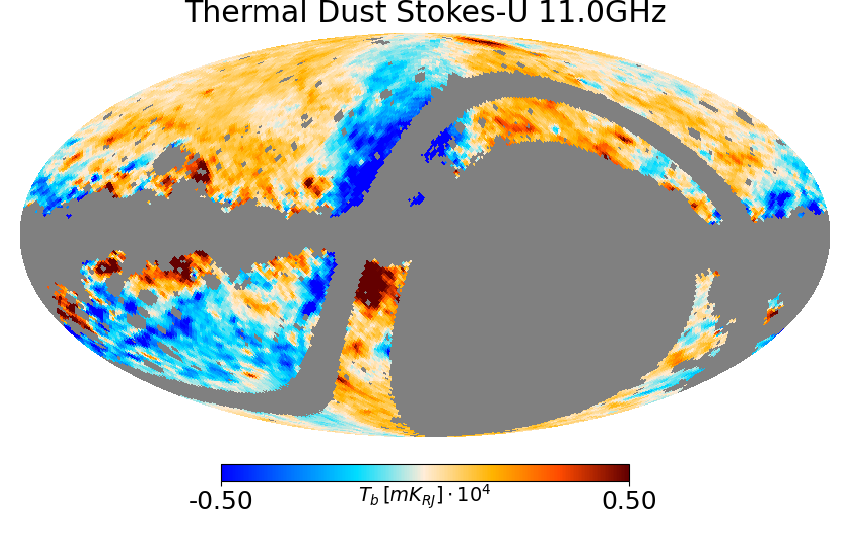}
    \includegraphics[width=0.32\textwidth]{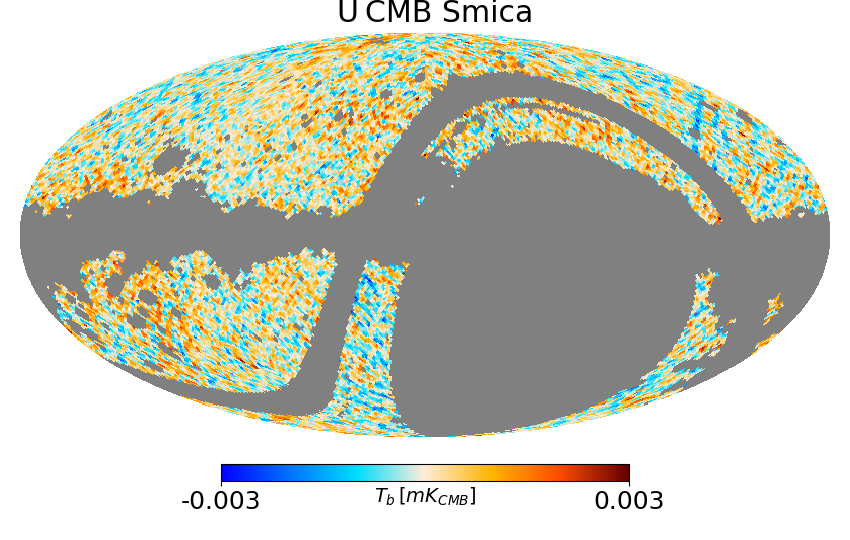}
    \caption{Polarization template maps at 11\,GHz, of Stokes $Q$ (top) and $U$ (bottom). They are, from left to right: synchrotron, thermal dust (multiplied by $10^4$ for display purposes) and the CMB. The maps are in units of mK Rayleigh-Jeans, and are mean corrected. The grey area represents the mask that is used for the analysis, which is a combination of the QUIJOTE sky coverage with the free-free and CMB mask, as described in Sect.~\ref{secch5:templates}.}
    \label{figch5:templates_pol}
\end{figure*}

The fitting is performed in intensity and polarization, using independently the $I$ map, and the $Q$ and $U$ Stokes parameters maps simultaneously for polarization, assuming a negligible \textit{Q} and \textit{U} correlation. All the templates are convolved to $1^\circ$ angular resolution and degraded to $N_{\mathrm{side}}=64$ in order to avoid pixel-to-pixel correlation, as we did for the data (see Sect.~\ref{sec:data}). The intensity and polarization templates are shown, respectively, in Figs.\,\ref{figch5:templates_int} and \ref{figch5:templates_pol}. A detailed description follows in this section.

\paragraph*{Synchrotron.}
The full-sky intensity map by \cite{haslam1982}, at $408$\,MHz, is dominated by synchrotron emission, and it is only marginally contaminated by free-free along the Galactic plane and in bright free-free sources (e.g., M42).  This makes the $408$\,MHz map a good tracer of diffuse synchrotron emission in intensity. We use the reprocessed version of this map by \cite{Remazeilles2015} as a template, and we scale\footnote{The frequency scaling of a template map is usually irrelevant for template fitting. Indeed, given the spatial morphology of the template, we fit for a global amplitude. However, in order to assign priors as explained in Sect.\,\ref{secch5:priors}, frequency scaling is needed.} it in frequency using a power-law spectrum assuming a spatially-constant spectral index $\beta_{\rm s}=-3.1$ across the full sky. 
In addition, as indicated by \cite{2012ApJ...750...17D}, the cosmic ray propagation length is energy dependent, and this results in a synchrotron radiation that is more extended around the  Galactic disk at $408$\,MHz compared with the higher frequencies (like QUIJOTE, \textit{Planck} and \textit{WMAP}). In order to trace this excess at low frequency, and following \cite{2012ApJ...750...17D} and \cite{PlanckHaze}, we adopt an elliptic Gaussian template centred in the Galactic centre, with extension $(\sigma_l,\sigma_b)=(\pm20^\circ, \pm5^\circ)$. The diffuse synchrotron and the disk-like synchrotron excess are fitted independently with two separate templates.

In polarization, we use the 2018 Stokes \textit{Q} and \textit{U} \texttt{Commander}\footnote{\texttt{Commander} is a software developed for the component separation of \textit{Planck} data. It consists of a pixel based Bayesian parametric method (MCMC Gibbs sampling algorithm), aimed to fit the parameters describing different Galactic foreground components. See \cite{Eriksen2004,Eriksen2008} for more details.} synchrotron solution \citep{Planck_diffcomp_sep_2018}, scaled to each central frequency with a power-law with a spectral index \mbox{$\beta=-3.1$,} which is assumed to be constant across the sky.

\paragraph*{Thermal dust and AME}
Thermal and AME are two distinct foreground components produced by dust grains. The thermal dust follows a modified black body spectrum that shows up mainly at high frequencies ($\nu$>100\,GHz), while the spinning dust is significant at intermediate frequencies ($10\,{\rm GHz} \lesssim \nu \lesssim 60\,{\rm GHz}$). The carriers of the AME have not been unequivocally identified yet, but the most accredited hypothesis to date is that AME is produced by the rotation of small dust grains (for a review see \citealp{DikinsonAME}). 

We could use two independent templates to fit thermal dust and AME, using the \texttt{Commander} solution \citep{pla2015X} for the two components. However, AME and thermal dust are highly correlated, and a simultaneous fit of the two components could be affected by strong degeneracy. In addition, we noticed that the \texttt{Commander} AME map presents an excess of emission with a shape similar to that of the Haze. There is the possibility that a fraction of the Haze emission leaked into this map. Moreover, \cite{pla2015XXV} reported that the degeneracy between the AME and free-free components could affect the stability of the \texttt{Commander} AME solution, due to the lack of low-frequency information. Therefore, in order to perform a blind and unbiased fit of the foregrounds we decided not to use the \texttt{Commander} AME map, fitting the combination of thermal dust and AME with a single template.  We adopt the 2015 \texttt{Commander} solution for thermal dust, scaled at each central frequency with the modified black body spectrum of thermal dust reported in \cite{pla2015X}. Due to the dust and AME correlation, this template will capture, in addition to the thermal dust component, the AME emission at intermediate frequencies ($\sim$20--60\,GHz).  Note also that, thanks to the fact that the AME ($\nu \lesssim 60$\,GHz) and the thermal dust ($\nu \gtrsim 800$\,GHz; \citealp{pla2015X}) emissions do not overlap in frequency, even if we use a single template to fit the two components, they are easily distinguishable in the frequency spectrum.

In this work we assume no polarized AME, which is well justified given the observational constraints that set the AME polarization to be $\lesssim$1\% (e.g., \citealp{Rubino12,W44,DikinsonAME}). Therefore no AME is fitted in polarization.
Thermal dust instead is typically 5--10\,\%  polarized \citep{Dickinson2011,pla2015X,Pla2016XXXVIII}.  We fit therefore the polarized dust emission using the $Q$ and $U$ 2018 \texttt{Commander} thermal dust maps \citep{Planck_diffcomp_sep_2018} as templates, after scaling to each central frequency as indicated in \cite{pla2015X}.

\paragraph*{Free-free.}
We construct the free-free intensity template using the H$\alpha$ map by \citet{Finkbeiner2003}. We correct the H$\alpha$ map for dust absorption by applying the methodology of \citet{Dickinson2003}, and using the reddening $E(B-V)$ map\footnote{\url{https://irsa.ipac.caltech.edu/data/Planck/release_1/all-sky-maps/previews/HFI\_CompMap\_DustOpacity\_2048\_R1.10/}} of \textit{Planck} \citep{Pla2013XI}. We assume uniform mixing between gas and dust by setting an effective dust fraction along the line of sight\footnote{We also tried $f_{\rm d}=0.33$, but this change did not affect the resulting Haze morphology and spectrum.}  $f_d=0.5$, an average electron temperature $T_e=7000$\,K across the full sky, and we scale the corrected H$\alpha$ map from Rayleigh (R) to $\mu$K, at each central frequency, by computing the conversion factor with Eq. (11) in \citet{Dickinson2003}. Despite these approximations, what is important here is to construct a good enough tracer of the spatial distribution of free-free emission, independently from the absolute scale. With this aim, applying a good correction of dust absorption is important.

This template provides a sufficiently good approximation of the free-free in the sky, except for the regions with high dust absorption. Furthermore, we expect large fluctuations of the gas temperature in the brightest H$\alpha$ regions, which can produce some inaccuracies in the template \citep{Dickinson2003, PlanckHaze}. In order to avoid such problematic regions, we mask the pixels with absorption larger than one magnitude ($2.51 \cdot E(B-V)>1$\,mag), or with H$\alpha$ intensity greater than 10\,R. The free-free has negligible polarization, therefore it is fitted only in intensity.

\paragraph*{CMB.}
The 2018 \texttt{SMICA}\footnote{Spectral  Matching  Independent  Component  Analysis (SMICA) is one of the methods that was implemented for the component-separation of \textit{Planck} data. It is based on a linear combination between the \textit{Planck} frequency channels, using weights that depend on the multipole. See \cite{Cardoso2008} for more details.} CMB map \citep{Planck_diffcomp_sep_2018} is subtracted from each frequency map, both in intensity and in polarization,  at $1^\circ$ angular resolution. As discussed in \cite{2012ApJ...750...17D}, the foreground contamination of the CMB map could produce a bias in the determination of the Haze spectrum. However, the last version of maps produced with the \textit{Planck} data provide now a high quality CMB map. We assume therefore that the CMB bias mentioned above is negligible as compared with other sources on uncertainty. In order to confirm that, we repeated the analysis using the \texttt{Commander} CMB map, obtaining compatible results on the Haze separation. 

\paragraph*{The Haze.}
Following \cite{WMAPHaze} and \cite{PlanckHaze}, we include  a template that approximately traces the emission of the Haze in the fitting of the intensity. Even if we do not have a precise characterization of the spatial distribution of the Haze, an approximated template is needed in order to avoid a bias in the fit of other foreground templates. We use a Gaussian ellipse in Galactic coordinates, centred in the Galactic centre, and with major axes perpendicular to the Galactic plane line ($b=0^\circ$). The minor and major axes are, respectively, $\sigma_{l}=15^\circ$ and $\sigma_{b}=25^\circ$. The template has the same unitary amplitude at different frequencies. 

\paragraph*{Monopole and dipole.}
In order to overcome any possible issue related with zero levels, we subtract the average value of the unmasked pixels from the maps and from the templates. In addition, we fit  a monopole component at each frequency in order to adjust any residual zero level mismatch.
Finally, from the residual maps at frequencies $\nu>40$\,GHz, we noticed a residual dipole pattern. For this reason, before applying the template fitting to these maps, we remove the residual dipole with the HEALPix routine \texttt{remove\_dipole}, after masking pixels with  $|b|<20^\circ$ to avoid Galactic contamination. 

\paragraph*{Mask.}
Following \citet{WMAPHaze} and \cite{PlanckHaze}, we mask all the regions where the templates can deviate from the real foreground emission. The mask includes, as described above for the free-free, the regions where the H$\alpha$ emission exceeds 10\,R, or where the dust extinction is larger than 1 magnitude. In addition, we mask the point sources from  the \textit{Planck} LFI catalog \citep{PLA2015XXVI}. We used the mask excluding the LFI compact sources that is available in the Planck Legacy Archive\footnote{The mask used in this work can be found in the PLA: \url{http://pla.esac.esa.int/pla/aio/product-action?MAP.MAP\_ID=LFI_Mask_PointSrc_2048_R2.00.fits }. Relevant information about the mask can be found in the PLA Explanatory Supplement at \url{https://wiki.cosmos.esa.int/planck-legacy-archive/index.php/Frequency_maps\#Masks}.} (PLA). Finally, in order to avoid any possible bias from foreground residuals in the CMB map, we mask the pixels that are outside the confidence region\footnote{The CMB mask used in this work is taken from the fits file containing the CMB map (SMICA, PR3-2018), downloaded from the PLA (\url{http://pla.esac.esa.int/pla}). Relevant information about the mask can be found in the PLA Explanatory Supplement at \url{https://wiki.cosmos.esa.int/planck-legacy-archive/index.php/CMB_maps\#SMICA}.} of the CMB map that we are using.

\subsubsection{Priors}\label{secch5:priors}
Our implementation of the template fitting procedure, which is described in Sect.~\ref{secch5:templ_fitting}, allows us to apply priors on the amplitudes of the foreground templates. The priors are introduced by the vector $a_{0}$, which contains the central values of the prior at frequency\footnote{For brevity in the notation, the subscript $\nu$ is not explicit, keeping in mind that the fitting is always performed at a given frequency.} $\nu$, and by the covariance matrix $C_{\rm a}$. The elements of the covariance matrix are defined as:
\begin{equation}
    C_{{\rm a},ij}=cov(a_i,a_j)=E\left[(a_i-a_{0i})(a_j-a_{0j})\right],
\end{equation}
where $E[\cdot]$ denotes the expected value operator, $a_0$ the expected amplitude, and the indices $i$ and $j$ indicate the foreground maps at the frequency $\nu$ (e.g., $i$=thermal dust, $j$=synchrotron, at 11\,GHz). 
The diagonal elements of $C_{\rm a}$ are:
\begin{equation}
    C_{{\rm a},ii}=cov(a_i,a_i)=\sigma_i^2,
\end{equation}
where $\sigma_i$ is our choice for the width of the Gaussian prior for the amplitude of the template $i$. We assign to the width of the priors the analytic uncertainty on $a_i$ that is obtained by the second derivative of the logarithm of the posterior in Eq. \ref{chi2}, neglecting the priors term ($C_{\rm a}^{-1}=0$).  It is:
\begin{equation}
    \sigma_i^2 = (\textbf{P}_i^T C_{\rm w}^{-1} \textbf{P}_i)^{-1},
    \label{eq:prior_width}
\end{equation}
where $\textbf{P}_i$ is the $i^{th}$ column of the templates matrix $\textbf{P}$, so it is simply the map of the $i^{th}$ template (e.g., $i$=thermal dust). 
The off diagonal elements of $C_{\rm a}$ are:
\begin{equation}
    C_{{\rm a},ij}=cov(a_i,a_j)=\rho_{ij}\cdot\sigma_i \sigma_j,
    \label{eq:ca_offdiag}
\end{equation}
where $\rho_{ij}$ is the correlation between the templates $i$ and $j$. It is known that different foreground mechanisms are spatially correlated (e.g., \citealp{2015JCAP...12..020C}), therefore $\rho_{ij} \ne 0$ and $C_{\rm a}$ is not diagonal. In this work, we assign average values of correlation between the intensity templates of the foregrounds, by computing $\rho_{ij}$ as:
\begin{equation}
    \rho_{ij} = \left<\frac{C_{\ell}^{\textbf{P}_i \times \textbf{P}_j}}{\sqrt{C_{\ell}^{\textbf{P}_i} \cdot C_{\ell}^{\textbf{P}_j})}}\right>_{2<\ell<100},
    \label{eq_rho}
\end{equation}
where $C_{\ell}^{\textbf{P}_i \times \textbf{P}_j}$ is the cross power spectrum between the template maps $i$ and $j$ (e.g., $i$=thermal dust, $j$=synchrotron, at 11\,GHz), while $C_{\ell}^{\textbf{P}_i}$ and $C_{\ell}^{\textbf{P}_j}$ are their auto power spectra. The level of correlation between templates is not the same at large and small angular scales. As $\rho_{ij}$ is a function of the multipole $\ell$, in order to provide an average level of correlation, we compute the mean value of $\rho_{ij}(\ell)$ in the multipole range $2<\ell< 100$. 

We computed the power spectra of Eq.~\ref{eq_rho} with the publicly available code {\sc Xpol}\footnote{\url{https://gitlab.in2p3.fr/tristram/Xpol}} \citep{Tristam2005}, and we used a mask of the full sky, excluding a band in Galactic latitude $|b|<5^\circ$ to mask the brightest Galactic plane emission. 
The averages in the multipole range $2<\ell<200$,  are $\rho_{s,d}=0.30$ for synchrotron and thermal dust, $\rho_{s,f}=0.14$ for synchrotron and free-free, and $\rho_{d,f}=0.26$ for thermal dust and free-free. In polarization we have $\rho_{s,d}=0.20$ for synchrotron and thermal dust, in agreement with \cite{2015JCAP...12..020C}, who measured a correlation $\rho=0.2$ between \textit{Planck} 353\,GHz and \textit{WMAP} 23\,GHz in the multipole range $30<\ell<200$.

Finally we define the central values of the priors. For synchrotron and free-free we use  $a_{0,s}=a_{0,f}=1$, since the template maps are specifically computed at each central frequency, and the expected emission by synchrotron and free-free are the template map themselves. 
For the fitting of the thermal dust and the AME we use a single template, which is the thermal dust of \texttt{Commander}, scaled at the corresponding central frequency, as described in Sect.~\ref{secch5:templates}. Here we assume that AME and the thermal dust are totally correlated, and that we can capture these two components with the same template, with an expected amplitude  $a_{0,d}=1+r$, where $r$ is an average AME to thermal dust ratio. We define $r$ as a representative value of the ratio between the \texttt{Commander} AME and the thermal dust maps, computed (following \citealp{pla2015X}) at the same central frequency $\nu$:
\begin{equation}
    r(\nu)=\left<\frac{\mathrm{AME}(\nu)}{\text{th-dust}(\nu)}\right>,
\end{equation}
where $<>$ indicates the median over the pixels enclosed in the mask described in Sect.~\ref{secch5:templates}. 
 We impose a prior on the total dust amplitude which is centred in $a_{0,d}=1+r$. 
For the rest of the templates, which are the Galactic ellipse of diffuse synchrotron, the monopole and the Haze, we do not want to impose any stringent prior. Therefore we assign to them $a_0=0$ and $\sigma \approx \infty$.

In polarization, we fit a synchrotron and a thermal dust template, separately in $Q$ and $U$. Similarly to intensity, the templates are computed to match the emission of the foreground at the corresponding central frequency, therefore we assign the expected central value with the prior  $a_{0,s}^{Q,U}=a_{0,d}^{Q,U}=1$. The width of the priors are computed with Eq. \ref{eq:prior_width}.
The off-diagonal elements of the covariance matrix are computed as in Eq. \ref{eq:ca_offdiag} and \ref{eq_rho}, giving $\rho_{s,d}^{Q,U}=0.2$.

\subsection{Polarization T-T plots} \label{secch5:tt-plots}
In order to analyze the polarization data with a different and independent technique, we use correlation plots, commonly called T-T plots. This methodology is widely used in the literature (e.g., \citealp{pla2015XXV,fsk2019}), therefore we applied it in order to reproduce results presented in previous works \citep{pla2015XXV,SpassHaze}, and extend them using the new QUIJOTE data. The specifics of the applied methodology are described as follows.

\subsubsection{T-T plots of $P_{\mathrm{MAS}}$} \label{secch5:tt-plots-pmas}
The low frequency polarized foregrounds are dominated by synchrotron radiation, which is described by a power-law spectrum:
\begin{equation}
    d_{\nu} = \left(\frac{\nu}{\nu_0}\right)^\beta \cdot d_{\nu_0},
\label{eq:sync_spec}
\end{equation}
where $d_{\nu}$ are the polarization data at frequency $\nu$, $d_{\nu_0}$ are the polarization data at a reference frequency $\nu_0$, and $\beta$ is the synchrotron spectral index.

It is possible, therefore, to derive the synchrotron spectral index across a coherent region with a simple correlation analysis between the polarized emission of two frequency maps. We can fit a linear dependence of $d_{\nu}$ as a function of $d_{\nu_0}$: 
\begin{equation}
    d_{\nu} =  m\cdot d_{\nu_0}+q,
    \label{eq:lin_fit}
\end{equation}
where $q$ is a relative offset, and the slope $m$ is related to the spectral index $\beta$ (with Eq.\,\ref{eq:sync_spec} and \ref{eq:lin_fit}) as:
\begin{equation}
    \beta= \frac{\ln(m)}{\ln(\nu/\nu_0)}.
    \label{eq:beta}
\end{equation}
The uncertainty on $\beta$ can be derived as the propagation of the uncertainty on $m$, $\sigma_m$, as:
\begin{equation}
    \sigma_{\beta} = \frac{\sigma_m}{m}\frac{1}{\ln(\nu/\nu_0)}.
    \label{eq:sigma_beta}
\end{equation}

This technique is commonly used to compute the spectral index of the polarization amplitude $P=\sqrt{Q^2+U^2}$,  in Rayleigh-Jeans temperature units. However, with $P$ being a positive definite quantity, it is affected by noise bias. Several techniques have been proposed to estimate an unbiased polarization amplitude (\citealp{MAS, Vidal_bias}). In this paper, we use the unbiased polarization amplitude $P_{\mathrm{MAS}}$ by applying the Modified Asymptotic estimator (MAS) presented in \cite{MAS}, as:
\begin{equation}
    P_{\rm MAS} = P-b^2\frac{1-e^{-P^2/b^2}}{2P},
    \label{eq:pmas}
\end{equation}
with
\begin{equation}
    b = \sqrt{(Q\sigma_U)^2+(U\sigma_Q)^2}/P,
\end{equation}
where $P$ is the noise biased polarization amplitude (as defined above), and $\sigma_Q$ and $\sigma_U$ represent the uncertainties on the  measured $Q$ and $U$ parameters.   The uncertainty on  $P_{\rm MAS}$ is given by:
\begin{equation}
    \sigma_{P_{\rm MAS}} = \sqrt{(Q\sigma_Q)^2+(U\sigma_U)^2}/P.
    \label{eq:err_pmas}
\end{equation}
This estimator is unbiased for pixels with signal-to-noise larger than 2. 

\subsubsection{T-T plots of \textit{Q} and \textit{U} combined projection}\label{sec:tt-plots-fsk}

In order to overcome problems related with polarization noise bias in the data, due to zero-level mismatch, and also to variation of the spectral index with the polarization angle of the emission, we apply the technique that was proposed in \cite{fsk2014}. The T-T plot method described in Sect.\,\ref{secch5:tt-plots} has been widely used in previous works (e.g., \citealp{pla2015XXV}), so we have also applied it for the sake of reproducing their results, however we believe that the \cite{fsk2014} method is more reliable, and hence we use that by default.

This methodology does not compute the polarization amplitude $P$, which is affected by noise bias, and allows to marginalize the result over the polarization angle. We make direct use of the \textit{Q} and \textit{U} Stokes maps that, after a projection into a rotated reference, are mixed to construct the data vector $d(\alpha)$: 
\begin{equation}
 d(\alpha) = Q \cos(2\alpha) + U \sin(2\alpha),
 \label{eq:QcUs}
 \end{equation}
where $\alpha$ is the rotation angle.
We can use the data $d(\alpha)$ and Eq.\,\ref{eq:beta} and \ref{eq:sigma_beta} to compute the spectral index as a function of $\alpha$, for a set of 18 angles distributed in the range $\alpha \in [0^{\circ},85^{\circ}]$, in steps of $5^{\circ}$. The resulting (strongly correlated) spectral indices $\beta_i=\beta(\alpha_i)$ are finally averaged with weights:
\begin{equation}
    \beta = \frac{\sum_{i=1}^{18}(\beta_i/\sigma^2_{\beta_i})}{\sum_{i=1}^{18}(1/\sigma^2_{\beta_i})}.
    \label{eq:beta_fin}
\end{equation}
Due to correlation of the estimated $\beta(\alpha_i)$ as a function of the angle, the statistical uncertainty on the final spectral index $\beta$ is taken to be the minimum uncertainty among the 18 measurements: 
\begin{equation}
    \sigma_{\beta}^{\mathrm{stat}} = \min_{i \in [1,18]}\left(\sigma_{\beta_i}\right).
\end{equation}
However, variations of the spectral index as a function of the polarization angle can induce an additional uncertainty on the determination of $\beta$ across a wide region. We can define an intrinsic uncertainty due to this effect as the standard deviation of the $\beta_i$ estimated at different rotation angles, as:
\begin{equation}
    \sigma_{\beta}^{\mathrm{int}} = \mathrm{std}_i\left(\beta_i\right).
\end{equation}
In order to account for the effect that dominates the uncertainty of the spectral index in each particular region (statistical or intrinsic uncertainty),  we adopt as a final uncertainty the maximum between the two estimates of the error:
\begin{equation}
    \sigma_{\beta} = \max\left(\sigma_{\beta}^{\mathrm{stat}}, \sigma_{\beta}^{\mathrm{int}}\right).
\end{equation}

In the process of estimating the spectral index with correlation plots of $d(\alpha)$, we perform the linear fit considering the uncertainties of $d(\alpha)$ in both axes, and, for each angle $\alpha$, we apply colour corrections (see Sect.\,\ref{secch5:data-overview}) in an iterative way, until the spectral index variations are lower than 0.001. The main results of this work, in polarization (Sect.~\ref{secch5:tt-plots-results}), are obtained by applying the methodology described in this section. However, in a few special cases, we compare the resulting spectral indices with those obtained with the more common methodology described Sect.~\ref{secch5:tt-plots-pmas}, the T-T plots of the polarization amplitude $P_{\mathrm{MAS}}$ in order to give strength to the reliability of the result, and show some possible sources of error. In addition, in order to check the robustness of the linear regression of the T-T plots for each angle $\alpha_i$, we compute the posterior distribution of
the spectral index parameters.  Appendix \ref{secA:ttplot_posterior} provides the details and the results of this last check.


\section{Results} \label{sec:results}
Here we report first the results obtained for the Haze with the methodology of template fitting. The aim is to compare with the results from \cite{PlanckHaze} in intensity (in Sect.~\ref{secch5:templatefitting-int}), and to present our results in polarization (in Sect.~\ref{secch5:templatefitting-pol}), which is the main novelty from this work. In this part of the analysis, after fitting the foreground templates across the full sky, we perform a detailed study of the residuals in regions of particular interest among those listed in Sect.~\ref{secch5:regions}.  

Subsequently, in Sect.~\ref{secch5:tt-plots-results} we show the results obtained with the correlation T-T plots in polarization, following the methodology described in Sect.~\ref{sec:tt-plots-fsk}. Also in this case, we concentrate the analysis on the regions that are presented in Sect.~\ref{secch5:regions}.  

As noted in \cite{mfiwidesurvey} (see Sect.~2.4.2 and Appendix~B) and in \cite{MFIcompsep_pol}, the filter that is applied to the QUIJOTE-MFI data to clean residual RFI contamination (so-called FDEC) removes from the maps a monopole term at constant declination. We have checked that the effect of the FDEC filter does not induce any significant bias on the results presented in this work.

\subsection{Intensity template fitting}\label{secch5:templatefitting-int}

\begin{table*}
\centering
\begin{tabular}{ccccccc}
\hline
& \multicolumn{6}{c}{I}  \\
  \cmidrule(lr){2-7}
Map  & Sync & Free-free & Dust & Disk & Mono & Haze \\ 
     &  &  &  &  [mK$_{\rm RJ}$] &  [mK$_{\rm RJ}$] &  [mK$_{\rm RJ}$] \\ \hline
QUIJOTE\,11 & 0.93 & 1.28 & 357.38 & 8.85 & \phantom{$-$}9.0$\times 10^{-3\phantom{0}}$ & 0.70 \\
QUIJOTE\,13 & 0.89 & 1.26 & 231.44 & 5.40 & \phantom{$-$}6.6$\times 10^{-3\phantom{0}}$ & 0.54 \\
\textit{WMAP}\,K-band & 1.07 & 0.85 & \phantom{0}44.37 & 0.39 & \phantom{$-$}3.5$\times 10^{-15}$ & 0.19 \\
\textit{Planck}\,30 & 1.07 & 0.87 & \phantom{0}18.02 & 0.03 & $-$2.3$\times 10^{-16}$ & 0.10 \\
\textit{WMAP}\,Ka-band & 0.98 & 0.85 & \phantom{0}\phantom{0}9.05 & 0.07 & $-$2.6$\times 10^{-16}$ & 0.07 \\
\textit{WMAP}\,Q-band & 0.89 & 0.88 & \phantom{0}\phantom{0}3.74 & 0.07 & $-$8.6$\times 10^{-17}$ & 0.02 \\
\textit{Planck}\,44 & 0.90 & 0.90 & \phantom{0}\phantom{0}2.69 & 0.05 & $-$4.3$\times 10^{-17}$ & 0.02 \\
\textit{WMAP}\,V-band & 0.86 & 0.83 & \phantom{0}\phantom{0}1.16 & 0.04 & $-$1.5$\times 10^{-16}$ & 0.01 \\
\textit{Planck}\,70 & 0.85 & 0.82 & \phantom{0}\phantom{0}1.02 & 0.01 & \phantom{$-$}6.4$\times 10^{-17}$ & 0.01 \\\hline
\end{tabular}
\begin{tabular}{ccc} \hline
 \multicolumn{3}{c}{Q,U}  \\
\cmidrule(lr){0-2}
 Sync & Dust & Mono \\
  &  &  [mK$_{\rm RJ}$] \\\hline
0.87 & -37.51 & 2.8$\times 10^{-3}$ \\
0.92 & \phantom{0}-8.59 & 1.6$\times 10^{-3}$ \\
0.98 & \phantom{-}\phantom{0}3.33 & 6.9$\times 10^{-5}$ \\
1.04 & \phantom{-}\phantom{0}0.29 & 8.1$\times 10^{-6}$\\
 0.95 & \phantom{-}\phantom{0}1.59 & 5.8$\times 10^{-5}$\\
1.01 & \phantom{-}\phantom{0}0.94 & 4.7$\times 10^{-5}$\\
1.04 & \phantom{-}\phantom{0}0.82 & 4.4$\times 10^{-5}$\\
1.01 & \phantom{-}\phantom{0}0.69 & 6.4$\times 10^{-5}$\\
1.31 & \phantom{-}\phantom{0}0.72 & 8.0$\times 10^{-3}$\\
\hline
\end{tabular}

\caption{Stokes \textit{I} (left) and \textit{Q,U} (right) template fitting coefficients, fitted as described in Sect.~\ref{secch5:templ_fitting}.}\label{tab:fit_ampls}
\end{table*}

We performed a template-fitting component separation using the intensity frequency maps of QUIJOTE, WMAP and \textit{Planck} (see Table\,\ref{tabch5:summary_dsataseta} and Sect. \,\ref{sec:data} for a more detailed description of the data). We show in the appendix (Fig.\,\ref{figch5:maps_int})  the CMB subtracted sky maps within the sky area used in this analysis, which is limited by the QUIJOTE sky coverage and by the mask of reliable foregrounds description (see Sect.~\ref{secch5:templates} for further details on the mask). 

The templates that are used for the component separation in intensity are shown in Fig.\,\ref{figch5:templates_int}. They are: synchrotron, free-free,  dust (thermal dust and AME are adjusted with the same template of thermal dust), a disk template\footnote{The reconstruction of the Haze signal does not change significantly if we exclude the Galactic diffuse disk and Haze templates from the fit.} for the Galactic plane diffuse synchrotron emission, and the Haze, as described in Sect.~\ref{secch5:templates}. The CMB is fixed and subtracted from the maps before the fitting.  The fitted amplitudes for these templates are reported in Table\,\ref{tab:fit_ampls}.

As a result of this simple component separation, we construct the residual map $R_{\nu}$, by subtracting the foreground templates $\textbf{P}_{\nu}$ scaled by the fitted amplitudes $a_{\nu}$ from the corresponding frequency map $d_{\nu}$. It is:
\begin{equation}
    R_{\nu}= d_{\nu}-a_{\nu}\cdot \textbf{P}_{\nu}.
\end{equation}
Ideally, the residual $R_{\nu}$ is a map of the noise at frequency $\nu$. However, the foreground templates may not perfectly trace the real foreground spatial structure, and some residual sky structure could leak in the residual map. In particular, we are interested in the Haze component, which we fit with an approximate Gaussian elliptic template centred in the Galactic centre. This template is not expected to trace perfectly the spatial distribution of the Haze, therefore part of it could remain as a residual. For this reason, following \cite{WMAPHaze} and \cite{PlanckHaze}, we construct a residual plus Haze map as:
\begin{equation}
    R_{\nu}^{H}= R_{\nu}+a_{\nu}^{H}\cdot \textbf{P}^{H}_{\nu},
    \label{eq:RH}
\end{equation}
where $\textbf{P}^{H}_{\nu}$ is the Haze template and $a_{\nu}^{H}$ is the fitted Haze amplitude. 

The residual maps can then be used to study the physical properties of the isolated emission of the Haze as compared with the global synchrotron emission. With this aim we define the total synchrotron map as the residual map, plus the fitted Haze and synchrotron as:
\begin{equation}
    R_{\nu}^{S}= R^H_{\nu}+a_{\nu}^{s}\cdot \textbf{P}^{s}_{\nu},
    \label{eq:RS}
\end{equation}
where $\textbf{P}^{s}_{\nu}$ is the synchrotron template, $a_{\nu}^{s}$ its amplitude at frequency $\nu$, and $R_{\nu}^H$ the residual plus Haze map (Eq.\,\ref{eq:RH}). 

We show the resulting maps in Fig.\,\ref{figch5:res_int}, and we  study the Haze spectrum, which is shown in Fig.\,\ref{fig:haze_spec_int} and \ref{fig:haze_integrated_spec_noqjt}. 

\subsubsection{Intensity Haze maps}\label{sec:templfit_I_map}

\begin{figure*}
    \centering

    \includegraphics[width=0.24\textwidth]{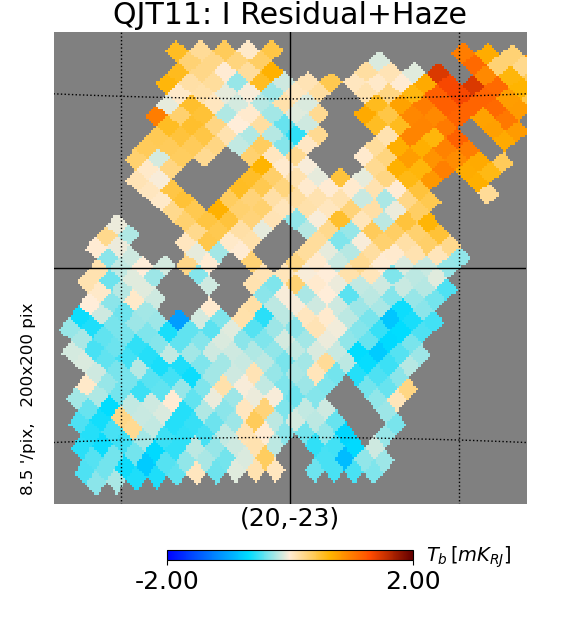}
    \includegraphics[width=0.24\textwidth]{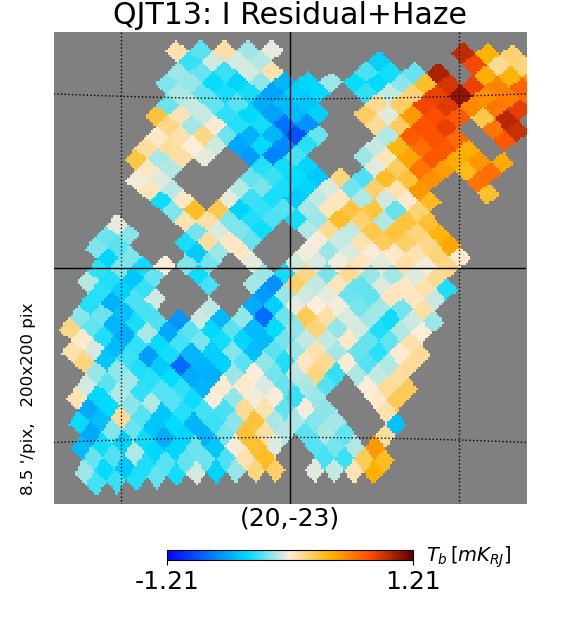}
    \includegraphics[width=0.24\textwidth]{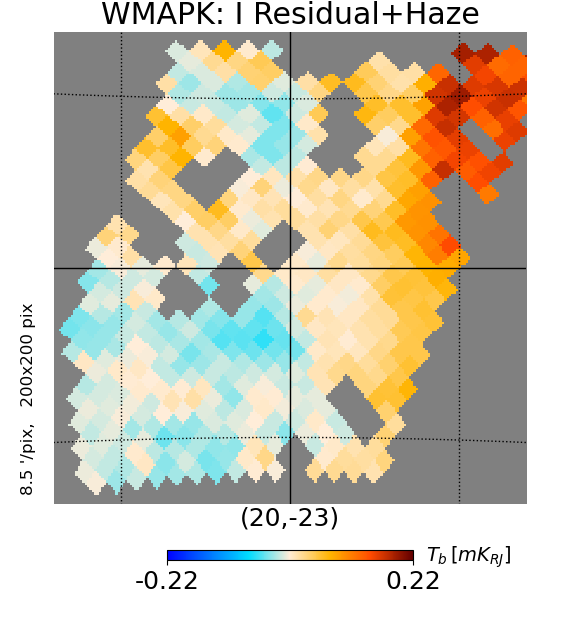}
    \includegraphics[width=0.24\textwidth]{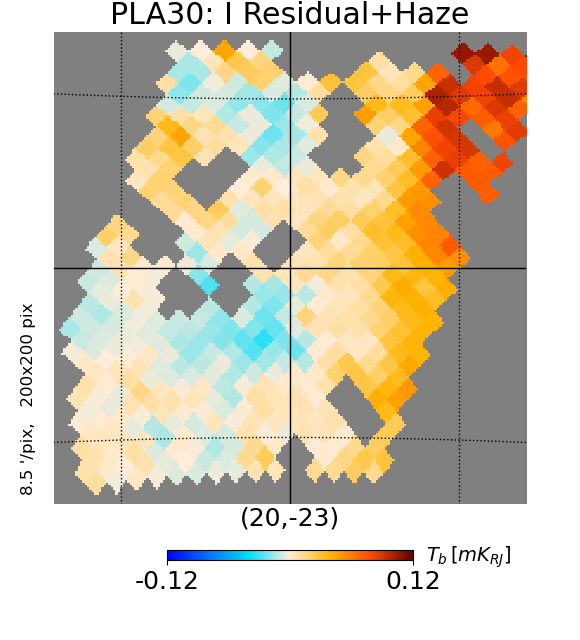}
    
    \caption{Residual plus Haze intensity maps in the southern Haze region (region 8, see Sect.~\ref{secch5:regions}), for, from left to right: QUIJOTE 11\,GHz, QUIJOTE 13\,GHz, \textit{WMAP} K-band, \textit{Planck} 30\,GHz. The grid is centred at coordinates $(l,b)=(20^{\circ},-23^{\circ})$ and is spaced by $10^{\circ}$ in Galactic latitude and longitude. The maps are in mK Rayleigh-Jeans temperature units, and the colour bar is scaled with a synchrotron-like power-law: 2\,mK$\cdot(\nu/11\,\mathrm{GHz})^\beta$, with $\beta=-3.0$.  }
    \label{figch5:res_int}
\end{figure*}

In Fig.\,\ref{figch5:res_int} we show the residual plus Haze maps ($R_{\nu}^H$) across region~8 (defined in Sect.~\ref{secch5:regions}), for several selected frequencies (QUIJOTE 11 and 13\,GHz, \textit{WMAP} K-band and \textit{Planck} 30\,GHz).
We observe that the bulk of the Haze component is detected in all the maps, including QUIJOTE.

Nulltest maps of QUIJOTE (see appendix \ref{secA:nulltets} and Fig.\,\ref{figch5:nulltest}) have been used to validate the sky origin of the observed signal. The nulltest maps do not show evident residual systematics, therefore the structures observed in the residual maps are associated with sky signal.

Beyond the Haze, we notice that the bottom part of the NPS (region 1), close to the Galactic centre at $(l,b)\sim(31.5^{\circ},\,16.5^{\circ})$, is visible in the residual maps of QUIJOTE.  This indicates that our templates do not perfectly match the base of the NPS region, and this could be associated with a synchrotron component with a spectrum that is different with respect to the sky average. Note that the NPS residual that we observe in this work corresponds to the region that \cite{Panopoulou2021} identified as possibly associated with Galactic centre activity. In contrast, the NPS emission at high galactic latitudes is usually ascribed to a nearby supernova shell. A detailed study of the NPS with QUIJOTE data is beyond the scope of this work and will be presented in Watson et al. (in preparation). 

\subsubsection{Intensity Haze spectrum}

Under the hypothesis that the Haze is synchrotron emission, both the Haze and the total synchrotron are characterized by a power-law spectrum (as in Eq.\,\ref{eq:sync_spec}), which is defined by two parameters: the amplitude and the spectral index $\beta$. 

We performed the measurement of the spectral index of the Haze and of the total synchrotron by fitting the SED of the signal within a selected area: region 8 in this case. We computed the average of the emission in the unmasked $R^H$ and $R^S$ pixels within the selected region.
The zero level must be properly set at each frequency $\nu$. We therefore fitted a linear slope to the pixel-to-pixel correlation plot of  $R^H_{\nu}$ against  $R^H_{22.8}$ (or $R^S_{\nu}$ against  $R^S_{22.8}$), given by:
\begin{equation}
R^{H,S}_{\nu} = m_{\nu} \cdot R^{H,S}_{22.8}+q_{\nu},
\label{eq:lin_fit_i}
\end{equation}
obtaining the relative offset to \textit{WMAP} K-band, $q_{\nu}$, and the slope $m_{\nu}$. This is done with a linear fit  accounting for errors in both axes,\footnote{For this fit we used the Orthogonal Distance Regression (ODR) \texttt{SciPy} package (\url{https://docs.scipy.org/doc/scipy/reference/odr.html}).} where the uncertainty is calculated as the standard deviation of the residual map in the selected area, propagated in quadrature with the uncertainty on the fitted amplitude of the templates.

The uncertainty on the SED points is given as the standard deviation of the residual map, scaled by the square-root of the number of averaged pixels, and summed in quadrature with the calibration uncertainty of each frequency map.
We assume a power-law behaviour for the spectrum of the Haze ($\left<R^H_{\nu}\right>-q^H_{\nu}$) and of the total synchrotron ($\left<R^S_{\nu}\right>-q^S_{\nu}$) at our frequencies, therefore we can write the linear relation of $\ln\left(\left<R^{H,S}_{\nu}\right>-q^{H,S}_{\nu}\right)$ against $\ln(\nu)$, as:
\begin{equation}
    \ln\left(\left<R^{H,S}_{\nu}\right>-q^{H,S}_{\nu}\right)=\beta^{H,S} \cdot \ln(\nu)+const.,
    \label{eq:fit_spect}
\end{equation}
whose slope provides the spectral index $\beta$.

\begin{figure} 
\includegraphics[width=\columnwidth]{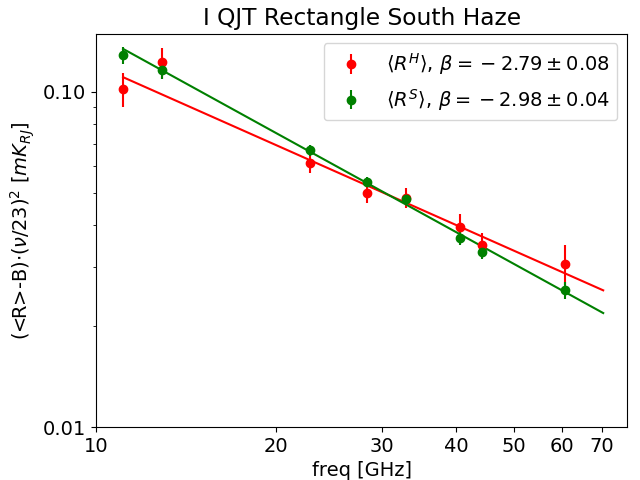}
\caption{Intensity SED of the Haze enclosed in region 8 (see Sect.~\ref{secch5:regions} and  Fig.\,\ref{figch5:res_int}).  
The data and the fit of the residual plus Haze spectrum (multiplied by three for display purposes) are shown in red, while the data and the fit of the total synchrotron are shown in green. 
}\label{fig:haze_spec_int}
\end{figure}

\begin{figure*}
\centering
\includegraphics[width=0.32\textwidth]{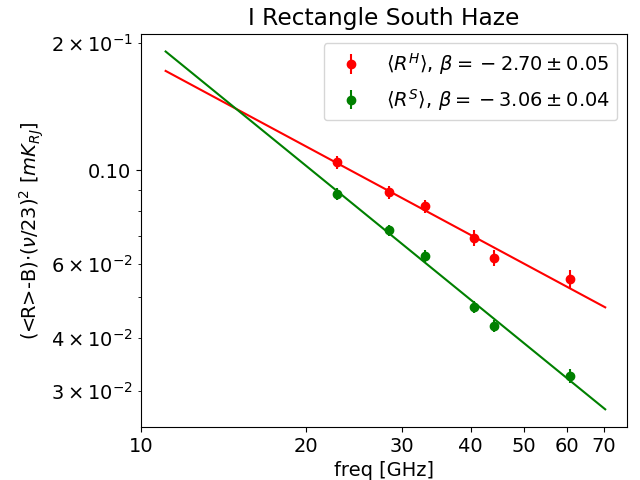}
\includegraphics[width=0.32\textwidth]{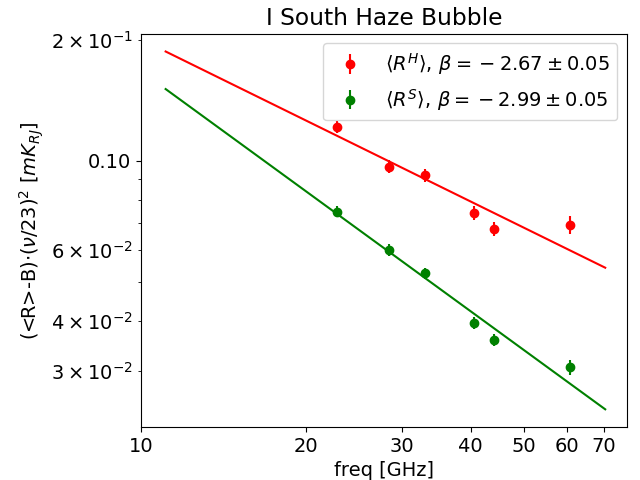}
\includegraphics[width=0.32\textwidth]{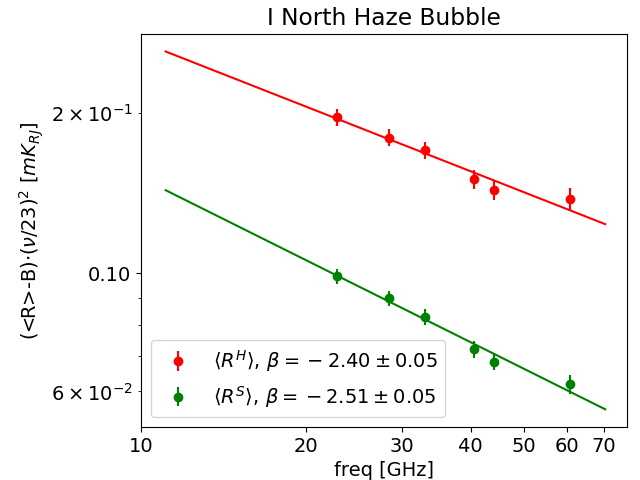}
\caption{Intensity SED of the rectangle enclosing the southern Haze region (region 7), in the \shb{} (region 9) and in the \nhb{} (region 5). In red is the spectrum of the residual plus Haze, and in green is the total synchrotron. } \label{fig:haze_integrated_spec_noqjt}
\end{figure*} 

Here we look at the southern Haze area (region 7) that has been identified by previous works \citep{WMAPHaze,PlanckHaze}. However, the sky observed by QUIJOTE does not cover the full area of region 7, and we restrict our analysis in the overlap with the QUIJOTE sky coverage (region 8), which is also shown on the right in Fig.\ref{figch5:res_int}.

The SED with the integrated spectrum in region 8 is given in the legend of  Fig.\,\ref{fig:haze_spec_int}. With a linear fit to these data,\footnote{The fit is performed with a MCMC sampling of the full posterior of the data, implemented with the \texttt{Python} \texttt{emcee} package (\citealp{Foreman-Mackey2013},  \url{https://emcee.readthedocs.io/en/stable/}).} we measure $\beta^H=-2.79\pm0.08$ and $\beta^S=-2.98\pm0.04$. 

We can observe that the spectrum of the Haze is flatter than that of the total synchrotron, with a difference in the spectral index of about $\Delta\beta=\beta^H-\beta^S = 0.26\pm0.13$. The difference has a significance of $2\sigma$.

Notice that if we remove the QUIJOTE data from the SED fit, the spectral indices  are $\beta^H=-2.76 \pm 0.12$ and $\beta^S=-3.02 \pm 0.06$, showing that QUIJOTE data does not significantly change the central value of the fit, but improves the precision with which the spectral indices are determined, by a factor 1.5.

The correlation plots mentioned above, which we performed to set the zero level for the SED, also provided an estimate of the spectral index as obtained from the slope of the linear fit $m_{\nu}$ with Eq.\,\ref{eq:beta}. We obtained $\beta^{H}=-2.78 \pm 0.19$ and $\beta^{S}=-2.97 \pm 0.05$.
This secondary measurement is consistent with the results that are obtained from the SED fitting, but the methodology is less precise.

We can also discuss the effect of using, instead of region 8, the broader region 7, which is the subject of the studies presented in \cite{PlanckHaze}.  Excluding QUIJOTE data and applying our methodology that uses priors to fit the various foregrounds, we measure the values $\beta^H=-2.70\pm0.05$ and $\beta^S=-3.06\pm0.04$ (see left panel in Fig.\,\ref{fig:haze_integrated_spec_noqjt}). We first notice that our measurement of the Haze spectrum in region 7 is slightly flatter than what we obtain in region 8 (by $\Delta \beta^H=0.06$), although they are consistent within the uncertainties.

In region 7, \cite{PlanckHaze} reports values of  $\beta^H=-2.56\pm0.05$ and $\beta^S=-3.1$. If we compare this with our results we can see that, in agreement with the \textit{Planck} paper, the Haze in region 7 emits with a flatter index than that of the total synchrotron, but there is a discrepancy in the recovered Haze spectral index. 
In order to test the origin of this discrepancy, we reproduced the results of \cite{PlanckHaze} by applying their same methodology, with no priors, excluding QUIJOTE data, and integrating the same southern Haze area (region 7). In this case, we obtain  $\beta^H=-2.47 \pm 0.06$ and $\beta^S=-3.19 \pm 0.04$, which is consistent with the \textit{Planck}'s results, showing that the main source of the observed difference is the use of priors, which results in a shift of the Haze spectral index towards steeper values, by $\Delta \beta^H=0.23$ in region 7.  
The use of priors in the pipeline of this work has been tested with simulations (as discussed in Sec. \ref{secch5:templ_fitting}), with which we noticed a clear improvement in the fitting of the foregrounds when compared with the case with no-priors. For this reason we finally applied priors in our analysis, despite the slightly different results in the Haze region as compared with previous works. 

\subsubsection{Spectra of regions 5, 7 and 9}
There are more regions that are interesting for the study of the Haze, but which are unfortunately not accessible by QUIJOTE, in the northern hemisphere. These are: the \shb{} (region 9), which is located in the southern sky and can only be partially observed with QUIJOTE, and the \nhb{} (region 5), which is observed by QUIJOTE but  coincides with a region with large residuals that are not fully understood. We studied these two regions by applying our template fitting methodology (with priors), using only \textit{WMAP} and \textit{Planck} data. 
We show their integrated spectra in the central and right panels in Fig.\,\ref{fig:haze_integrated_spec_noqjt}. In the \shb{} (region 9) we obtain the spectral indices  $\beta^H=-2.67\pm0.05$ and $\beta^S=-2.99\pm0.05$, which are compatible with the already discussed results for the rectangle enclosing the southern Haze (region 7; left panel in Fig.\,\ref{fig:haze_integrated_spec_noqjt}). In the \nhb{}, instead, we obtain a flatter Haze spectrum, with $\beta^H=-2.40\pm0.05$, and also a flatter total synchrotron spectrum, it being $\beta^S=-2.51\pm0.05$. We detect a significant difference between the spectral index of the North and South Haze bubbles in intensity, with the spectrum of the northern bubble flatter than that in the South.  As in other regions, the Haze component is flatter than the total synchrotron, but in the northern bubble  the total synchrotron spectrum is also significantly flatter than that in other regions. Interestingly, as we report later (Sect.~\ref{secch5:tt-plots-results}), the polarization between 23\,GHz and 30\,GHz shows the same behaviour, with the northern bubble having a flatter spectrum than the southern one. In addition, the polarization spectral index of the \nhb{} is compatible with that of the total synchrotron in intensity, while the polarization spectral index of the \shb{} is between the intensity $\beta^H$ and $\beta^S$ .

\subsection{Polarization template fitting}\label{secch5:templatefitting-pol}

We applied the template fitting procedure in polarization, by fitting a synchrotron and thermal dust component to the \textit{Q} and \textit{U} frequency maps simultaneously across the full unmasked sky (see Sect.~\ref{secch5:templ_fitting} for a detailed description of the methodology).  The CMB is fixed and subtracted from the maps before the fitting. The resulting fitted amplitudes are reported in Table\,\ref{tab:fit_ampls}.

We computed the residual polarization amplitude maps as:
\begin{equation}
    R_{{\rm P},\nu}=\sqrt{R_{Q,\nu}^2+R_{U,\nu}^2},
    \label{eq:p_res}
\end{equation} 
with\footnote{We drop the specification of $\nu$ subscript for brevity.} $R_{Q,U}=R^0_{Q,U}-(Q_{0},U_{0})$. Here, $R^0_{Q,U}$ are the residual \textit{Q},\textit{U} maps  obtained after subtracting the fitted foregrounds from the original \textit{Q},\textit{U} frequency maps. $Q_0$,$U_0$ are constant offsets to be subtracted to $R^0_{Q,U}$ in order to adjust the zero level across frequencies.  $Q_0$ and $U_0$ are obtained with T-T plots of $R^0_{Q,U}$ at frequency $\nu$ with respect to the residual map at $\nu=22.8$\,GHz (\textit{WMAP}\,K-band), across the unmasked sky pixels.
At this stage  we do not attempt to debias the polarization amplitude maps, so $R_{\rm P}$ could be marginally affected by noise bias.

We also define the fitted polarization amplitude synchrotron map as:
\begin{equation}
    S_{{\rm P},\nu} = \sqrt{ S_{Q,{\nu}} ^2+ S_{U,{\nu}}^2},
    \label{eq:P_sync}
\end{equation}
where $S_{Q,U}=a^{s}_{Q,U} (Q^s,U^s)$ are the fitted synchrotron maps, with
$Q^s$,$U^s$ being the polarization synchrotron template maps, and $a_Q^{s}$,$a_U^{s}$ the correspondent fitted amplitudes. 

Finally, we define the residual plus synchrotron map as:
\begin{equation}
    R^S_{{\rm P}, \nu} =\sqrt{\left(R^S_{{\rm Q}, \nu}\right)^2+\left(R^S_{{\rm U}, \nu}\right)^2},
 \label{eq:P_res_plus_sync}
\end{equation}  
where $R^S_{{\rm Q,U}}=R_{Q,U}+S_{Q,U} -(Q'_0,U'_0)$ are the $Q,U$ residual plus synchrotron maps,
with $Q'_0,U_0'$ adjusting the relative zero levels across frequencies, computed with T-T plots across the unmasked sky pixels with respect to \textit{WMAP}\,K-band (22.8\,GHz).

\subsubsection{Polarization residual maps}\label{secch5:templatefitting-pol_maps}

\begin{figure*}
    \centering
    \includegraphics[width=0.35\textwidth]{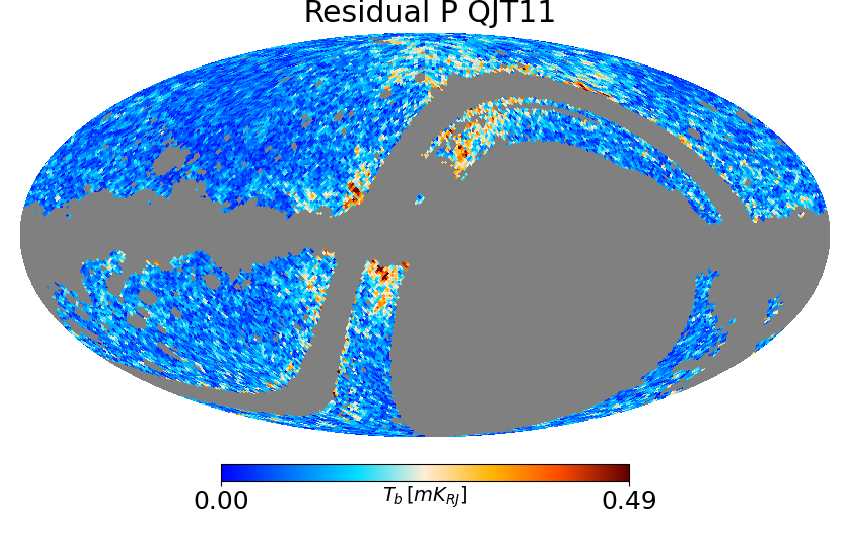}
    \includegraphics[width=0.35\textwidth]{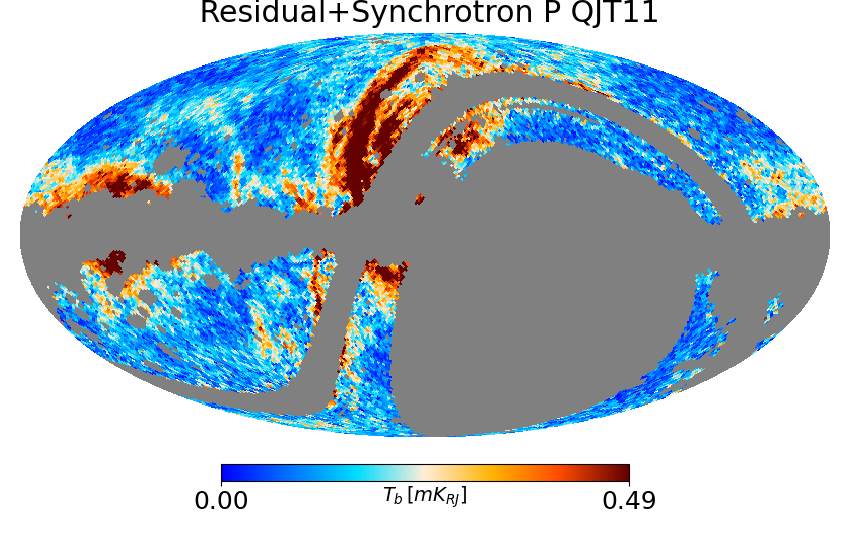}
    \includegraphics[width=0.22\textwidth]{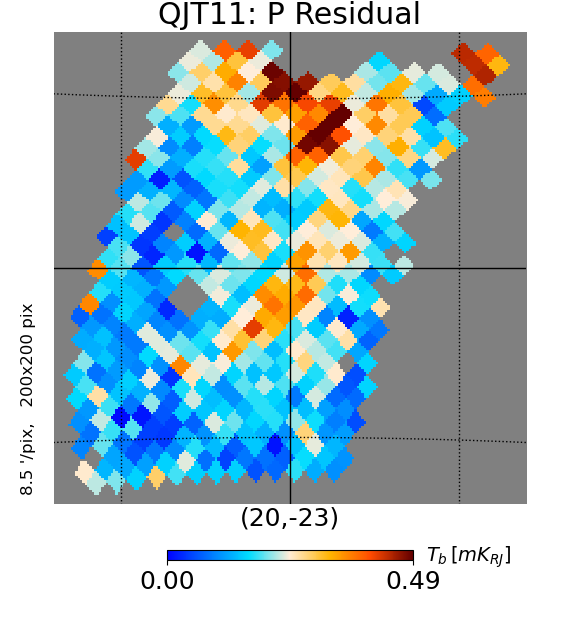}

    \includegraphics[width=0.35\textwidth]{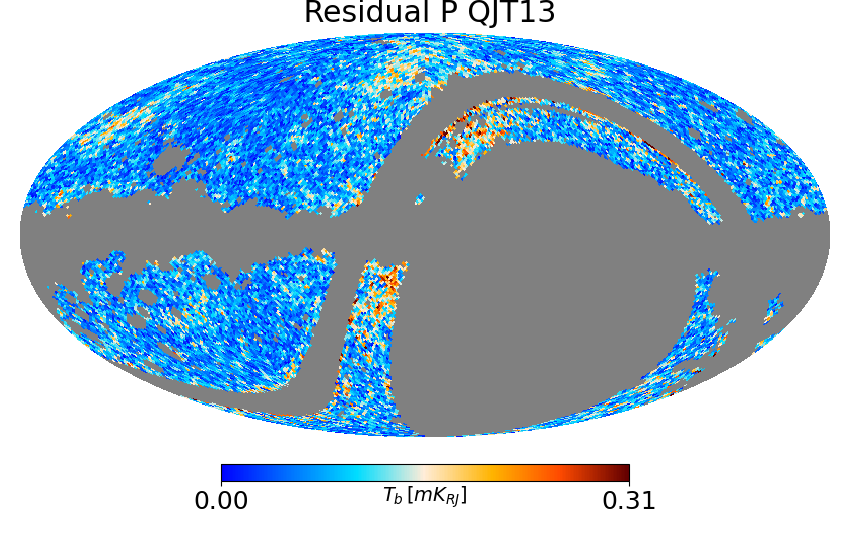}
    \includegraphics[width=0.35\textwidth]{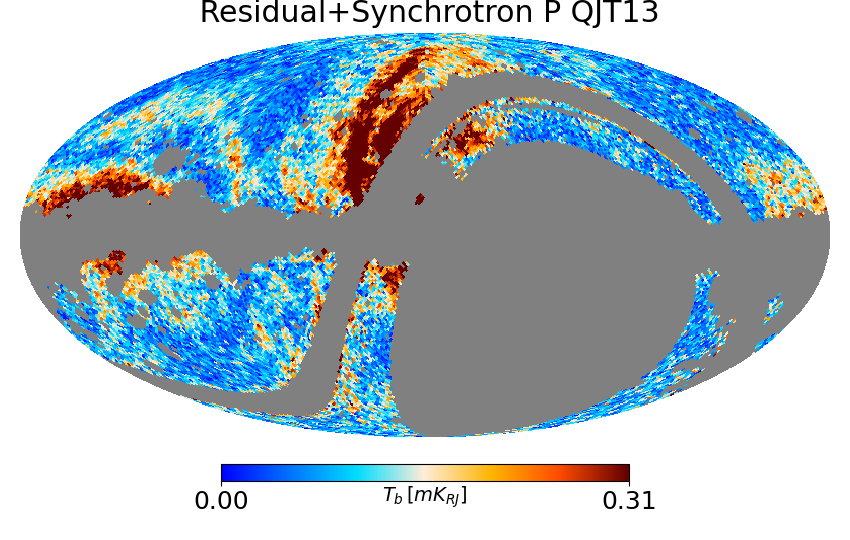}
    \includegraphics[width=0.22\textwidth]{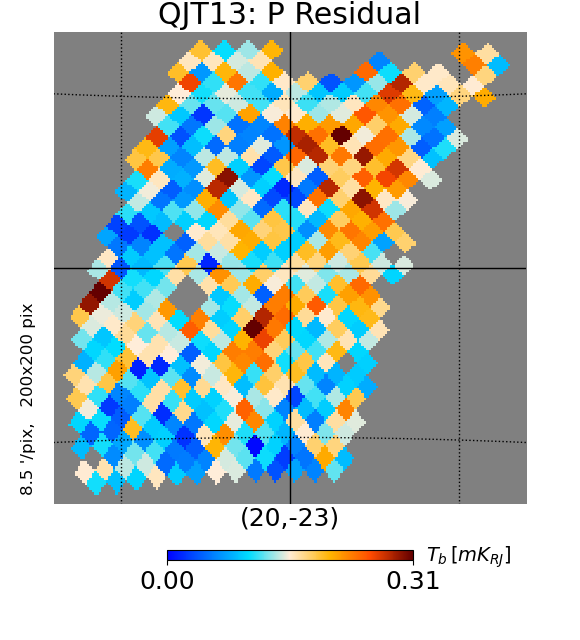}
    
    \includegraphics[width=0.35\textwidth]{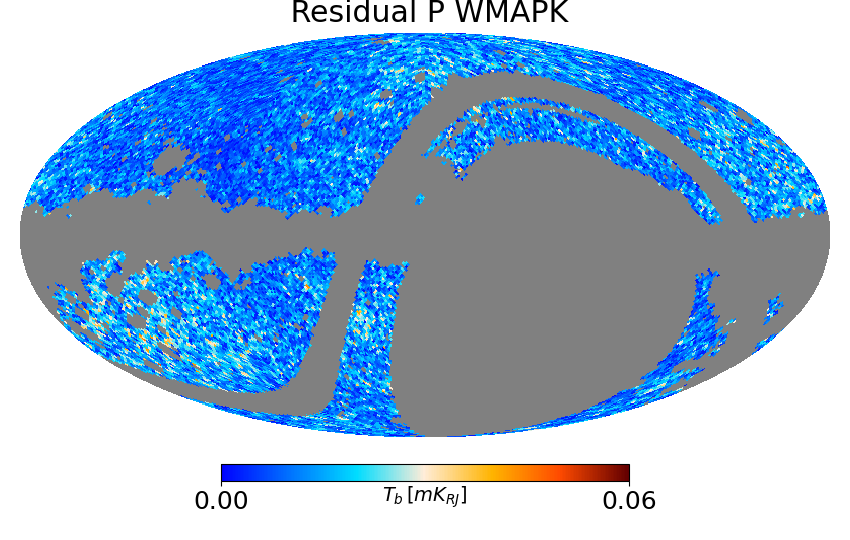}
    \includegraphics[width=0.35\textwidth]{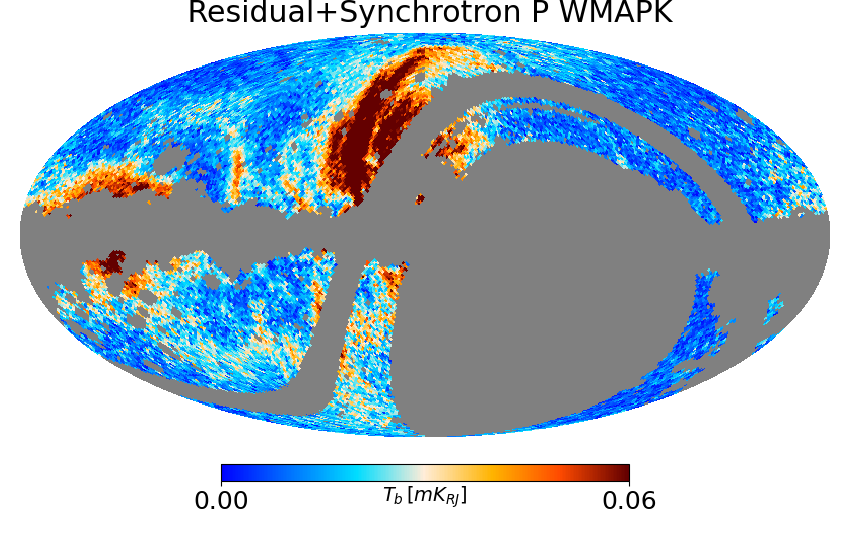}
    \includegraphics[width=0.22\textwidth]{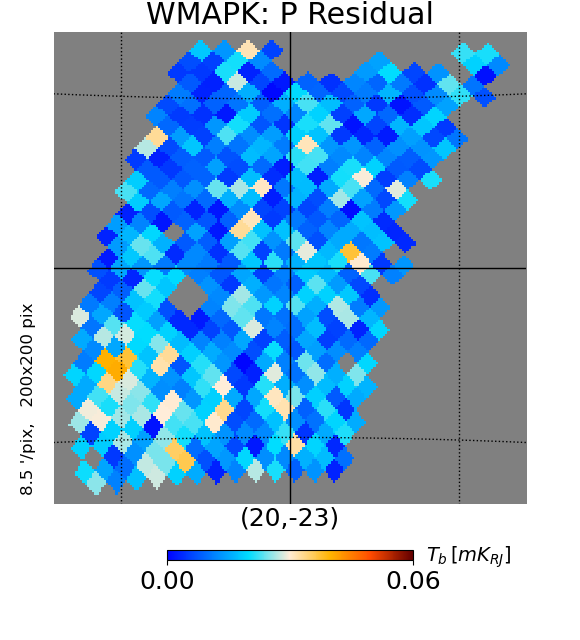}
    
    \includegraphics[width=0.35\textwidth]{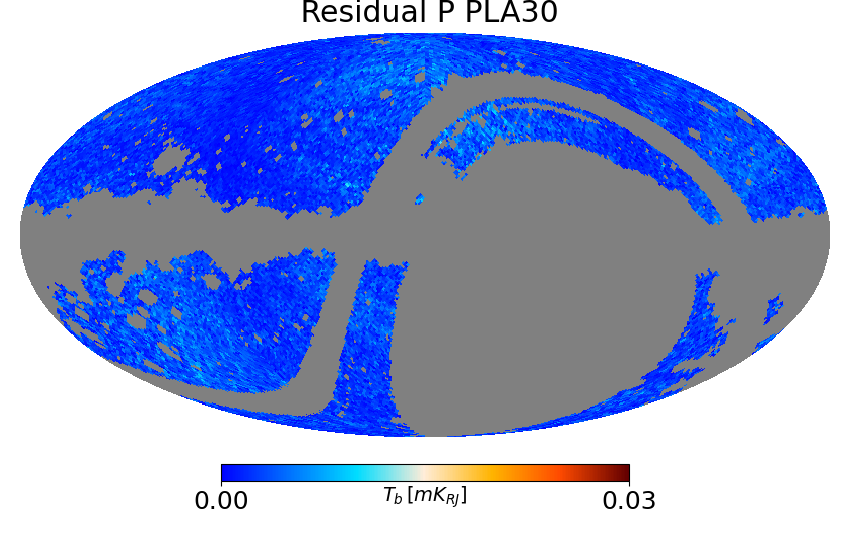}
    \includegraphics[width=0.35\textwidth]{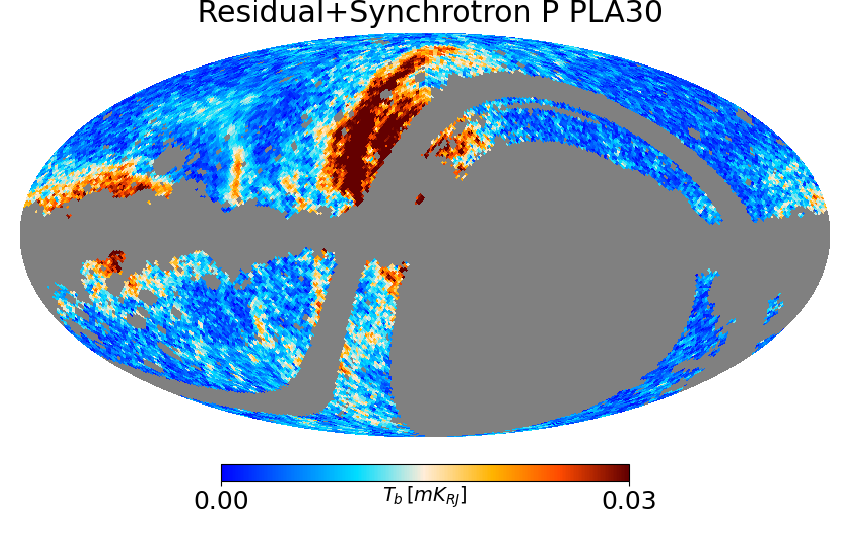}
    \includegraphics[width=0.22\textwidth]{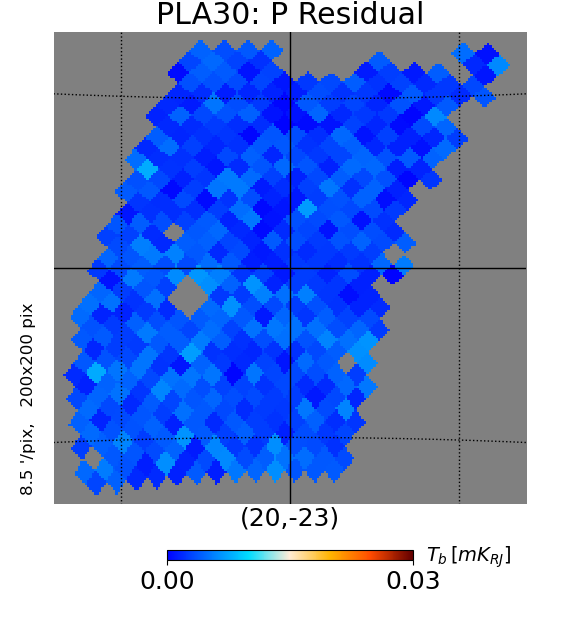}
    \caption{Residual (left, Eq.\,\ref{eq:p_res}) and residual plus synchrotron (centre, Eq.\,\ref{eq:P_res_plus_sync}) polarization amplitude maps of, from top to bottom,  QUIJOTE 11\,GHz, QUIJOTE 13\,GHz, \textit{WMAP} K-band, \textit{Planck} 30\,GHz. The right column figures show the residual $P$ maps zoomed in the southern Haze region (region 8, see Sect.~\ref{secch5:regions}, same grid as the right panels in Fig.\,\ref{figch5:res_int}). The maps are in mK Rayleigh-Jeans temperature units, and the colour bar is scaled with a synchrotron-like power-law: $0.5\,{\rm mK}\cdot(\nu/11\,\mathrm{GHz})^\beta$, with $\beta=-3.0$.} \label{fig:res_p}
\end{figure*} 

Fig.\,\ref{fig:res_p} shows maps of the residual polarization amplitude  $R_{\rm P}$ (left, see Eq.\,\ref{eq:p_res}) and of the residual plus synchrotron $R^{\rm S}_{\rm P}$ (centre, see Eq.\,\ref{eq:P_res_plus_sync}) of QUIJOTE 11 and 13\,GHz, of \textit{WMAP} K-band and of \textit{Planck} 30\,GHz. Similar to Fig.\,\ref{figch5:res_int}, we also show a zoom-in of the $R_{\rm P}$ maps across region 8 (right). We can observe that the \textit{WMAP} K-band and \textit{Planck} 30\,GHz residual maps are mostly noise with potentially some low level systematics. In particular, the residual polarization map of \textit{Planck} 30\,GHz has very low values as compared with the other residual maps. This is due to the fact that for the template fitting procedure we use the \texttt{Commander} synchrotron solution  (see Sect.~\ref{secch5:templates}), which strongly relies, by construction, on the 30\,GHz \textit{Planck} polarization data. 

The QUIJOTE polarization residual maps, instead, show structures that can be associated with residual sky signal. The detection of similar structures was not possible in previous works based only on \textit{WMAP} or \textit{Planck} data. QUIJOTE data is now providing hints of a detection of a previously unknown polarized diffuse signal. 
Indeed we can observe, at 11\,GHz and 13\,GHz, evident structures across the full Haze area, towards the South in region 8, but also towards the North reaching high Galactic latitudes ($b\sim85^{\circ}$). 
We also detect residual signal in the lower part of the NPS, close to the Galactic plane (at $(l,b)\sim(31.5^{\circ},\,16.5^{\circ})$, bottom of region 1), which is seen also in the intensity residual maps (see Sect.\,\ref{sec:templfit_I_map}). We refer to  Watson et al. (in preparation) for a detailed study of the NPS using QUIJOTE data. 

In order to validate the sky origin of the observed polarization excesses, we analyzed noise maps of QUIJOTE obtained with nulltests (as shown in appendix \ref{secA:nulltets}, Fig.\,\ref{figch5:nulltest}), showing that the noise level can not explain the observed residuals, which are therefore ascribed to sky signal. 

This kind of residuals could possibly be originated by spatial variations of the synchrotron spectral index, which has not been taken into account in the fitting procedure. We tested this hypothesis by repeating the analysis allowing the synchrotron spectral index to vary across the sky. We used for this purpose the synchrotron spectral index map extracted by \cite{MFIcompsep_pol}, which is derived from a pixel-based component separation (\texttt{B-SeCRET}, \citealp{delahoz2020}) using data from QUIJOTE, \textit{WMAP} and \textit{Planck}. In this case we recover similar residual polarization maps as those shown in Fig.\,\ref{fig:res_p}, concluding that the observed residuals are not attributable to spatial variations of the synchrotron index. On the other hand, they could be due to a curvature of the spectrum at low frequencies ($\nu<23$\,GHz), across the area where we observe a positive residual. 

\subsubsection{Polarization residual spectrum}\label{secch5:templatefitting-pol_spect}

\begin{figure*}
\includegraphics[width=0.49\textwidth]{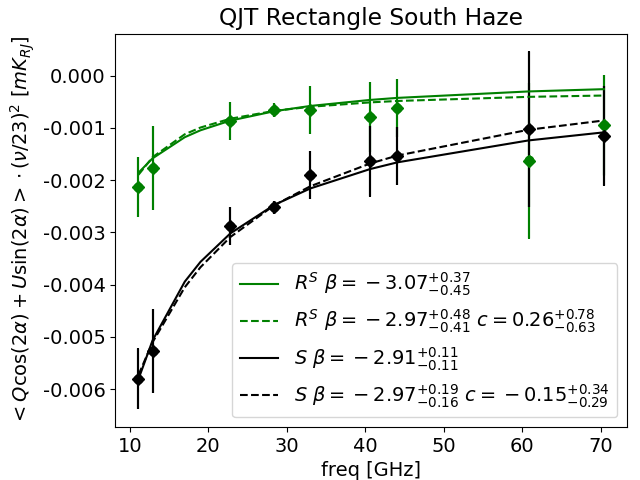}
\includegraphics[width=0.49\textwidth]{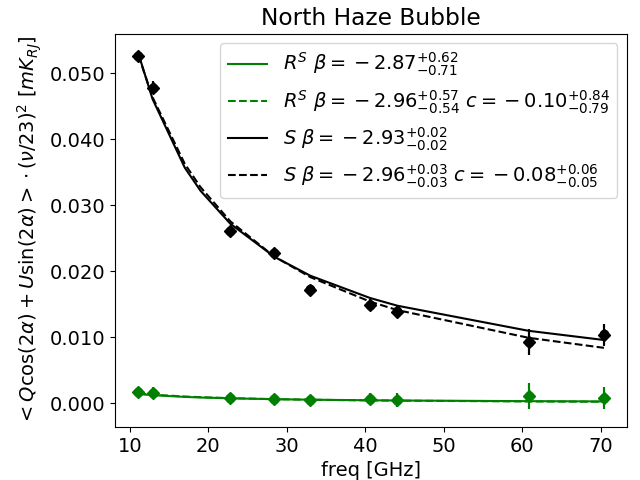}

\caption{Polarization SED after template fitting in the rectangle enclosing the southern Haze area in the overlap with the QUIJOTE sky (region 8, left), and in the \nhb{} (region 5, right). The green points and lines represent the averaged and colour-corrected residual plus synchrotron spectrum ($R^{\rm S}$, e.g., Eq.\,\ref{eq:P_res_plus_sync}), fitted with a simple (thick green line) and modified (dashed green line) power-law. The same, in black color, is for synchrotron ($S$, e.g., Eq.\,\ref{eq:P_sync}). The fitted spectra indices are reported in the legend.}\label{fig:p_spect_templfit}\end{figure*} 

Following the same procedure that is applied in intensity, we computed the spectrum of maps integrated in several selected regions. In this case, as stated in Sect.~\ref{secch5:templates}, we do not perform the fit of an independent Haze template, because the projection of the Haze in the Stokes \textit{Q} and \textit{U} maps is unknown. Therefore, if the data contain a Haze component that is not identified as synchrotron with the sky average spectral index, or as thermal dust (even if it is a very minor component at these frequencies), it will be revealed in the residual maps $R_{\rm Q},\,R_{\rm U}$, or $R_{\rm P}$ defined in Eq.\,\ref{eq:p_res} and shown in Fig.\,\ref{fig:res_p}. We therefore look  for a polarized Haze component in the residual plus synchrotron spectrum (e.g., Eq.\,\ref{eq:P_res_plus_sync}), by comparing it with the spectrum of the synchrotron alone (e.g., Eq.\,\ref{eq:P_sync}).  

We computed the spectrum of the combination of Stokes \textit{Q} and \textit{U} parameters with a sinusoidal function, as defined in Eq.\,\ref{eq:QcUs}, projecting them in the direction of the polarization angle $\alpha$ of the region, and averaging the resulting signal within the selected region. This allows us to overcome problems related with noise bias of the polarization amplitude. 

The representative projection angle $\alpha$ in the region is determined by inverting the median value of $\sin(2\alpha)$, which is a continuum function when the angle has a discontinuity (at $\alpha \pm 90^{\circ}$). The angle is computed using the \textit{WMAP} K-band data, and it is used for all the other frequencies.  We use $\alpha = 71.5^{\circ}$ in region 8  and $\alpha =36.3^{\circ}$ in region 5.


In Fig.\,\ref{fig:p_spect_templfit} we show the spectrum of the average \textit{Q} and \textit{U} combination for the polarized synchrotron ($S$, in black) and for the residual plus synchrotron ($R^S$, in green) as a function of the frequency, within two different regions: the \nhb{} (region 5) and the southern Haze area (region 8). 

The \textit{Q} and \textit{U} uncertainties ($\sigma_Q\,,\sigma_U$) for the \textit{WMAP} and \textit{Planck} data points are estimates of the scatter of the residual $R_Q$ and $R_U$ maps, between pixels enclosed in the region being examined.  For  QUIJOTE, instead, the residual maps show an evident signal contribution, therefore we derive $\sigma_Q$ and $\sigma_U$ as the standard deviation of the null-test $Q$ and $U$ maps shown in appendix \ref{secA:maps}, within the selected region.
Finally, the \textit{Q} and \textit{U} uncertainties are normalized by the square-root of the number of unmasked pixels, are summed in quadrature with the corresponding calibration uncertainty, and are propagated through Eq.\,\ref{eq:QcUs}.

We fit to the synchrotron and residual plus synchrotron spectra the amplitude $A$, the spectral index $\beta$, and the curvature $c$ of a modified power-law (as in e.g., \citealp{Kogut2012}):
\begin{equation}
    d(\nu) = A\left(\frac{\nu}{\nu_0}\right)^{\beta+c \ln{\frac{\nu}{\nu_0}}},
    \label{eq:mod_power_law}
\end{equation} 
where $\nu_0=23$\,GHz a reference frequency. The range of frequency used is $11 \lesssim \nu \lesssim 70$\,GHz.
The fit is performed with a MCMC sampling of the full posterior of the data with \texttt{emcee}.
For the fit of the residual plus synchrotron  we applied a flat prior on the spectral index $-4<\beta<-2$, and a Gaussian prior to the curvature parameter, with a  width $\sigma_c=1$ and central values $\mu_c=0$ (dashed green line in Fig.\,\ref{fig:p_spect_templfit}). The synchrotron alone instead is fitted with no priors.
For comparison we also fit the $A$ and $\beta$ parameters for a simple power-law, given by Eq.\,\ref{eq:mod_power_law} with $c=0$. The fitted spectra in this case are shown as thick lines in Fig.\,\ref{fig:p_spect_templfit}, and the respective $\beta$ are reported in the legend. No priors are applied in this case.

It can be observed in region 8 (left panel in Fig.\,\ref{fig:p_spect_templfit}) that the spectral index of a simple power-law  for the residual plus synchrotron is $\beta=-3.07^{+0.37}_{-0.45}$, and for the synchrotron it is $\beta=-2.91\pm0.11$. The two spectral indices are compatible within the uncertainties. When including the curvature parameter in the fit, the estimated $\beta$ are in even better agreement, with $\beta=-2.97^{+0.48}_{-0.41}$ for the residual plus synchrotron and  $\beta=-2.97^{+0.19}_{-0.16}$ for the synchrotron alone. Although the residual plus synchrotron shows slight preference for a positive value of $c$, and the synchrotron alone shows a preference for negative $c$, curvature is not detected with this methodology.

In region 5 (see right panel in Fig.\,\ref{fig:p_spect_templfit})  the spectral index of a simple power-law for the residual plus synchrotron is $\beta=-2.87^{+0.62}_{-0.71}$, and for the synchrotron alone it is $\beta=-2.93\pm0.02$. The two spectral indices are compatible within the uncertainties. Also in this case, when including the curvature parameter in the fit, the estimated spectral indices are in better agreement, with $\beta=-2.96^{+0.57}_{-0.54}$ for the residual plus synchrotron, and $\beta=-2.96^{+0.03}_{-0.03}$ for the synchrotron alone.  Although both the residual plus synchrotron and the synchrotron alone show slight preference for values of $c<0$, curvature is not detected with this analysis.

To summarize, no clear differences between the synchrotron and residual plus synchrotron spectral indices are detected. The curvature is obtained to be compatible with zero given the large error bars, especially on the spectral index of the residuals plus synchrotron. However, estimates of the curvature on these two regions are also presented in \cite{MFIcompsep_pol}, where a negative curvature is detected at high significance using the parametric component separation method \texttt{B-SeCRET} \citep{delahoz2020}, although there is not enough statistical evidence to favour the curvature against the single power-law model.



An independent but complementary analysis of the polarization spectrum is shown in the next section, where we performed a detailed analysis with T-T plots in polarization. 

\subsection{T-T plots of Haze polarized plumes and spurs}\label{secch5:tt-plots-results}

\begin{figure*}
    \centering
    \includegraphics[width=0.32\textwidth]{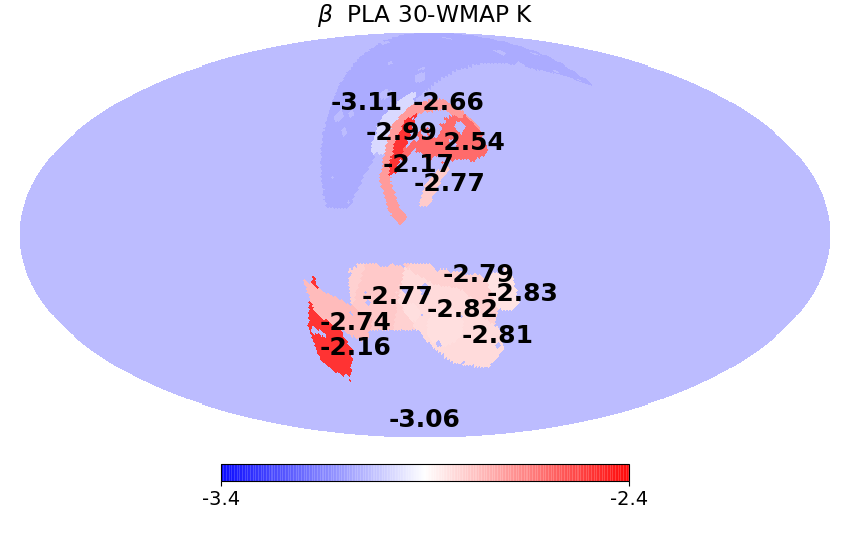}
    \includegraphics[width=0.32\textwidth]{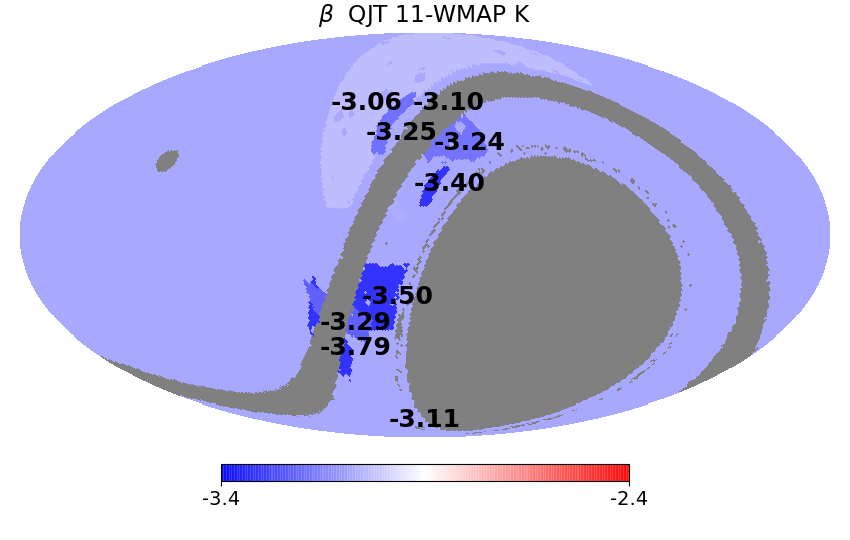}
    \includegraphics[width=0.32\textwidth]{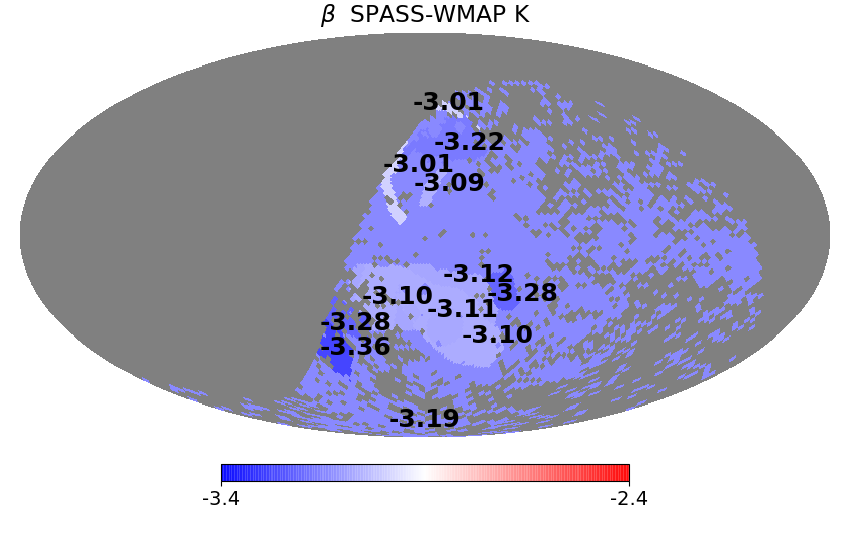}
    \includegraphics[width=0.32\textwidth]{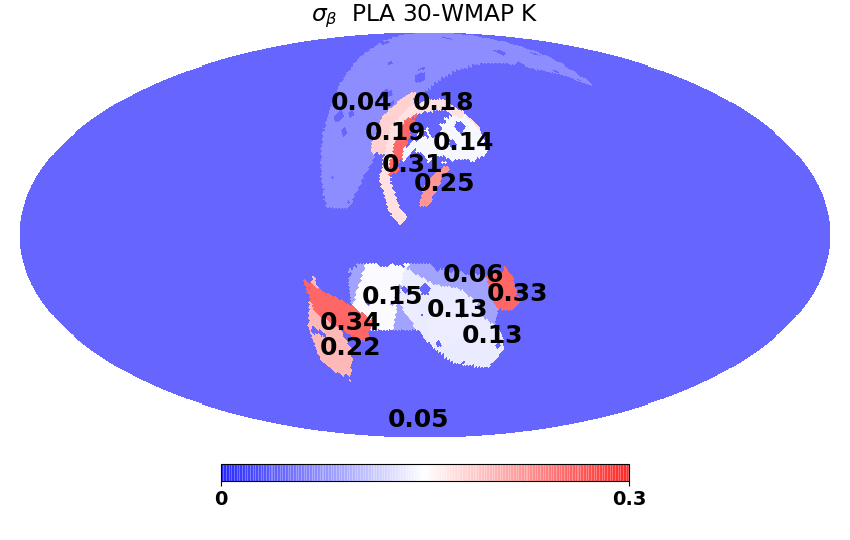}
    \includegraphics[width=0.32\textwidth]{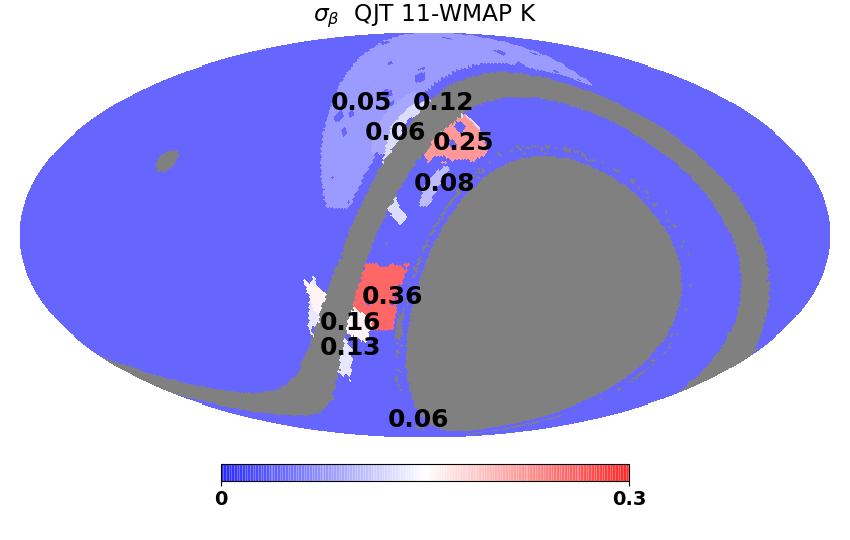}
    \includegraphics[width=0.32\textwidth]{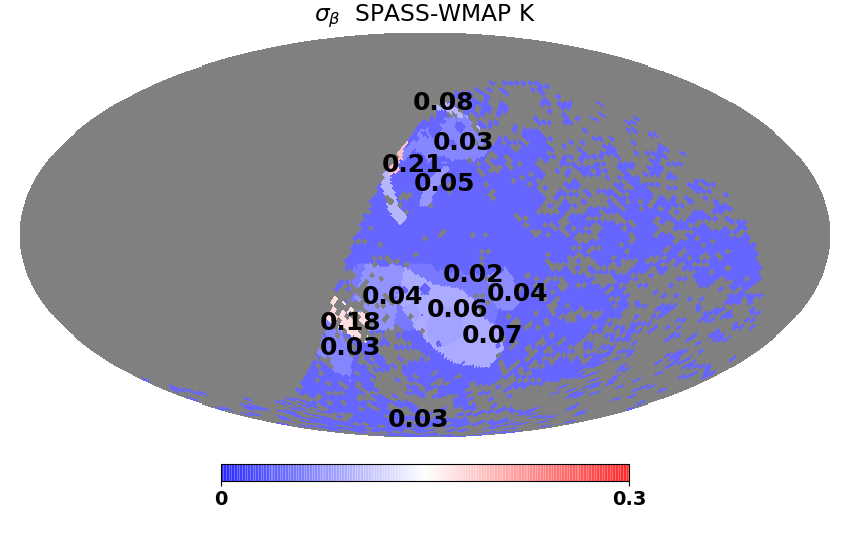}
    \caption{Polarization spectral indices in the selected regions (top) and uncertainties (bottom), obtained with a T-T plots analysis based on data from \textit{Planck} 30\,GHz (left), QUIJOTE 11\,GHz (center), and S-PASS 2.3\,GHz (right), with \textit{WMAP} K-band as a pivot.}
    \label{fig:betas_ttplot}
\end{figure*}

With the aim of studying the Haze region in polarization with a different approach to that presented in Sect.~\ref{secch5:templatefitting-pol}, we performed a correlation T-T plot analysis as described in Sect.~\ref{sec:tt-plots-fsk}, in the regions presented in Sect.~\ref{secch5:regions}. In this analysis, we also include the S-PASS data at 2.3\,GHz, corrected for Faraday rotation as described in Appendix~\ref{sec:FR}. We computed the spectral indices between the frequency pairs:
\begin{itemize}
    \item 23--30\,GHz (\textit{WMAP} K-band -- \textit{Planck} 30\,GHz) 
    \item 11--30\,GHz (QUIJOTE 11\,GHz -- \textit{Planck} 30\,GHz) 
    \item 11--23\,GHz (QUIJOTE 11\,GHz -- \textit{WMAP} K-band) 
    \item 2.3--30\,GHz (S-PASS 2.3\,GHz -- \textit{Planck} 30\,GHz)
    \item 2.3--23\,GHz (S-PASS 2.3\,GHz -- \textit{WMAP} K-band)
    \item 2.3--11\,GHz (S-PASS 2.3\,GHz -- QUIJOTE 11\,GHz)

\end{itemize}
A summary of the results is reported in Table\,\ref{tab:betas_ttplots}, and a graphical representation of the estimated spectral indices and uncertainties is shown in Fig.\,\ref{fig:betas_ttplot} for three selected frequency cases. In order to validate our results, we present a detailed analysis of the posterior distribution of the T-T plots in  appendix \ref{secA:ttplot_posterior}.

By looking at Fig.\,\ref{fig:betas_ttplot} (or Table\,\ref{tab:betas_ttplots}), we can notice that the Haze in polarization appears as two extended and slightly asymmetric bubbles (region 5,7--10), surrounded and connected to the Galactic plane with filaments and spurs (region 2, 3, 4, 6, 11, 12, 13). Our interpretation is that the regions 2--13 are related to the Haze, or in general to emission related to activity of the Galactic centre. Indeed, our measurements show that the spectral index of these regions  is flat at high frequencies (23--30\,GHz) and uniformly moves towards steeper values at lower frequencies (11--23\,GHz and 2.3--23\,GHz). The typical spectral indices of the Haze regions at 23--30\,GHz are $-2.8 \lesssim \beta \lesssim -2.6$, while at lower frequencies they became steeper, being $-3.2 \lesssim \beta \lesssim -3.0$ at 11--23\,GHz and at 2.3--23\,GHz.  

We quote for comparison the average spectral indices of the full-sky available from each survey combined with the mask described in Sect.\,\ref{secch5:templates},  and of the NPS (region 1), which is a widely studied region, currently modeled as synchrotron emission originating from the expanding shell of a nearby supernova explosion (e.g., \citealp{pla2015XXV,Panopoulou2021} and Watson et al. (in preparation).  We can notice that the spectral indices of the full-sky and of the NPS at 23--30\,GHz are steeper than those of the Haze associated regions. Instead, at 11--23\,GHz we observe the opposite behaviour: the full sky and NPS spectral indices are flatter than those of the Haze associated regions.

We will extend the discussion of these results in Sec\,\ref{sec:discussion}, where we provide an overview and an interpretation of the measurements obtained with different methodologies.

\begin{table*}
\begin{tabular}{cccccccc}
\hline
Region & Description & $\beta$ 23--30 & $\beta$ 11--23 & $\beta$ 11--30 & $\beta$ 2.3--23 & $\beta$ 2.3--30 & $\beta$ 2.3--11 \\ \hline
0 & Full high-latitudes sky & $-$3.06 $\pm$ 0.05 & $-$3.11 $\pm$ 0.06 & $-$3.12 $\pm$ 0.04 & $-$3.19 $\pm$ 0.03 & $-$3.19 $\pm$ 0.03 & $-$3.42 $\pm$ 0.12 \\
1 & NPS & $-$3.11 $\pm$ 0.04 & $-$3.06 $\pm$ 0.05 & $-$3.08 $\pm$ 0.06 & - & - & - \\ \hline
2 & Ext Haze Filament & $-$2.99 $\pm$ 0.19 & $-$3.25 $\pm$ 0.06 & $-$3.20 $\pm$ 0.05 & - & - & - \\
3 & Haze Filament & $-$2.66 $\pm$ 0.18 & $-$3.10 $\pm$ 0.12 & $-$3.01 $\pm$ 0.12 & $-$3.01 $\pm$ 0.08 & $-$2.96 $\pm$ 0.05 & $-$3.09 $\pm$ 0.19 \\
4 & Int Haze Filament & $-$2.17 $\pm$ 0.31 & $-$3.40 $\pm$ 0.80 & $-$3.10 $\pm$ 1.08 & $-$3.01 $\pm$ 0.21 & $-$2.92 $\pm$ 0.32 & - \\
5 & North Haze Bubble & $-$2.54 $\pm$ 0.14 & $-$3.24 $\pm$ 0.25 & $-$3.10 $\pm$ 0.18 & $-$3.22 $\pm$ 0.03 & $-$3.18 $\pm$ 0.01 & $-$3.24 $\pm$ 0.11 \\
6 & GCS & $-$2.77 $\pm$ 0.25 & $-$3.40 $\pm$ 0.08 & $-$3.30 $\pm$ 0.10 & $-$3.09 $\pm$ 0.05 & $-$3.08 $\pm$ 0.06 & $-$2.94 $\pm$ 0.10 \\
7 & Rectangle South Haze & $-$2.79 $\pm$ 0.06 & $-$3.50 $\pm$ 0.24 & $-$3.31 $\pm$ 0.13 & $-$3.12 $\pm$ 0.02 & $-$3.10 $\pm$ 0.02 & $-$3.05 $\pm$ 0.02 \\
8 & QJT Rectangle South Haze & $-$2.77 $\pm$ 0.15 & $-$3.50 $\pm$ 0.36 & $-$3.32 $\pm$ 0.23 & $-$3.10 $\pm$ 0.04 & $-$3.11 $\pm$ 0.04 & $-$3.17 $\pm$ 0.07 \\
9 & South Haze Bubble & $-$2.82 $\pm$ 0.13 & $-$3.50 $\pm$ 0.09 & $-$3.26 $\pm$ 0.15 & $-$3.11 $\pm$ 0.06 & $-$3.09 $\pm$ 0.04 & $-$2.97 $\pm$ 0.07 \\
10 & South Haze Bubble clean & $-$2.81 $\pm$ 0.13 & $-$3.54 $\pm$ 0.09 & $-$3.33 $\pm$ 0.13 & $-$3.10 $\pm$ 0.07 & $-$3.09 $\pm$ 0.05 & $-$2.96 $\pm$ 0.07 \\
11 & eRosita West & $-$2.16 $\pm$ 0.22 & $-$3.79 $\pm$ 0.13 & $-$3.43 $\pm$ 0.11 & $-$3.36 $\pm$ 0.03 & $-$3.24 $\pm$ 0.05 & $-$3.41 $\pm$ 0.26 \\
12 & unknown residual & $-$2.74 $\pm$ 0.34 & $-$3.29 $\pm$ 0.16 & $-$3.23 $\pm$ 0.17 & $-$3.28 $\pm$ 0.18 & $-$3.31 $\pm$ 0.11 & $-$3.22 $\pm$ 0.15 \\
13 & eRosita East & $-$2.83 $\pm$ 0.33 & - & - & $-$3.28 $\pm$ 0.04 & $-$3.28 $\pm$ 0.05 & - \\
\hline
\end{tabular}
\caption{Polarization spectral indices in the selected regions obtained with a T-T plots analysis based on data from \textit{Planck} 30\,GHz, \textit{WMAP} K-band, QUIJOTE 11\,GHz, and S-PASS 2.3\,GHz. For the determination of the spectral indices we use the methodology described in Sect.~\ref{sec:tt-plots-fsk}.}
\label{tab:betas_ttplots}
\end{table*}

\section{Summary and Discussion} \label{sec:discussion}

We discuss here the results presented in the previous section, and summarize what we obtained in some specific regions, particularly in the southern Haze area (regions 7--10), in the \nhb{} (region 5), and in North Haze filament (region 3), in intensity and polarization, and with different methodologies.

\subsection{South Haze area}

Previous studies of the Haze emission in intensity have been concentrating in the area below the Galactic centre (region 7 in this work) because of the apparently little complexity and low foregrounds contamination of the intensity signal at \textit{WMAP} and \textit{Planck}-LFI frequencies. However, the S-PASS polarization data (see Fig.\,\ref{figch5:maps_pol_ang_data}) provided a more detailed picture of the area, showing an extended polarized plume in the south (region 9), but also localized contaminated areas that appear to be depolarized, and whose location is indicated in Fig.\,\ref{fig:regions_ttplot_data}. These depolarized areas are, in particular, region "A" identified by \cite{gradP_SPASS} and G353.34, a nearby supernova remnant (see Tab.\,\ref{tab:regions}). Moreover, S-PASS data in polarization show that almost the full southern bubble is affected by Faraday rotation at low frequencies. Indeed, from the polarization angle maps shown in Fig.\,\ref{figch5:maps_pol_ang_data}, we can observe that the polarization angle across the \shb{ (region 9)} has a transition from positive to negative values when comparing the high (23\,GHz and 30\,GHz) and low (2.3\,GHz) frequencies.
In this work, according to these considerations, we identified several regions in the area below the Galactic centre (region 7, 8, 9, 10 - see Sect.~\ref{secch5:regions}) and we studied them with different methodologies, including also the new QUIJOTE data, both in intensity and polarization. 

First of all, we reproduced the analysis of the Haze in intensity by using \textit{Planck} and \textit{WMAP} data in region 7, and applying a similar technique to that in \cite{PlanckHaze}. We obtained a spectrum of the Haze in region 7, using only \textit{Planck} and \textit{WMAP} data (Fig.\,\ref{fig:haze_integrated_spec_noqjt}), with $\beta^H=-2.70\pm0.05$, and of the total synchrotron with $\beta^S=-3.06\pm0.04$. We repeated the same analysis in region 9, which encloses the brightest part of the \shb{}, obtaining $\beta^H=-2.67\pm0.05$ and $\beta^S=-2.99\pm0.05$. From these results we can notice that the intensity Haze spectrum in region 9 (the \shb{}) is consistent with the spectrum in region 7, although the latter is more extended. 

The main aim of this work is the characterization of the Haze with the QUIJOTE data at lower frequencies (e.g., 11 and 13\,GHz). Since QUIJOTE is a ground based experiment located in the northern hemisphere it does not cover the southern sky area enclosing the \shb{} (region 9). However, with QUIJOTE data, we have access to a fraction of region 7, which we call region 8 in this work. In Sect.~\ref{secch5:templatefitting-int} we presented the intensity analysis in this restricted area, including the low frequency QUIJOTE data, at 11 and 13\,GHz. A Haze component is detected in  region 8 as shown in Fig.\,\ref{figch5:res_int}.  The observed excess of diffuse signal is detected with $\sim 9 \sigma$ confidence level, at 11\,GHz.  We computed the spectrum of the emission in this region, as shown in Fig.\,\ref{fig:haze_spec_int}, obtaining a spectral index of the Haze $\beta^H=-2.79\pm0.08$ and of the total synchrotron $\beta^H=-2.98\pm0.04$. The spectrum of the Haze in region 8 is flatter than the total synchrotron by $\Delta \beta = 0.19 \pm 0.09$, with the difference significant at 2\,$\sigma$. The central value of the Haze spectral index in region 8 ($\beta=-2.79 \pm 0.08$) is slightly steeper than that obtained with \textit{WMAP} and \textit{Planck-LFI} data alone in region 7 ($\beta=-2.70 \pm 0.05$), but the difference is not significant.

A similar analysis is also performed, for the first time, in polarization. A map of the polarization residuals is shown in Fig.\,\ref{fig:res_p}, where we can observe residuals across the southern Haze area at QUIJOTE frequencies. The average residual signal in region 8 exceeds the noise level with high significance. The residual structure observed in polarization is slightly displaced with respect to that detected in intensity. The residual plus synchrotron component in region 8 has a spectrum at $11\lesssim \nu \lesssim 70$\,GHz that, if fitted with a simple power-law, has $\beta=-3.07^{+0.37}_{-0.45}$, which is consistent with that of the isolated synchrotron within the large uncertainty, which has $\beta=-2.91\pm0.11$, as shown in Fig.\,\ref{fig:p_spect_templfit}. The difference is $\Delta\beta = 0.16\pm0.43$. 
However when fitting a modified power-law with curvature to the residual plus synchrotron there are hints for a positive curvature, although with low significance.


More solid hints of a positive curvature are observed with the T-T plots analysis shown in Fig.\,\ref{fig:betas_ttplot}, with which we observe a steepening of the polarized emission within the \shb{} at low frequencies (2.3 and 11 GHz). It is evident from Fig.\,\ref{fig:betas_ttplot} that the spectral indices in region 7, 8 and 9, so in the  whole South Haze complex, are flat ("red") at 23--30\,GHz and steep ("blue") at 11--23\,GHz and 2.3--23\,GHz.  This low frequency steepening behaviour, however, is not only valid for the South Haze, but for the full complex associated with the Galactic centre, represented by regions 2--13. The only region where the low frequency steepening is not observed is the NPS, which indeed is thought to be a distinct component from the Haze (see e.g., \citealp{pla2015XXV,Panopoulou2021}) or from activity of the Galactic centre in general. 

We mentioned also about two depolarized spots at 2.3\,GHz, corresponding to region "A" and to the nearby supernova remnant G353.34. In order to check that the determination of the spectral index of the \shb{} is not affected by the presence of these two extra structures in the area, we repeated the T-T plot by masking region "A" and G353.34 (region 10 in Fig.\,\ref{fig:regions_ttplot_data}). As reported in Table\,\ref{tab:betas_ttplots} and in Fig.\,\ref{fig:betas_ttplot}, we obtained $\beta=-2.81\pm0.13$ at 23--30\,GHz and $\beta=-3.10\pm0.07$ at 2.3--30\,GHz, which are in perfect agreement with the spectral indices computed with the same T-T plots methodology in the whole \shb{}, which are $\beta=-2.82\pm0.13$ at 23--30\,GHz and $\beta=-3.11\pm0.06$ at 2.3--30\,GHz. We conclude that the depolarized regions across the South Haze do not bias the spectral index determination of the bubble.


\subsection{\nhb{}}
The \nhb{} (region 5) is the region, among those studied in this paper, with the flattest spectral index. From the analysis with the intensity data, in the range of frequencies 23-60\,GHz\footnote{We do not include QUIJOTE low frequency intensity data here, due to the not well understood structures in the residual, which could be due to atmospheric 1/f noise.} (Fig.\,\ref{fig:haze_integrated_spec_noqjt}, right panel) we measured a spectral index of the Haze  $\beta^H=-2.40\pm0.05$, and of the Haze plus synchrotron $\beta^S=-2.51\pm0.05$. Both $\beta^H$ and $\beta^S$ are far from the typical sky average synchrotron spectral index $\beta \approx -3$, meaning that, in this area and frequency range, the emission of the Haze is dominant over the Galactic diffuse synchrotron. The flat spectral index in this region could also be due to residual free-free emission, which is bright in this  area. However, we can compare this result with the spectral index in polarization between 23 and 30\,GHz, where there is no contamination from free-free emission. We obtained, with the T-T plots in the \nhb{}, a spectral index $\beta=-2.54 \pm 0.14$ between 23 and 30\,GHz. We emphasize that the results derived from T-T plots in polarization are compatible with those derived from the intensity template fitting for the total synchrotron emission, at \textit{WMAP} and \textit{Planck}-LFI frequencies, within  $1\sigma$.  We therefore infer that the observed flat intensity and polarization spectral index in the \nhb{} can be ascribed to the synchrotron emission produced by the Haze component, which dominates over the typical (steeper) synchrotron in this region. 

On the other hand, with the polarization template fitting analysis, we observe that, similarly to region 8, the spectrum of the residual plus synchrotron component fitted with a simple power-law at $11\lesssim \nu \lesssim70$\,GHz in region 5 is compatible,  within the large uncertainties, with that of the synchrotron alone, as shown in Fig.\,\ref{fig:p_spect_templfit} (left panel). The difference of the spectral indices is $\Delta\beta = 0.06\pm0.65$.

However, T-T plots have shown that the spectral index of the total emission between 23 and 30\,GHz is significantly flatter than that at lower frequencies, especially when including 2.3\,GHz data in the analysis, providing hints of a detection of curvature of the spectrum across this region.
This behaviour could be originated by a double electron population that generates the polarized synchrotron signal in region 5: one with a flat ($\beta\sim-2.5$) spectra index that dominates in the frequency range 20\,GHz\,--\,44\,GHz, and one with a steeper spectrum ($\beta \sim -3.2$) that emerges at $\nu < 20$\,GHz. QUIJOTE data provide a characterization that is compatible with that presented by \cite{SpassHaze} based on S-PASS data, fitting well with the interpretation presented in \cite{Crocker2015}. 
According to \cite{Crocker2015}  the observed emission is produced by: i) shock re-accelerated young cosmic-rays electrons that are responsible for the flat (or hard) synchrotron emission of the microwave Haze; ii) an old population of cosmic-rays electrons that escape the contact discontinuity of the shock, and emit the steeper synchrotron radiation observed in the S-PASS plume; iii) colliding hadrons enclosed in the contact-discontinuity surface that radiate the $\gamma$-rays, which is what we observe in the \textit{Fermi} bubbles.  A $\gamma$-ray component of IC emision, from the same electrons that radiate the microwave Haze, is also present, but it is subdominant. This model agrees with the results obtained in this work for both the North and South polarized lobes.


\subsection{Comparison between South and North Haze bubbles}

An interesting consideration is connected with the recent results presented by \cite{jew2020}, who computed with a novel technique the spectral indices of the North and South Haze bubbles between 30 and 44\,GHz, using \textit{Planck} data. They reported a difference between the polarization spectral index of the two bubbles, being $\beta=-2.36\pm0.09$ in the North and $\beta=-3.00\pm0.05$ in the South Haze.\footnote{Note that the regions studied in \cite{jew2020} do not perfectly match with ours. They integrated two approximately symmetric bubbles in the north and in the south corresponding to the $\gamma$-ray \textit{Fermi} bubbles, while we restrict our analysis to the brightest region of the plumes as observed at low frequency (2.3\,GHz), in polarization.}  In this work, we measure an asymmetry of the spectral indices of the northern and southern Haze bubbles in intensity, in the frequency range 23-60\,GHz, consistent with what \cite{jew2020} found in polarization. We obtain a total synchrotron index $\beta=-2.51\pm0.05$ in the \nhb{}, and $\beta=-2.99\pm0.05$ in the \shb{}, as shown in Fig.\,\ref{fig:haze_integrated_spec_noqjt} using only \textit{WMAP} and \textit{Planck} data. There is consistency between our total synchrotron intensity spectrum and the polarization spectrum at  30-44\,GHz measured by \cite{jew2020}. 

In addition, our results with T-T plots confirm that the asymmetry between the North and South Haze bubbles is also seen in polarization, at 23-30\,GHz: the spectral index across the \nhb{} is $\beta = -2.54\pm0.14$, and in the \shb{} it is $\beta=-2.82\pm0.13$. The \shb{} has a steeper spectrum than the \nhb{}, both in intensity and polarization. Interestingly, at lower frequencies, this trend is inverted. The T-T plots between 2.3\,GHz and 23\,GHz show that the polarization spectrum of the \nhb{} ($\beta=-3.22\pm0.03$) is slightly steeper than that in the \shb{} ($\beta=-3.11\pm0.06$), with a difference $\Delta \beta = 0.11 \pm 0.07$ $(1.6\sigma$).

\subsection{North Haze filament}
The North Haze filament (region 3) is an interesting case of study. It corresponds to the structure identified by \cite{Vidal2015} and \cite{pla2015XXV} as the filament surrounding the northern \textit{Fermi} bubble in $\gamma$-rays, and the microwave Haze in the north. In Sect.~5.5 of \cite{pla2015XXV}, a measurement of the spectral index of the filament using \textit{Planck} 30\,GHz and \textit{WMAP} K-band data is reported. They measured $\beta=-2.54\pm0.16$ with T-T plots of the unbiased $P_{\mathrm{MAS}}$ maps at 30 and 23\,GHz, with a methodology that is essentially the same as that described in Sect.~\ref{secch5:tt-plots-pmas}. In this work, we performed the measurement in the same region, but using a different technique as described in Sect.~\ref{sec:tt-plots-fsk}, obtaining $\beta=-2.66\pm0.18$ (see Table\,\ref{tab:betas_ttplots} and Fig.\,\ref{fig:betas_ttplot}).

In the attempt of reproducing the result of \cite{pla2015XXV} with T-T plots of $P_{\mathrm{MAS}}$, we identified a possible source of bias that can introduce significant differences in the estimate of the $\beta$. Although the polarization amplitude $P_{\mathrm{MAS}}$ is not affected by noise bias if computed as in Eq.\,\ref{eq:pmas}, it is a positive quantity, and, in regions with low signal to noise
the determination of the spectral index with the classical T-T plot methodology (Sect.~\ref{secch5:tt-plots-pmas}) can be biased. In addition, when allowing uncertainties in both axes, the correlation between zero-level and slope is strong.
For example, in the northern Haze filament (region 3), we obtained $\beta=-2.84 \pm 0.08$ using a T-T plot of the $P_{\mathrm{MAS}}$ maps, where we allowed to fit both the slope and an offset between the maps.  However, if the zero level of $P_{\mathrm{MAS}}$ is correctly set, as it is in this case where we first adjust the relative offset of the $Q$ and $U$ maps with the reference, we can force the intercept of the T-T plot slope to be zero. In this case, we obtain a value of the spectral index $\beta=-2.67\pm0.03$, which is inconsistent with the previous result, where the intercept was free to vary, but which is in full agreement  with the result reported in Sect.~\ref{secch5:tt-plots-results} (Table\,\ref{tab:betas_ttplots}), obtained with the method described in Sect.~\ref{sec:tt-plots-fsk} based on \cite{fsk2014}. 

\begin{figure}
    \centering
    \includegraphics[width=0.48\textwidth]{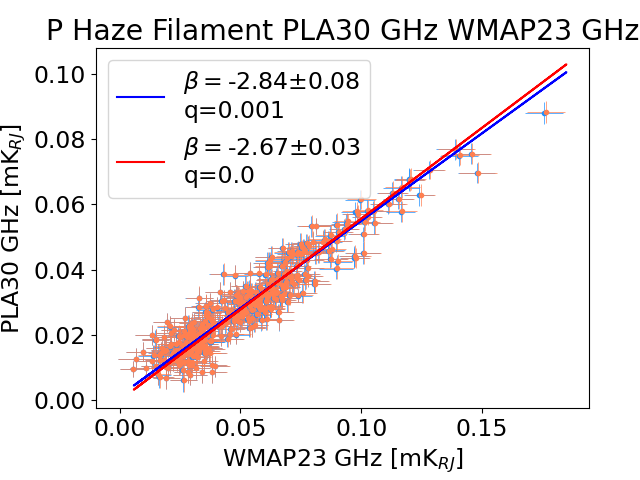}
    \caption{T-T plot with $P_{\mathrm{MAS}}$ in the North Haze filament (region 3), comparing the case where we fix or do not fix the offset of the linear fit, to illustrate the effect of Eddington bias.}\label{fig:tt_plot_filament}
    \label{fig:ttplot_mas_filament}
\end{figure}

In Fig.\,\ref{fig:tt_plot_filament} we show the T-T plot of the unbiased polarization amplitude in the filament (region 3). The blue line is a fit of the points for both the slope and offset. The red line is the fit over the same data points, but where the intercept is forced to be zero.  We can clearly see that the red and blue lines have a different slope, and that therefore the recovered spectral indices are also different.  The small difference between the red and blue data points is due to the different colour correction that is applied to the same initial data, given that the spectral index resulting from the two fitting methodologies is different. Our final conclusion is that T-T plot of the positive definite unbiased polarization amplitude ($P_{\mathrm{MAS}}$) could be affected by noise bias, which can be mitigated by fitting the T-T plot with the intercept fixed to zero. The  method applied in this work (described in  Sect.~\ref{sec:tt-plots-fsk} and based on \citealp{fsk2014,fsk2019}), instead, is not significantly biased, since it uses  T-T plots of a combination of \textit{Q} and \textit{U} data, which can be both positive and negative avoiding zero level problems. 

\section{Conclusions}\label{sec:conclusions}

We derived the spectral properties of the microwave Haze with different methodologies, in intensity and in polarization. For this we used, for the first time, new Haze observations from the QUIJOTE experiment, at 11 and 13\,GHz. In combination we used the publicly available S-PASS (2.3\,GHz), \textit{WMAP} (23--61\,GHz) and \textit{Planck}-LFI (30--70\,GHz) data. 


We computed the spectrum of the Haze after applying template fitting. We used the intensity data of QUIJOTE, \textit{WMAP} and \textit{Planck}-LFI. In particular we measured the Haze spectrum in the South, in the overlap with the QUIJOTE sky coverage, within region 8.   We obtained a synchrotron spectrum with spectral index $\beta=-2.79\pm0.08$ (see Fig.\,\ref{fig:haze_spec_int}) at frequencies 11--60\,GHz. As a general trend, we find that the spectrum of the Haze component in intensity is flatter than the typical diffuse synchrotron emission. Our results in the whole south Haze area (region 7), however, are in slight tension with those obtained by \cite{WMAPHaze} and \cite{PlanckHaze} in the same region, as we  obtain a spectral index of the Haze that is steeper than that obtained from previous works (e.g., $\beta = -2.70\pm0.05$ from this work and $\beta \sim -2.56\pm0.05$ from \citealp{PlanckHaze}, at frequencies 23--60\,GHz).

We also studied the intensity spectrum of the Haze in the North and South bubbles (regions 5 and 9), but excluding QUIJOTE data that are not available or possibly affected by noise artifacts. We measured a synchrotron emission from a Haze component with a spectral index at 23--60\,GHz of $\beta^H=-2.40\pm 0.05$ in the \nhb{} and $\beta^H=-2.67\pm0.05$ in the \shb{}, as shown in Fig.\,\ref{fig:haze_integrated_spec_noqjt}. We measured for the first time in intensity a difference of the spectral index in the North and South bubbles, in the frequency range 23--60\,GHz, with a significance of $\approx 4\sigma$. This behaviour is also observed in polarization in this work between 23 and 30\,GHz, in agreement with previous results by \cite{jew2020} between 30 and 44\,GHz.

In polarization, we studied the spectra of the Haze-related structures with both a template fitting (see Fig.\,\ref{fig:res_p} and \ref{fig:p_spect_templfit}) and correlation T-T plot technique (see Table\,\ref{tab:betas_ttplots} and  Fig.\,\ref{fig:betas_ttplot}) providing spectral indices for different frequency pairs. After fitting a synchrotron and dust component to the QUIJOTE maps, we observed residual polarized structures that are inconsistent with the expected noise level with high significance (see Fig.\,\ref{fig:res_p}). This result can be interpreted as a hint of curvature of the synchrotron spectral index across the area where we observe the positive residuals. However a study of the spectral properties in regions 5 and 8 does not have sufficient signal-to-noise to show clear differences between the synchrotron and residual plus synchrotron spectral indices (see Fig.\,\ref{fig:p_spect_templfit}). We also tried to constrain a curvature parameter, which is obtained to be compatible with zero given the large error bars.

However, our results based on T-T plots (Fig.\,\ref{fig:betas_ttplot}) show flat spectrum regions across the Haze area at 23--30\,GHz, and an evident steepening at low frequencies, in agreement with \cite{SpassHaze}. 
We observed in Fig.\,\ref{fig:betas_ttplot} that the Haze-related structures (regions 2--13) are significantly flat ($-2.8 \lesssim \beta \lesssim -2.6$)  compared with the sky average synchrotron ($\beta \sim-3$) at 23--30\,GHz, while at lower frequencies (11--30\,GHz and 2.3--30\,GHz) the spectrum of the Haze steepens significantly ($-3.2 \lesssim \beta \lesssim -3.0$). On the other hand the NPS (region 1), which is thought to be a nearby supernova shell not related with the Haze structures, shows the opposite behaviour: its spectral index does not show significant differences across the frequencies presented in this work.


Our results in polarization are compatible with those presented in \cite{SpassHaze}. They can be therefore interpreted with the model presented in \cite{Crocker2015}, according to whom young cosmic ray electrons enclosed in the contact discontinuity of a shock generated by Galactic centre nuclear activity radiate the flat synchrotron of the Haze, while older cosmic ray electron escaping the contact discontinuity produce the steeper synchrotron observed in the S-PASS lobes.  However, in intensity, we do not observe a change of the Haze spectral index as we do see in polarization. The intensity spectrum in region 8 is well characterized by a single power-law with $\beta^H\sim-2.8$. Further investigation is needed to understand this behaviour. Possibly the use of stellar absorption like in \cite{Panopoulou2021}, or the modelling of the magnetic field and of the cosmic rays as proposed by the IMAGINE Consortium \citep{Boulanger2018} could help to formulate a more comprehensive interpretation  of this complex area.

\section*{Acknowledgements}
The QUIJOTE experiment is being developed by the Instituto de Astrofisica de Canarias (IAC),
the Instituto de Fisica de Cantabria (IFCA), and the Universities of Cantabria, Manchester and Cambridge.
We thank the staff of the Teide Observatory for invaluable assistance in the commissioning and operation of QUIJOTE.
Partial financial support was provided by the Spanish Ministry of Science and Innovation 
under the projects AYA2007-68058-C03-01, AYA2007-68058-C03-02,
AYA2010-21766-C03-01, AYA2010-21766-C03-02, AYA2014-60438-P,
ESP2015-70646-C2-1-R, AYA2017-84185-P, ESP2017-83921-C2-1-R,
AYA2017-90675-REDC (co-funded with EU FEDER funds),
PGC2018-101814-B-I00, 
PID2019-110610RB-C21, PID2020-120514GB-I00, IACA13-3E-2336, IACA15-BE-3707, EQC2018-004918-P, the Severo Ochoa Programs SEV-2015-0548 and CEX2019-000920-S, the
Maria de Maeztu Program MDM-2017-0765, and by the Consolider-Ingenio project CSD2010-00064 (EPI: Exploring
the Physics of Inflation). We acknowledge support from the ACIISI, Consejeria de Economia, Conocimiento y 
Empleo del Gobierno de Canarias and the European Regional Development Fund (ERDF) under grant with reference ProID2020010108.
This project has received funding from the European Union's Horizon 2020 research and innovation program under
grant agreement number 687312 (RADIOFOREGROUNDS).
This research made use of computing time available on the high-performance computing systems at the IAC.
We thankfully acknowledge the technical expertise and assistance provided by the Spanish Supercomputing Network 
(Red Espa\~nola de Supercomputaci\'on), as well as the computer resources used: the Deimos/Diva Supercomputer, located at the IAC.
FG acknowledges funding from the European Research Council (ERC) under the European Union’s Horizon 2020 research and innovation programme (grant agreement No 101001897). 
EdlH acknowledge partial financial support from the \textit{Concepci\'on Arenal Programme} of the Universidad de Cantabria. 
%
FP acknowledges support from the Spanish State Research Agency (AEI) under grant number PID2019-105552RB-C43.
BR-G acknowledges ASI-INFN Agreement 2014-037-R.0.
DT acknowledges the support from the Chinese Academy of Sciences President's International Fellowship Initiative, Grant N. 2020PM0042.
This work has made use of S-band Polarisation All Sky Survey (S-PASS) data. 
Some of the results in this paper have been derived using the {\sc HEALPix}  \citep{Gorski2005} and {\sc healpy} \citep{healpy} packages. We also use {\tt Numpy} \citep{numpy}, and {\tt Matplotlib} \citep{matplotlib}.
%

\section*{DATA AVAILABILITY}
The QUIJOTE raster scan data used in this paper are property
of the QUIJOTE Collaboration and can only be shared on 
request to the corresponding authors. The QUIJOTE wide-survey maps will be made publicly available in the first QUIJOTE data
release, as detailed in \citet{mfiwidesurvey}. Ancillary data employed in the analysis for this paper are publicly available and can be accessed as referred along the paper text.



\bibliographystyle{mnras}
\bibliography{biblio_haze,quijote} 


\clearpage

\appendix


\section{Intensity and polarization maps used in the analysis}\label{secA:maps}

\begin{figure*}
    \centering
    \includegraphics[width=0.32\textwidth]{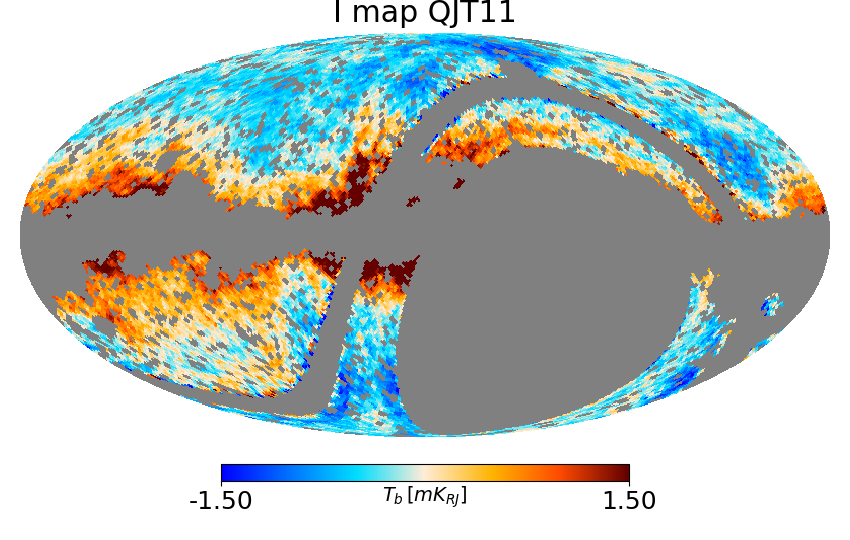}
    \includegraphics[width=0.32\textwidth]{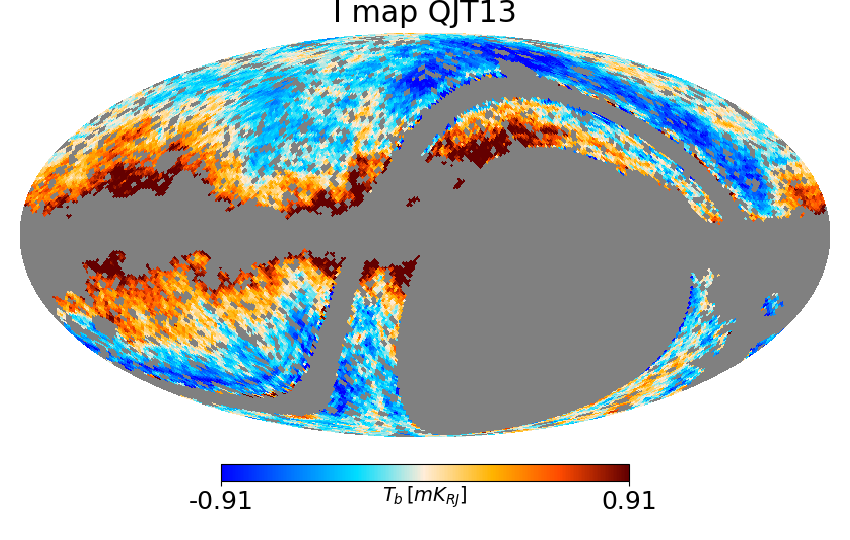}
    \includegraphics[width=0.32\textwidth]{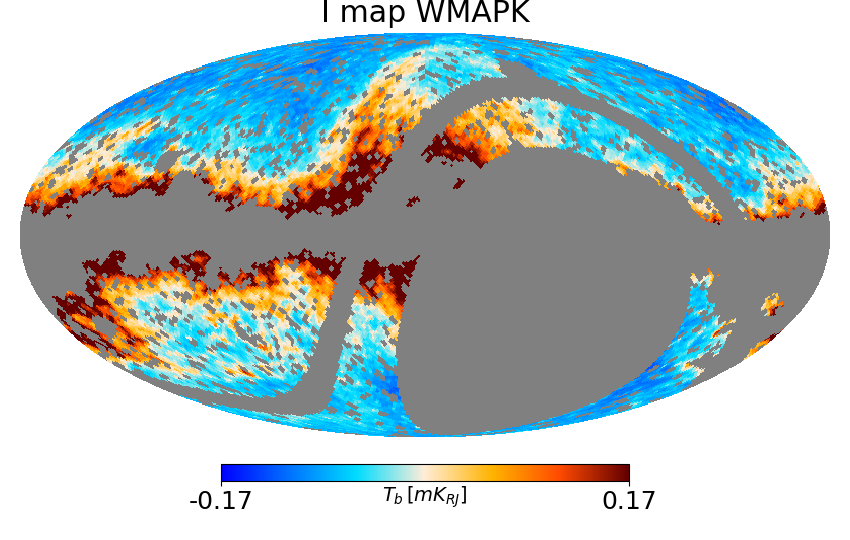}
    \includegraphics[width=0.32\textwidth]{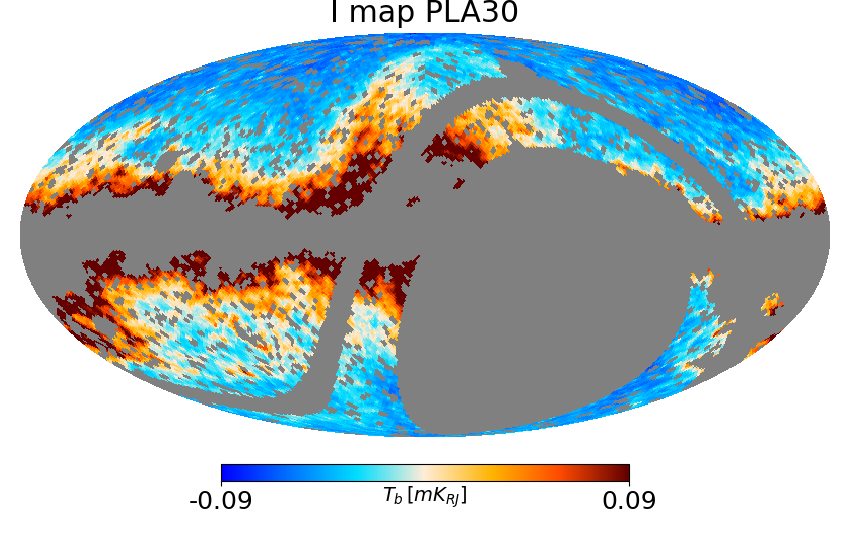}
    \includegraphics[width=0.32\textwidth]{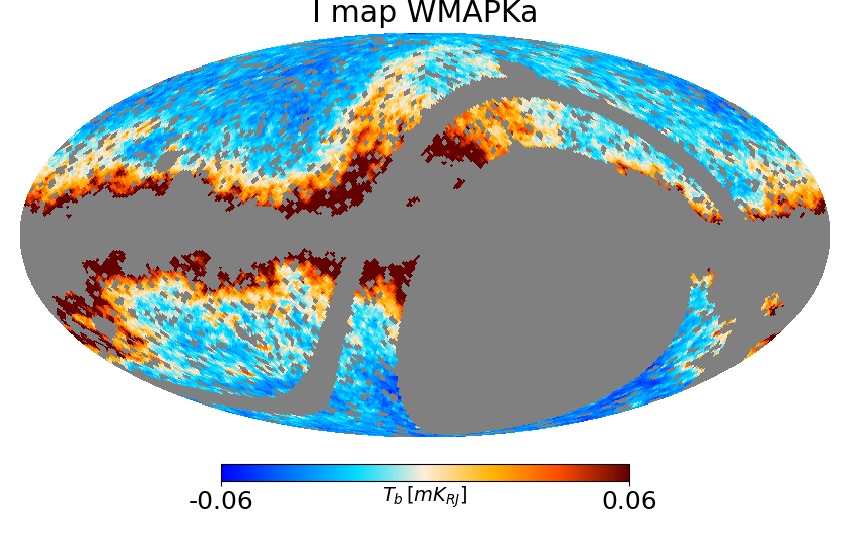}
    \includegraphics[width=0.32\textwidth]{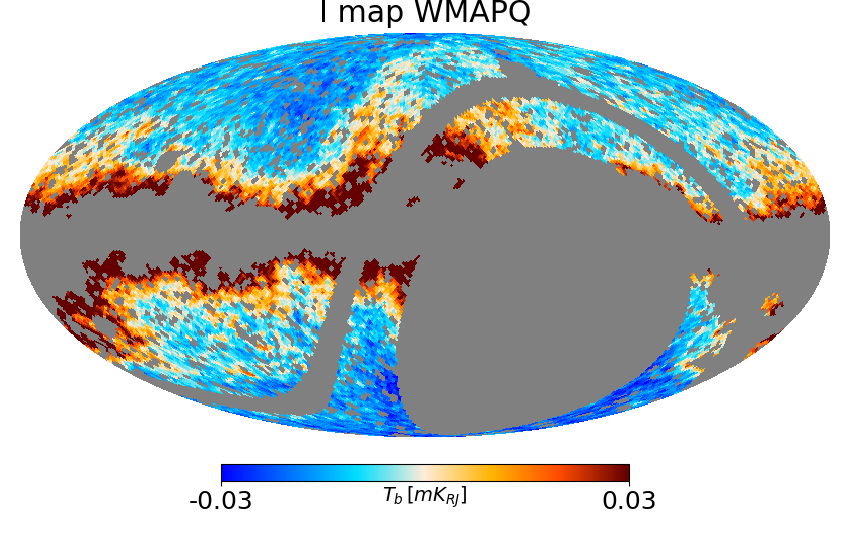}
    \includegraphics[width=0.32\textwidth]{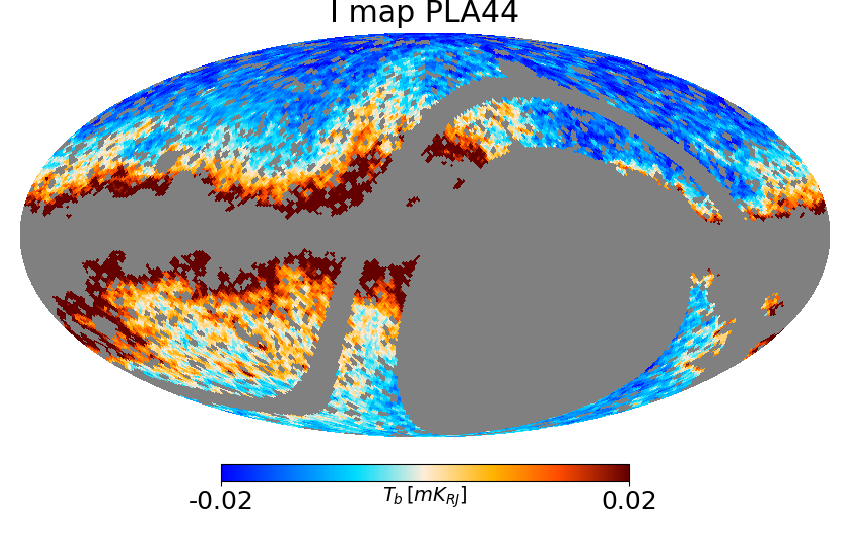}
    \includegraphics[width=0.32\textwidth]{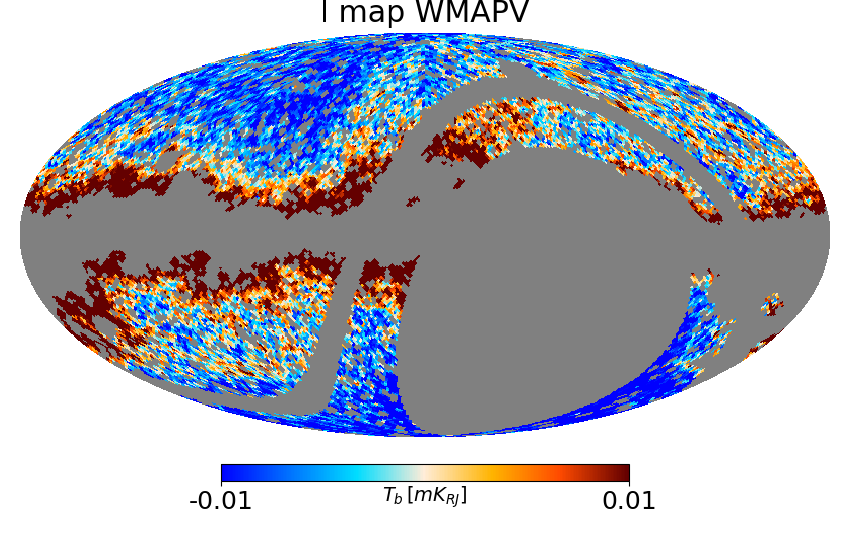}
    \includegraphics[width=0.32\textwidth]{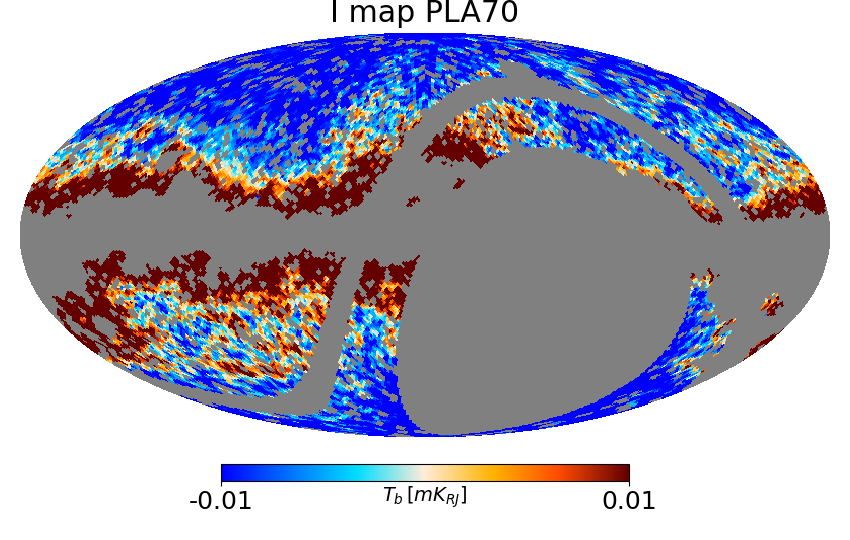}
    \caption{Intensity maps ordered in frequency: QUIJOTE 11\,GHz, QUIJOTE 13\,GHz, \textit{WMAP} K-band, \textit{Planck} 30\,GHz, \textit{WMAP} Ka-band, \textit{WMAP} Q-band, \textit{Planck} 44\,GHz, \textit{WMAP} V-band, \textit{Planck} 70\,GHz. The maps are in units of mK Rayleigh-Jeans, and the colorbar range values are scaled with a synchrotron-like power-law: $1.5\cdot(\nu/11\mathrm{\,GHz})^\beta$, with $\beta=-3$. The grey area represents the mask that is used for the analysis, which is a combination of the QUIJOTE sky coverage with the free-free and CMB mask, as described in Sect.\,\ref{secch5:templates}.}
    \label{figch5:maps_int}
\end{figure*}

\begin{figure*}
    \centering
    \includegraphics[width=0.32\textwidth]{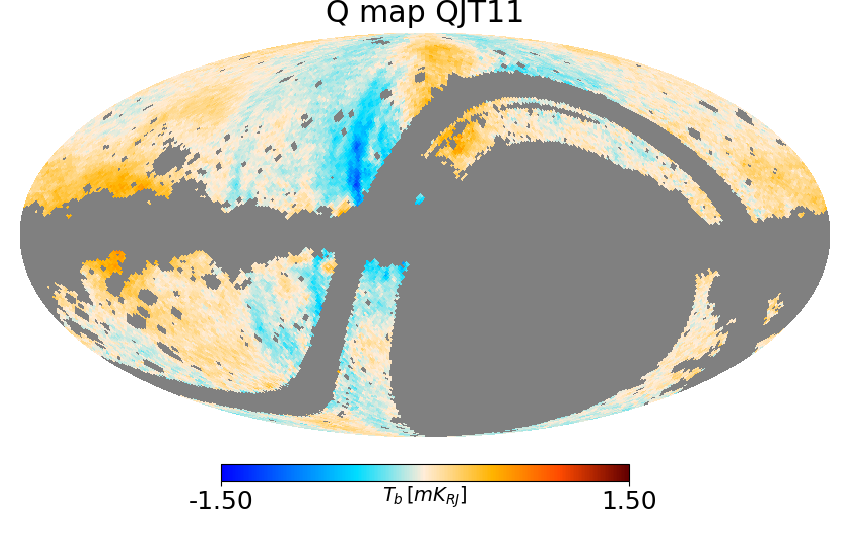}
    \includegraphics[width=0.32\textwidth]{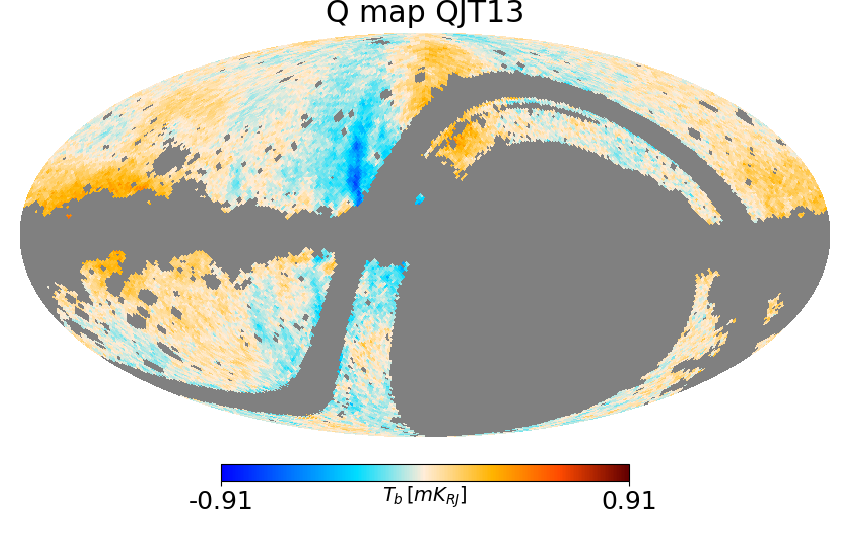}
    \includegraphics[width=0.32\textwidth]{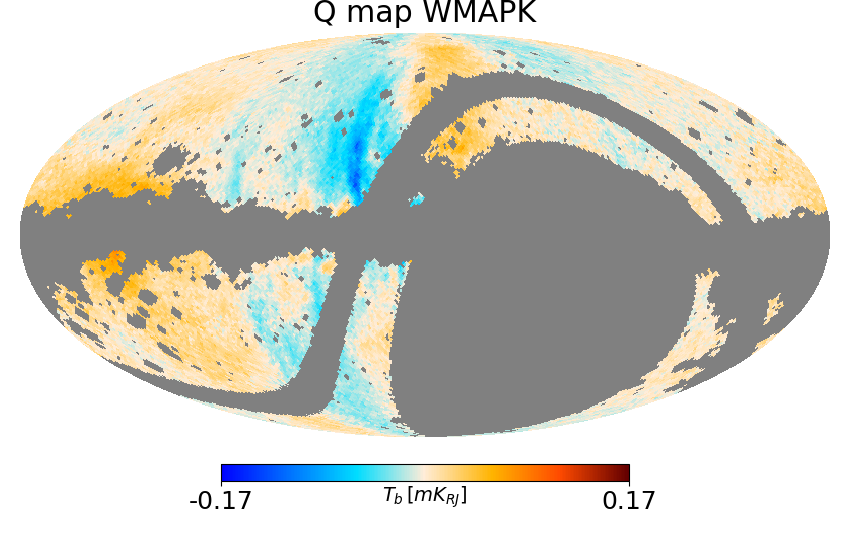}
    \includegraphics[width=0.32\textwidth]{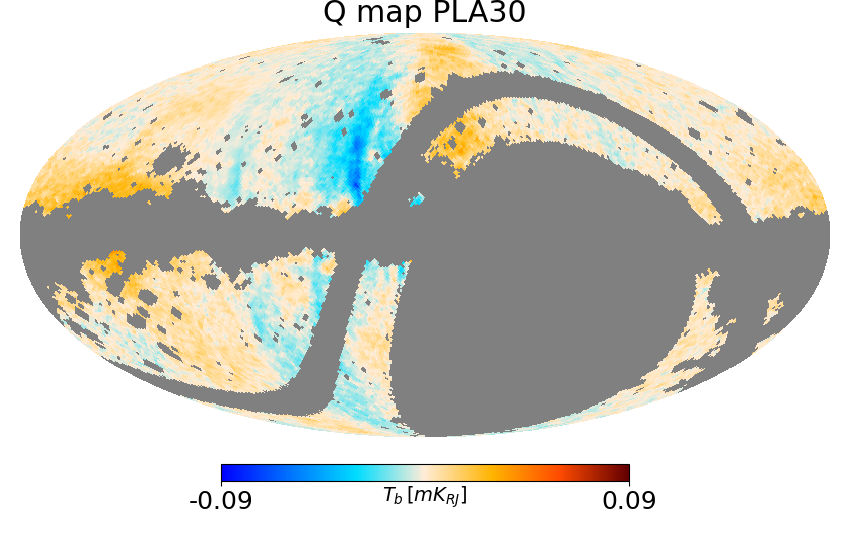}
    \includegraphics[width=0.32\textwidth]{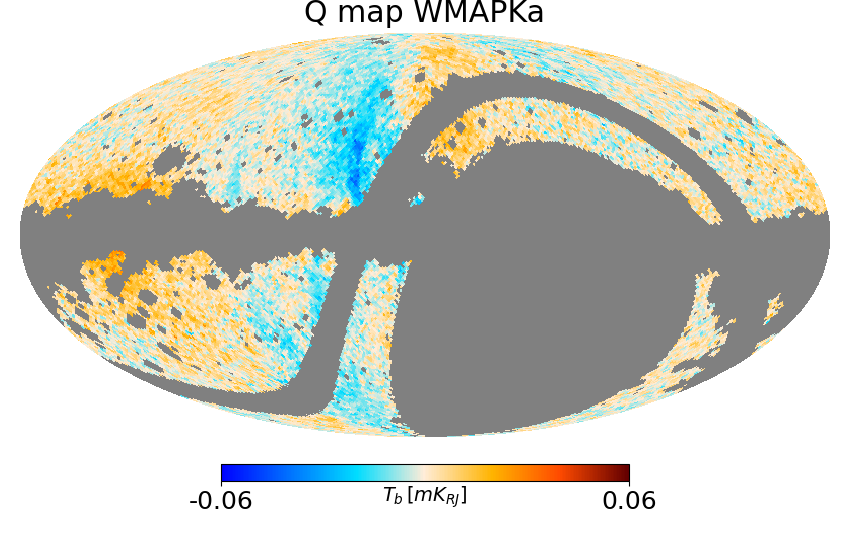}
    \includegraphics[width=0.32\textwidth]{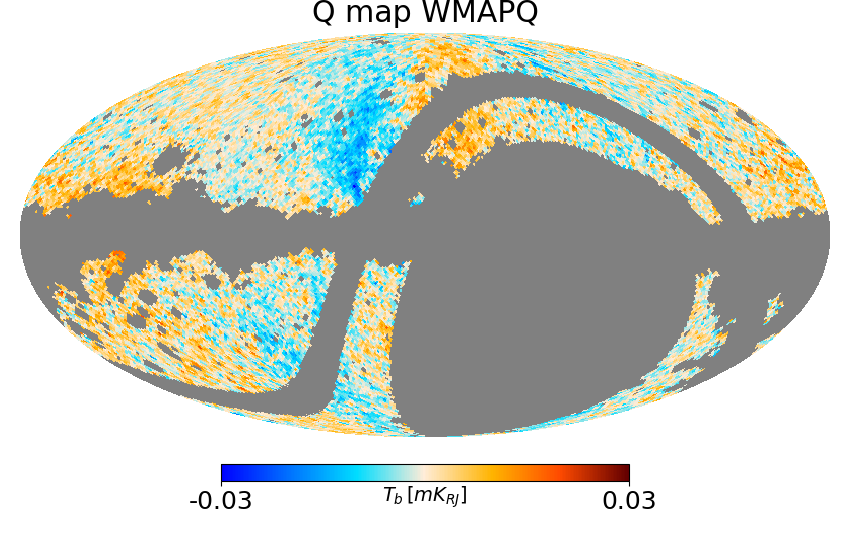}
    \includegraphics[width=0.32\textwidth]{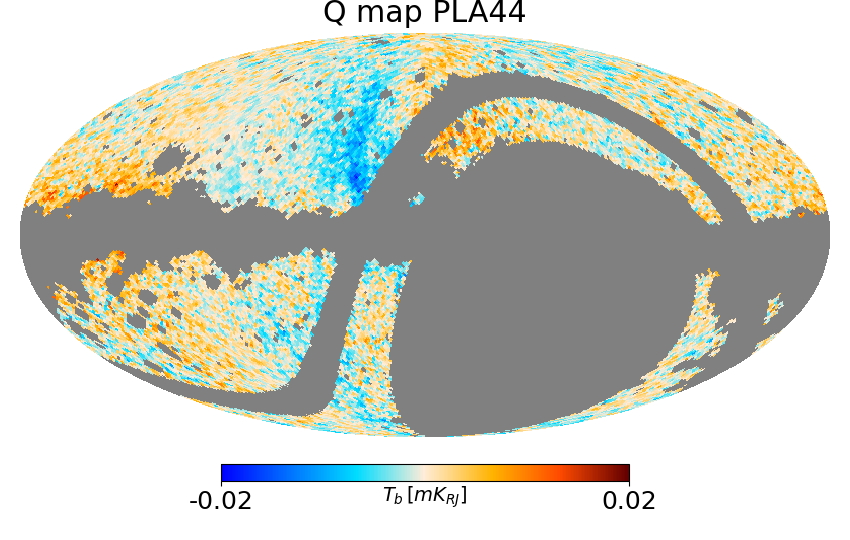}
    \includegraphics[width=0.32\textwidth]{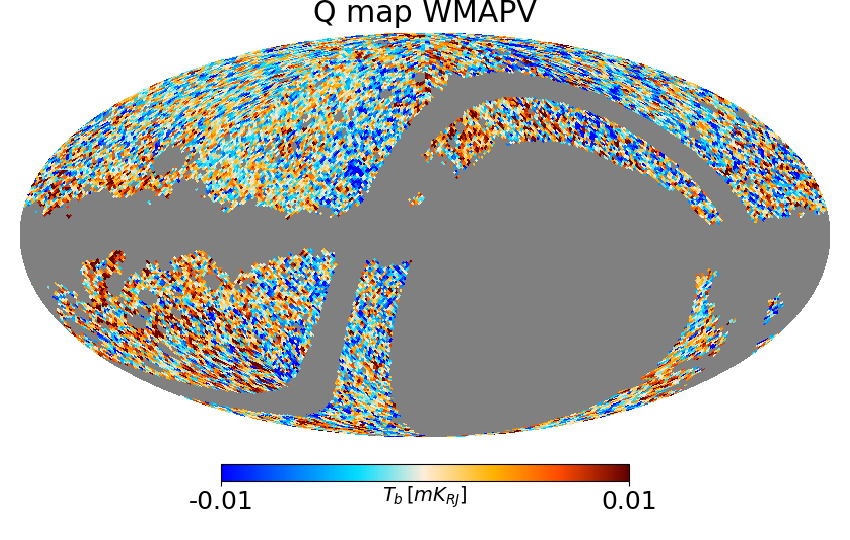}
    \includegraphics[width=0.32\textwidth]{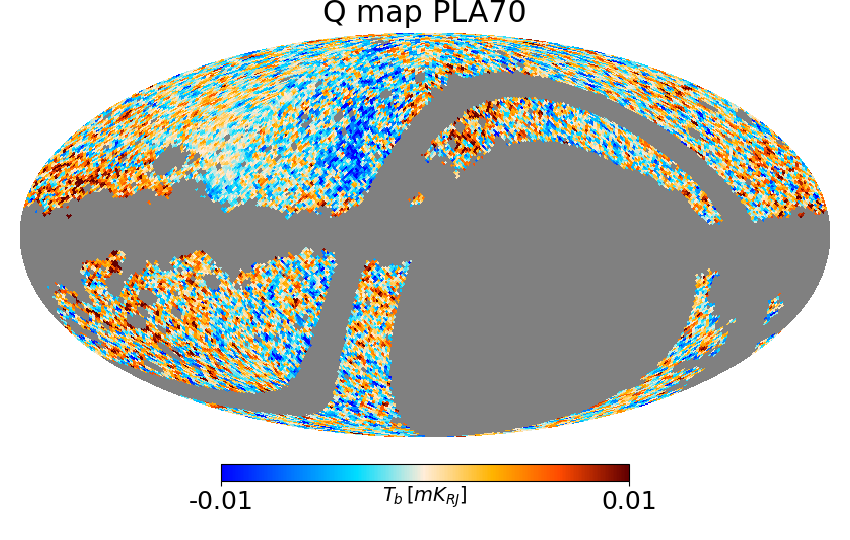}
    \caption{Same as in Fig.\,\ref{figch5:maps_int} for Stokes $Q$ maps.}
    \label{figch5:maps_q}
\end{figure*}

\begin{figure*}
    \centering
    \includegraphics[width=0.32\textwidth]{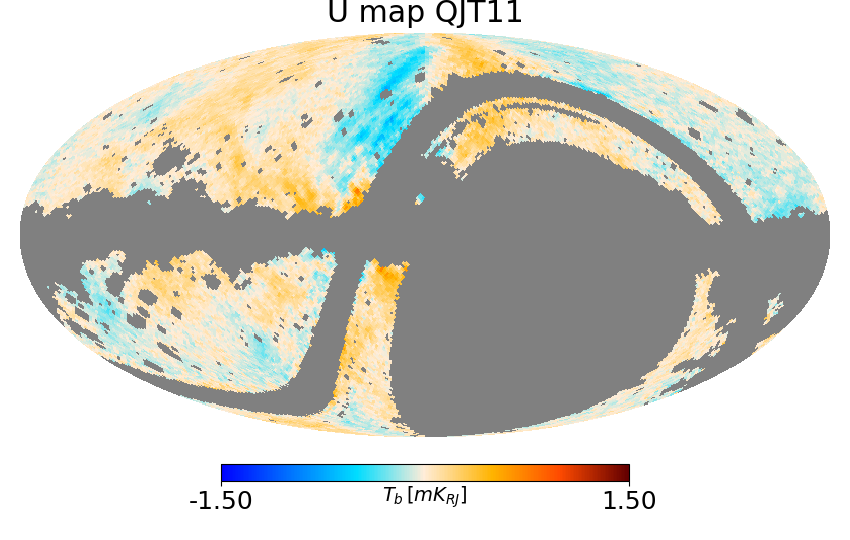}
    \includegraphics[width=0.32\textwidth]{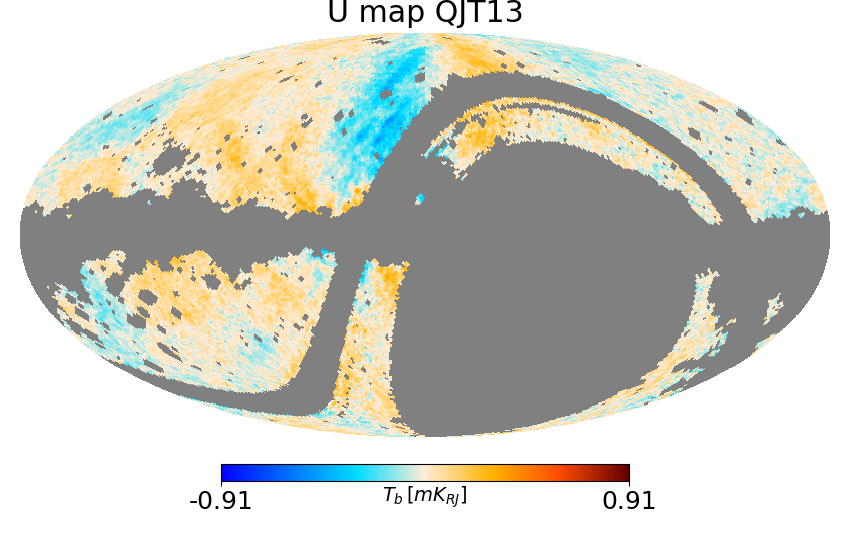}
    \includegraphics[width=0.32\textwidth]{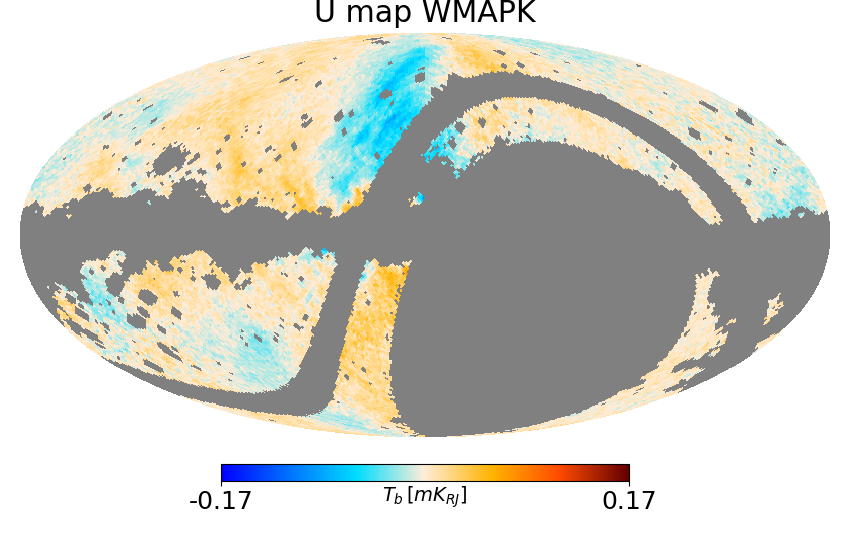}
    \includegraphics[width=0.32\textwidth]{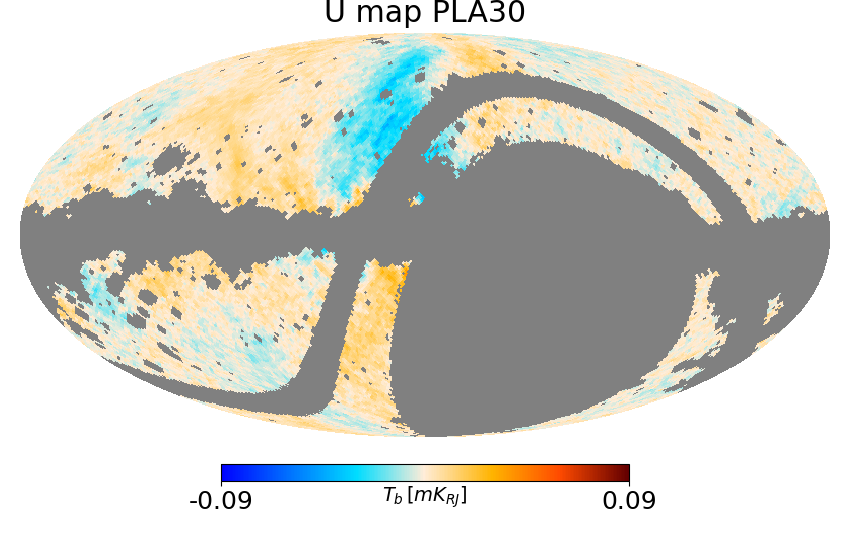}
    \includegraphics[width=0.32\textwidth]{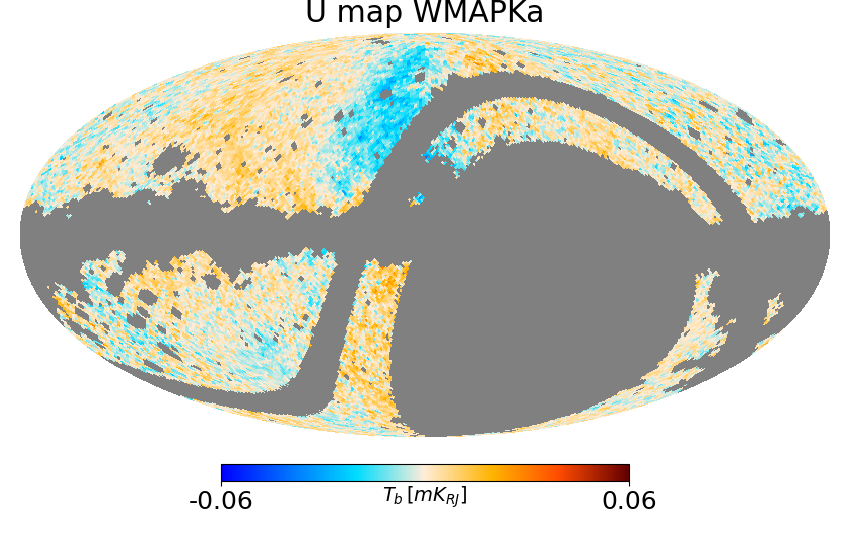}
    \includegraphics[width=0.32\textwidth]{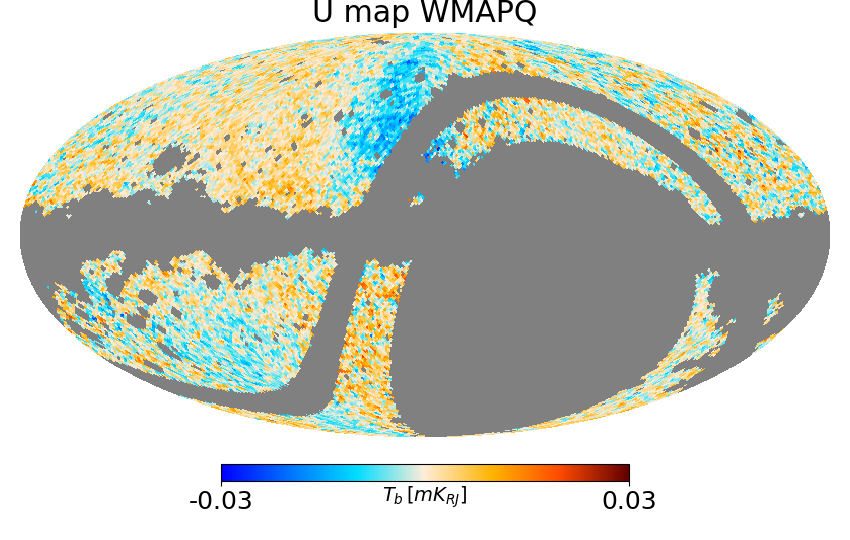}
    \includegraphics[width=0.32\textwidth]{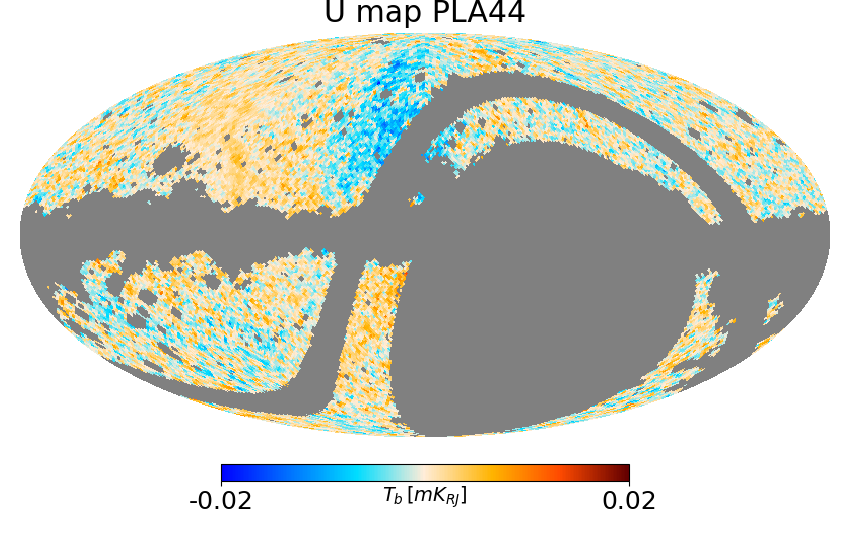}
    \includegraphics[width=0.32\textwidth]{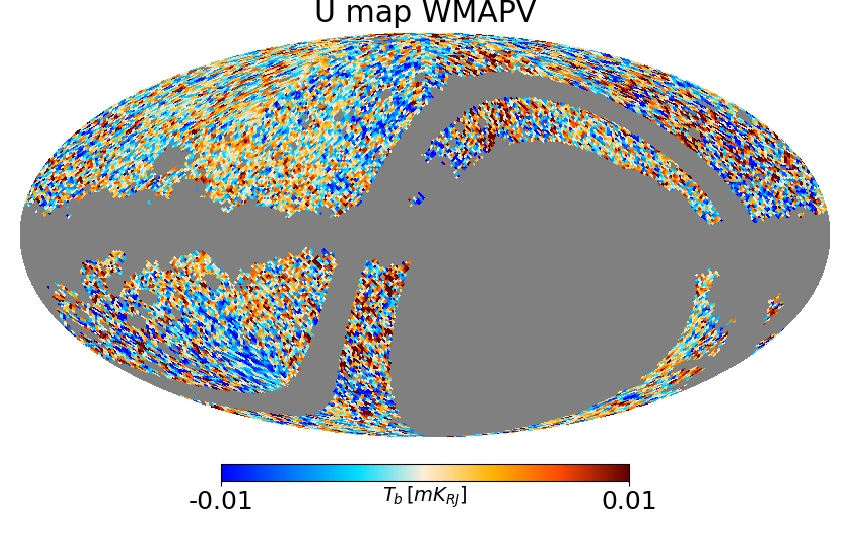}
    \includegraphics[width=0.32\textwidth]{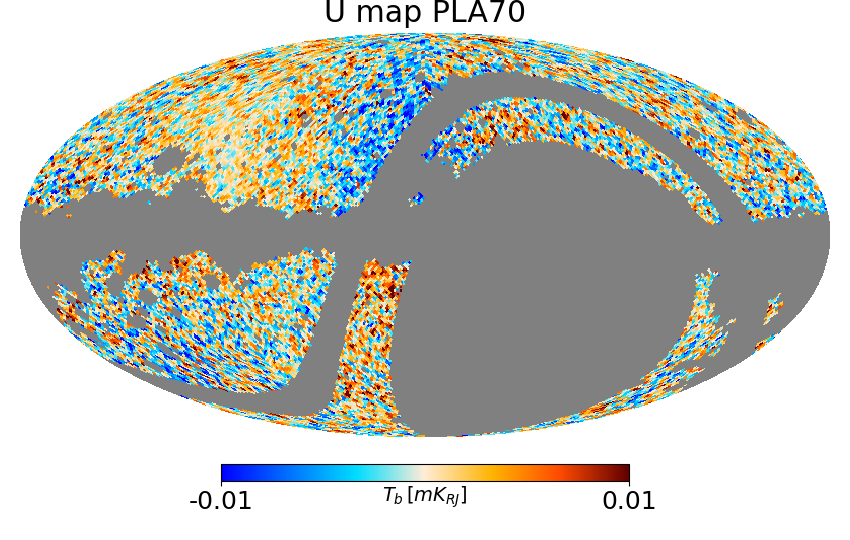}
    \caption{Same as in Fig.\,\ref{figch5:maps_int} for Stokes $U$ maps.}
    \label{figch5:maps_u}
\end{figure*}

Figs.\,\ref{figch5:maps_int}, \ref{figch5:maps_q} and \ref{figch5:maps_u} show the \textit{I}, \textit{Q} and \textit{U} maps that have been used for the first part of the study presented in this paper, which applies a template fitting technique (see Sect.~\ref{secch5:templ_fitting}) to the QUIJOTE, \textit{WMAP} and \textit{Planck}-LFI maps (results in Sect.~\ref{secch5:templatefitting-int} and \ref{secch5:templatefitting-pol}). 

\begin{figure*}
    \centering
    \includegraphics[width=0.32\textwidth]{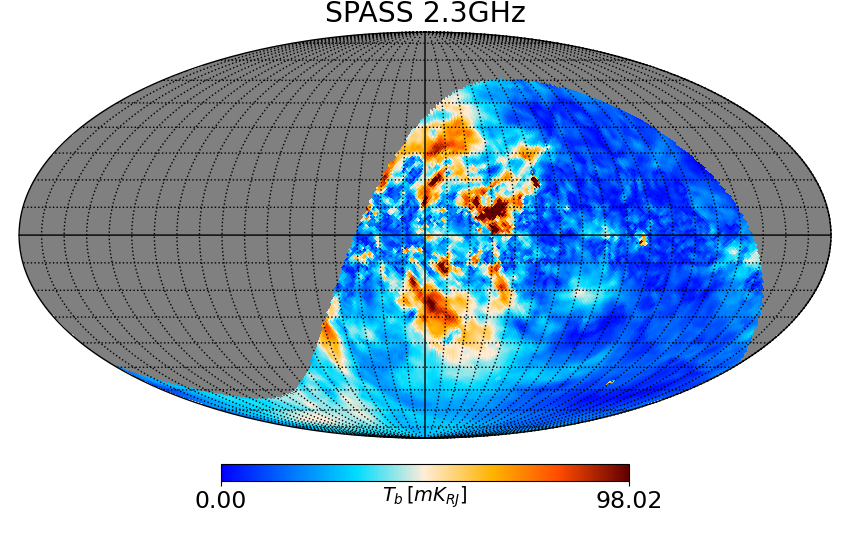}
    \includegraphics[width=0.32\textwidth]{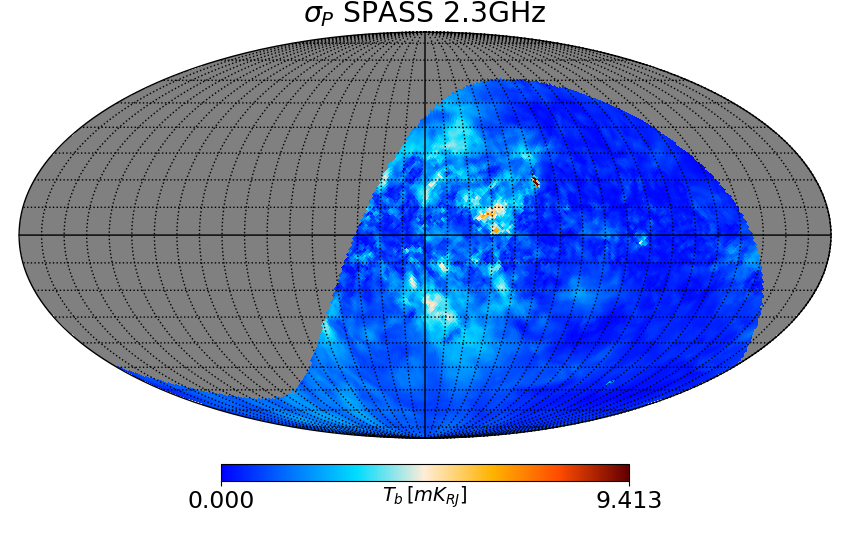}
    \includegraphics[width=0.32\textwidth]{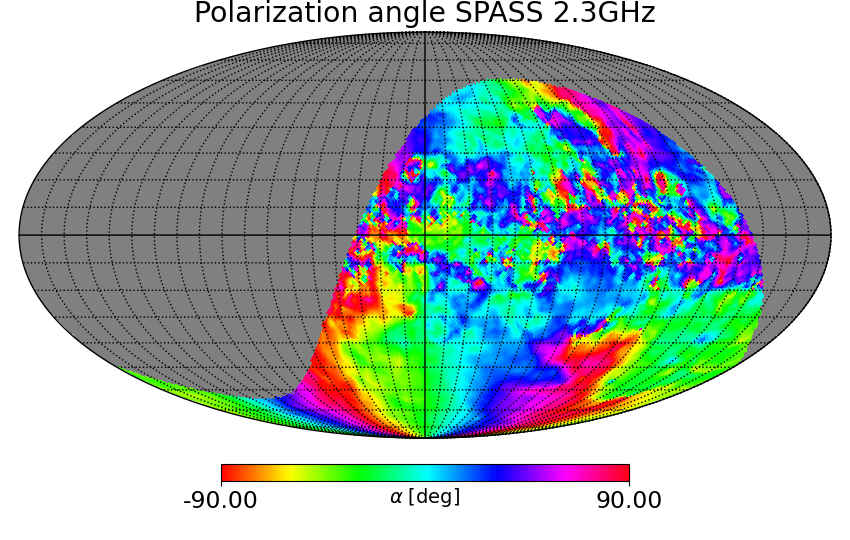}

    \includegraphics[width=0.32\textwidth]{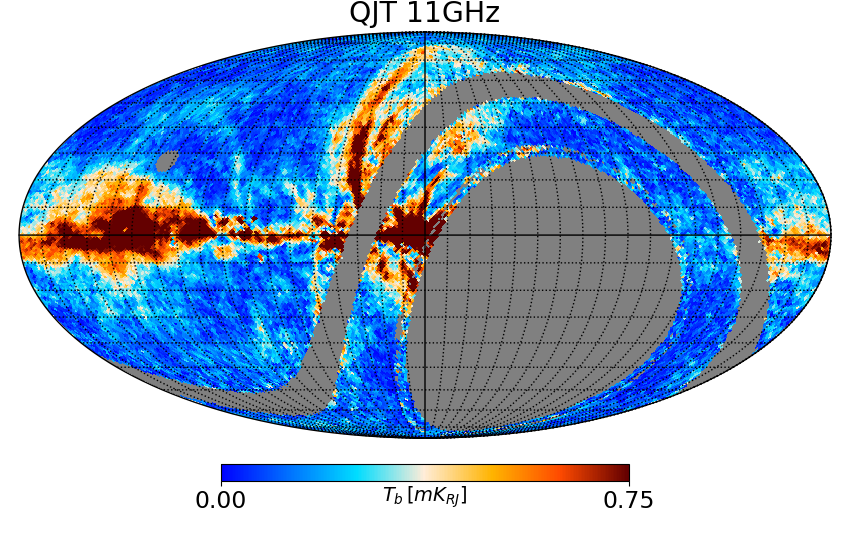}
    \includegraphics[width=0.32\textwidth]{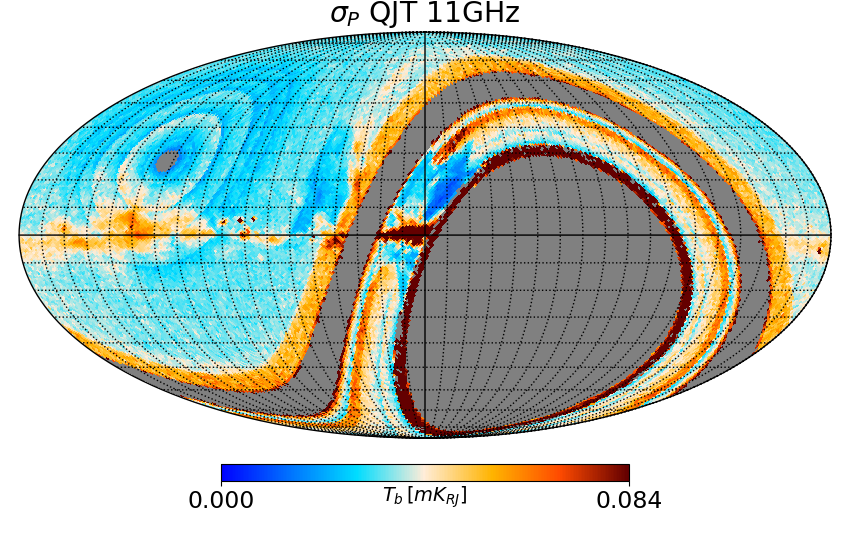}
    \includegraphics[width=0.32\textwidth]{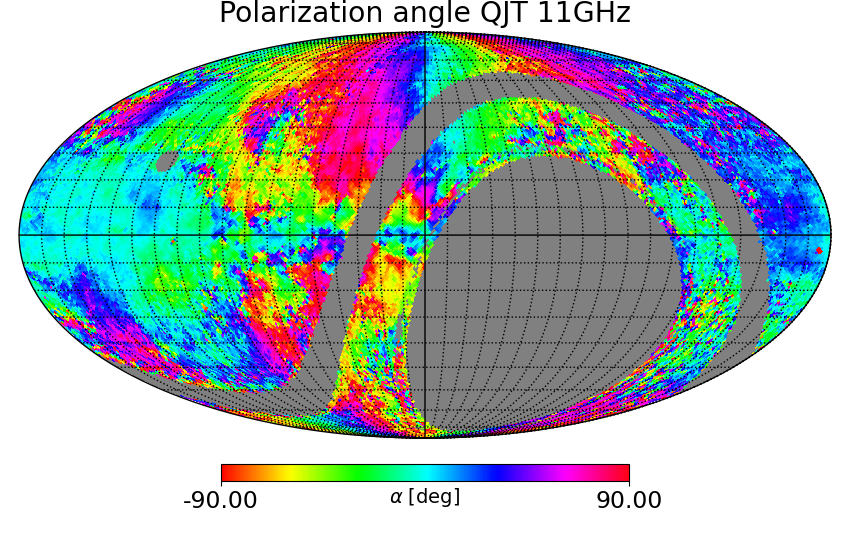}

    \includegraphics[width=0.32\textwidth]{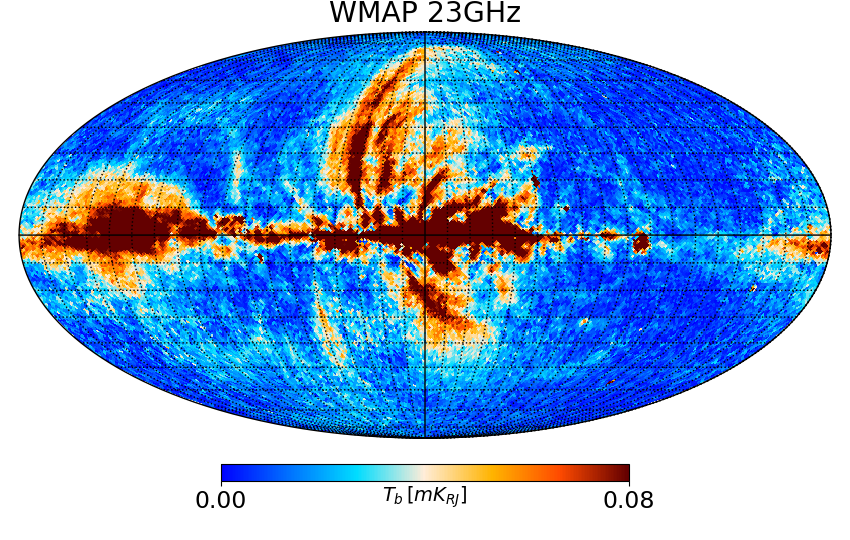}
    \includegraphics[width=0.32\textwidth]{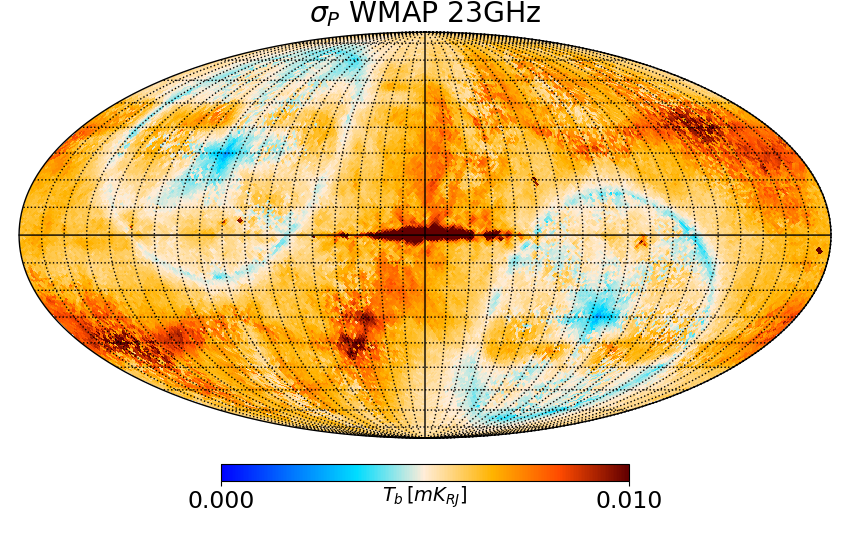}
    \includegraphics[width=0.32\textwidth]{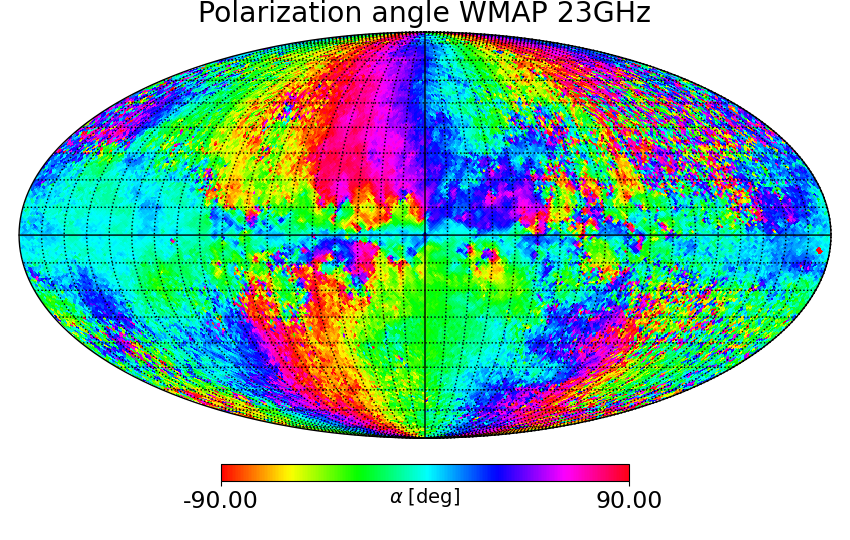}

    \includegraphics[width=0.32\textwidth]{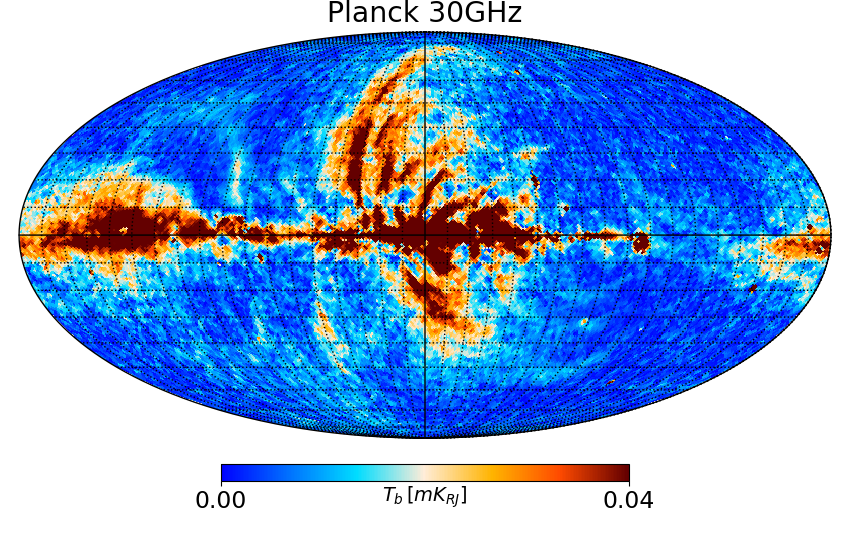}
    \includegraphics[width=0.32\textwidth]{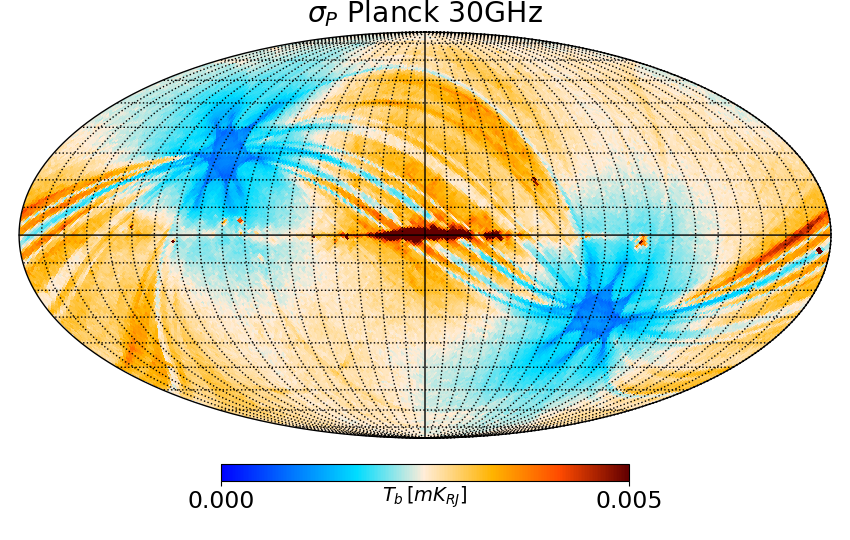}
    \includegraphics[width=0.32\textwidth]{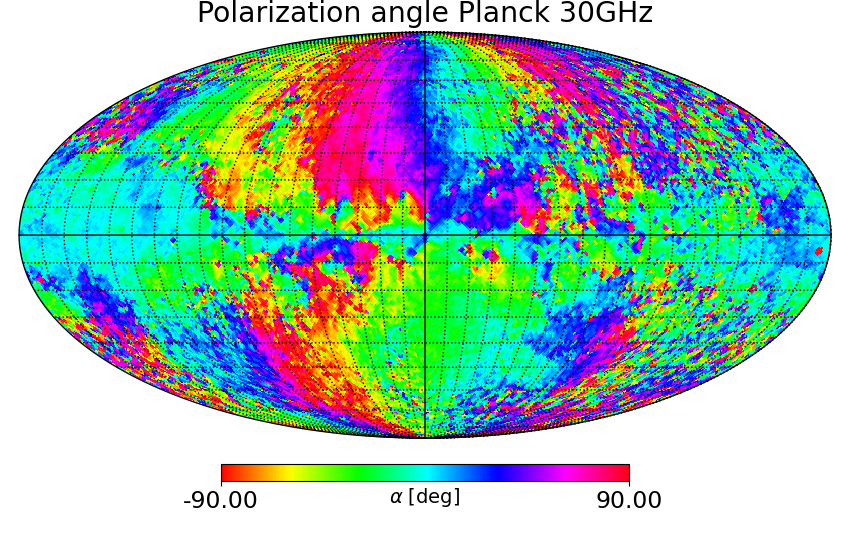}
    
    \caption{Debiased $P$ maps (left), corresponding error maps including calibration uncertainty (centre), and polarization angle (right) of S-PASS at 2.3\,GHz, QUIJOTE at 11\,GHz, \textit{WMAP} at 23\,GHz, and \textit{Planck} at 30\,GHz. }     \label{figch5:maps_pol_ang_data}
\end{figure*}

Fig.\,\ref{figch5:maps_pol_ang_data} shows the debiased polarization amplitude $P_{\mathrm{MAS}}$ maps computed as described in Sect.~\ref{secch5:tt-plots} (Eq.\,\ref{eq:pmas}), the corresponding uncertainties (centre) computed with Eq.\,\ref{eq:err_pmas} and accounting (in quadrature) for the calibration uncertainty, and the polarization angle (right). We show maps at the four frequencies that we selected to perform a study of the polarization spectral index analysis with T-T plots, which are, from top to bottom: S-PASS at 2.3\,GHz, QUIJOTE 11\,GHz, \textit{WMAP} K-band and \textit{Planck} 30\,GHz. 

\begin{figure*}
    \centering
    \includegraphics[width=0.32\textwidth]{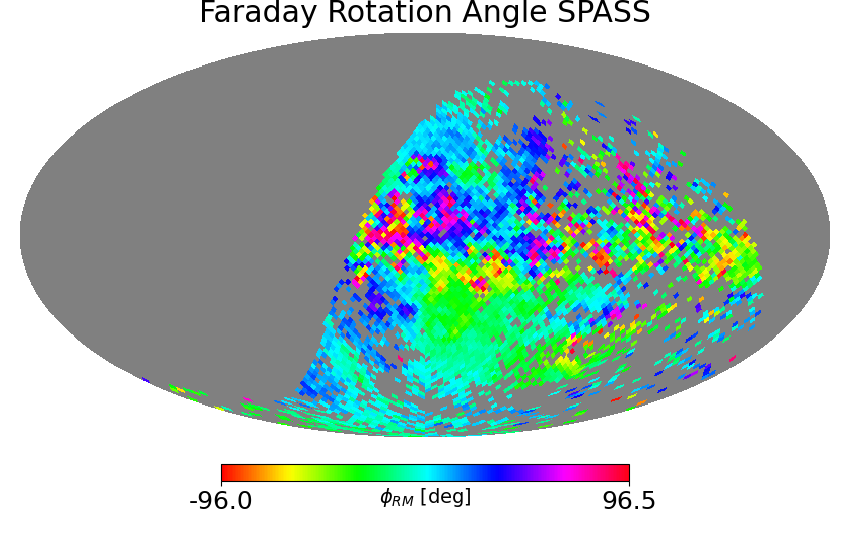}
    \includegraphics[width=0.32\textwidth]{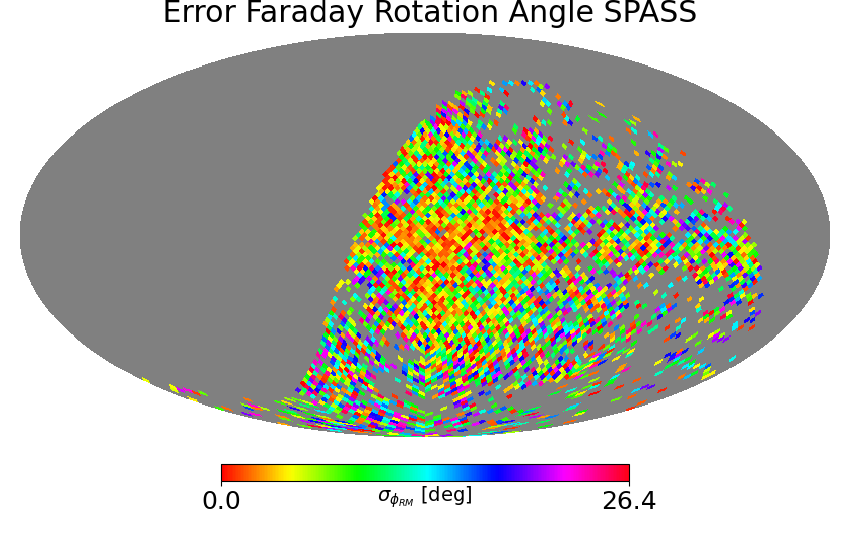}
    \includegraphics[width=0.32\textwidth]{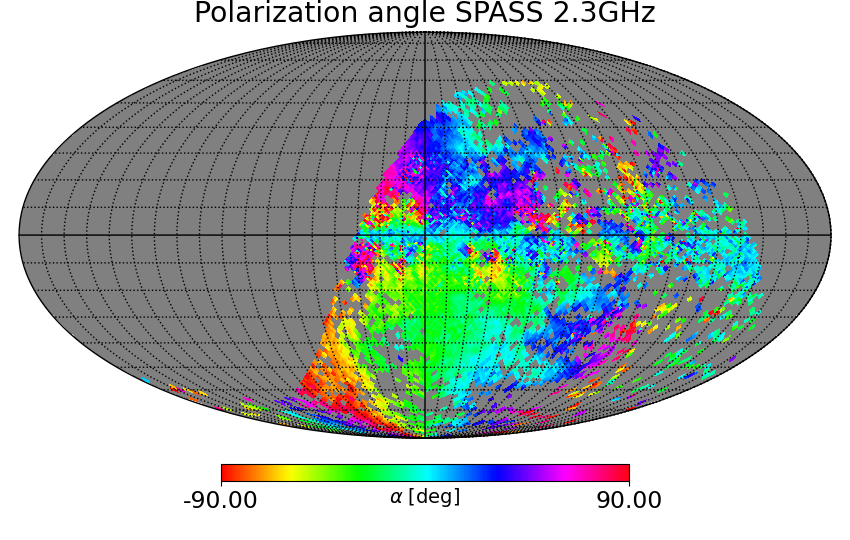}
    \caption{S-PASS Faraday rotation angle (left) and uncertainty (centre), in units of degrees, and following the CMB convention on polarization angles. The right panel shows the S-PASS polarization angle map after correcting the Faraday rotation.} 
    \label{fig:SPASS_FRang}
\end{figure*}

Fig.~\ref{fig:SPASS_FRang} shows the Faraday rotation angle map at 2.3\,GHz (left), and the corresponding uncertainty (centre), obtained from the rotation measure map derived from S-PASS data as $\phi_{FR}=RM\cdot\lambda^2$ (see Sect.~\ref{sec:FR}). The blank pixels are those where no $RM$ is provided, and they are therefore excluded from the analysis. 
The same figure also shows, on the right, the S-PASS Faraday rotation corrected polarization angle map, which is very similar to the polarization angle maps at higher frequencies shown in the right panels of Fig.\,\ref{figch5:maps_pol_ang_data}. 

\section{Faraday rotation correction to S-PASS}\label{sec:FR}

Low frequency photons suffer the effect of Faraday rotation and depolarization along their path across a magnetized interstellar medium, before they reach the observer. \cite{SPASSmaps}, \cite{gradP_SPASS} and \cite{fsk2019} discussed in detail these effects in the S-PASS data, which clearly shows depolarization in the Galactic plane, and rotation also at high Galactic latitudes. As we anticipated in Sect.~\ref{secch5:data-overview}, in this work we account for these effects by correcting the Faraday rotation and masking regions with evident depolarization. 

In order to correct for the Faraday rotation at 2.3\,GHz, we applied a backwards rotation of the \textit{Q} and \textit{U} maps of S-PASS, by the Faraday rotation angle $\phi_{FR}=RM\cdot\lambda^2$, where $RM$ is the rotation measure map delivered by the S-PASS collaboration\footnote[1]{\url{https://sites.google.com/inaf.it/spass/healpix-maps}} \citep{SPASSmaps} and $\lambda$ is the S-PASS observed wavelength. We produced an independent RM measurement using S-PASS, QUIJOTE, \textit{WMAP} and \textit{Planck} data. We obtained results that are consistent with the S-PASS RM map across the North and South Haze bubbles. 

The rotation is applied as follows:
\begin{equation}
\left(
\begin{array}{c}
     Q' \\
     U'
\end{array} \right) = \left(
\begin{array}{cc}
      \cos(2\phi_{RM}) & \sin(2\phi_{RM}) \\
      -\sin(2\phi_{RM}) &  \cos(2\phi_{RM})
\end{array} \right)
\left(
\begin{array}{c}
     Q \\
     U
\end{array} \right),
\label{eq:FR}
\end{equation}
where \textit{Q} and \textit{U} are the original S-PASS maps, and \textit{Q'} and \textit{U'} are the corrected ones.\footnote[2]{Note that, following what is now common practice in the CMB field, we apply the CMB convention for the polarization angle, while the S-PASS maps are delivered with the IAU convention. 
This inverts the sign of the \textit{U} map, and therefore also rotates the polarization angle in the opposite direction relative to North. } The final uncertainty of the \textit{Q'} and \textit{U'} maps is the propagation of the uncertainty of \textit{Q}, \textit{U} and $\phi_{RM}$, through Eq.\,\ref{eq:FR}:
\begin{align}
    \sigma^2_{Q'}=&\sigma_Q^2 \cos^2(2\phi_{RM})+\sigma_U^2 \sin^2(2\phi_{RM})\\
    &+4\sigma_{\phi_{RM}}^2(U\cos(2\phi_{RM})-Q\sin(2\phi_{RM}))^2\notag
\end{align}
\begin{align}
    \sigma^2_{U'}=&\sigma_Q^2 \sin^2(2\phi_{RM}) +\sigma_U^2 \cos^2(2\phi_{RM})\\
    &+4\sigma_{\phi_{RM}}^2(U\sin(2\phi_{RM})+Q\cos(2\phi_{RM}))^2\notag.
\end{align}
The pixels where no $RM$ is provided are excluded from the analysis. This correction however is not sufficiently accurate at low Galactic latitudes where Faraday rotation angle could be larger than 90\,deg, but our analysis is focused on the diffuse emission far from the Galactic plane, therefore this is not a critical issue for the stability of the results.

We show in appendix \ref{secA:maps} (Fig.~\ref{fig:SPASS_FRang}) the Faraday rotation angle map at 2.3\,GHz (left). The corresponding uncertainty is also shown in the same figure (centre), as well as the S-PASS polarization angle map after correcting for Faraday rotation (right). This last map can be compared with the polarization angle maps at higher frequencies (right panels in Fig.\,\ref{figch5:maps_pol_ang_data}), showing that, after applying the correction, the spatial distribution of the S-PASS polarization angles is very similar to that of \textit{WMAP} and \textit{Planck}.

\section{Null tests}\label{secA:nulltets}
\begin{figure*}
    \centering
    \includegraphics[width=0.48\textwidth]{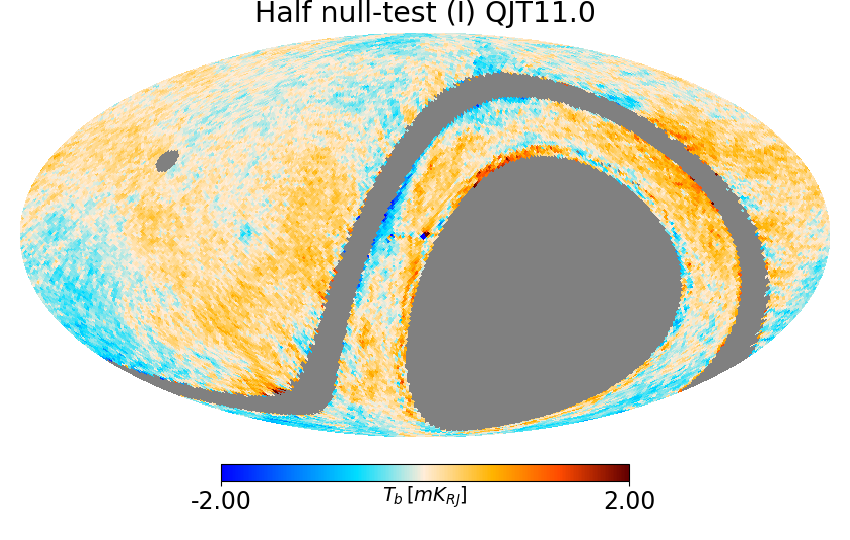}
    \includegraphics[width=0.48\textwidth]{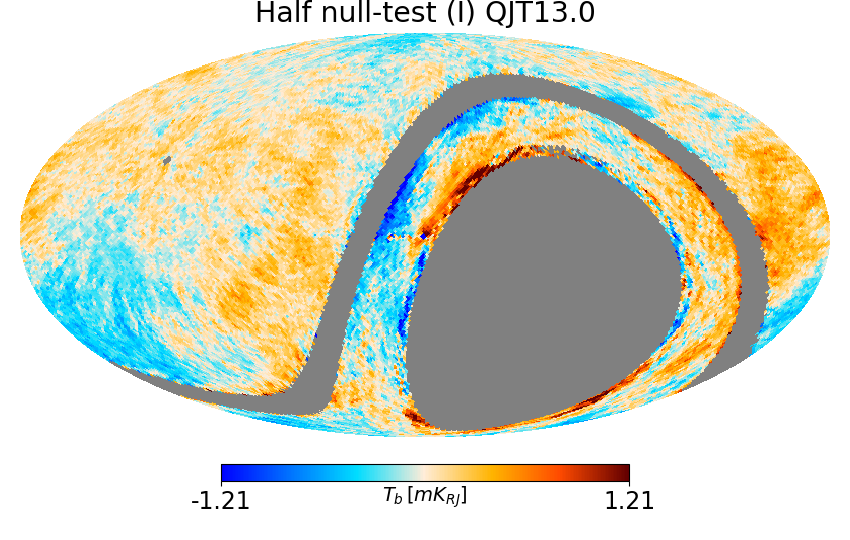}
    \includegraphics[width=0.48\textwidth]{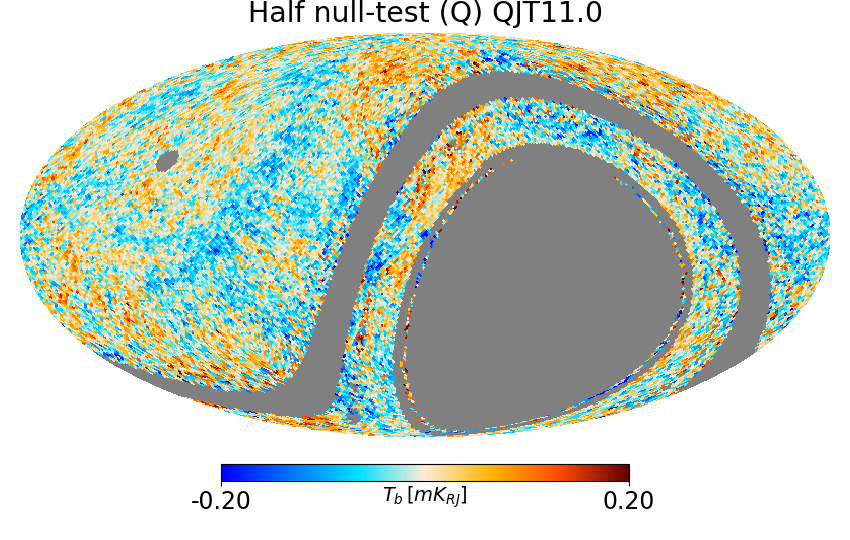}
    \includegraphics[width=0.48\textwidth]{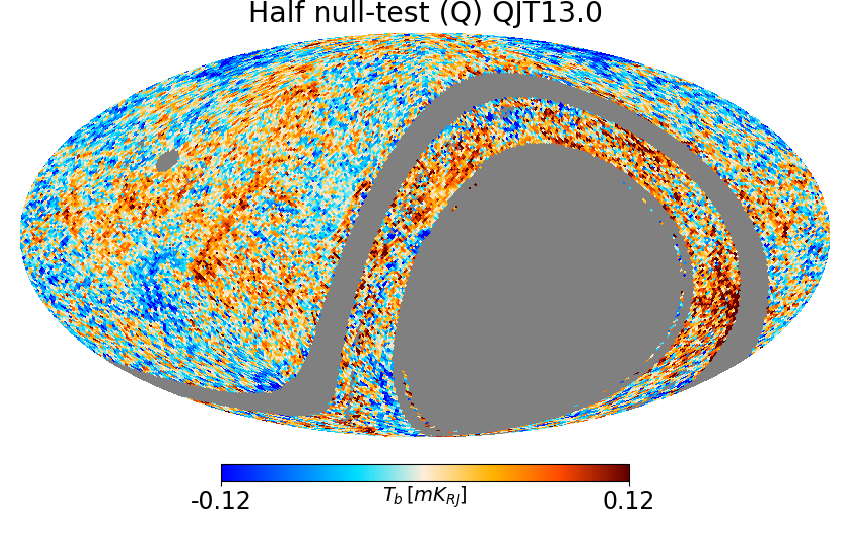}
    \includegraphics[width=0.48\textwidth]{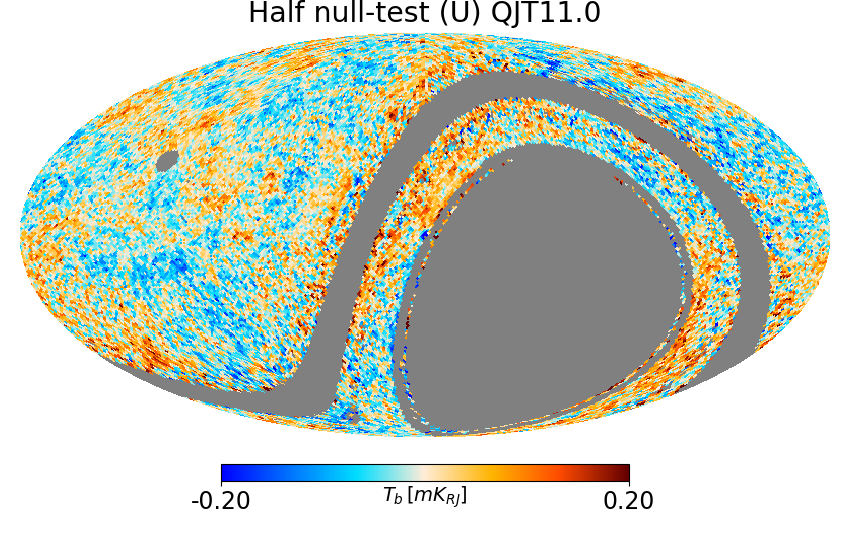}
    \includegraphics[width=0.48\textwidth]{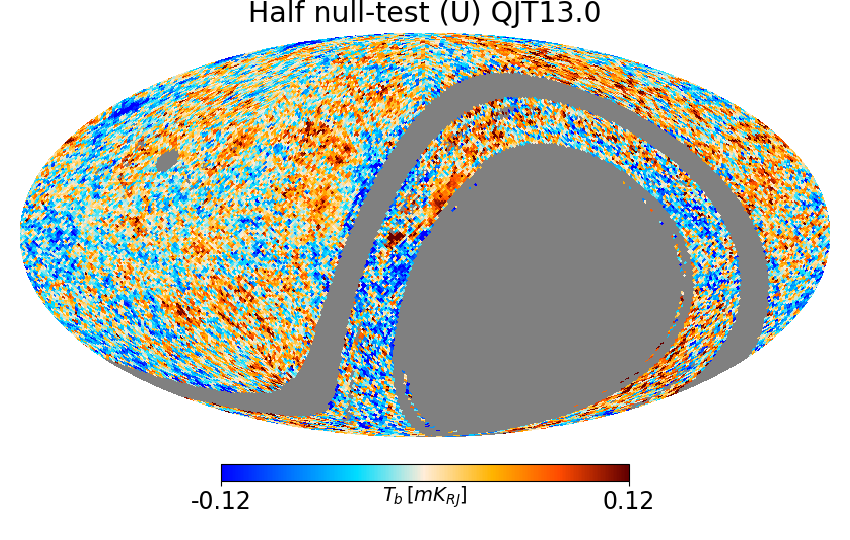}
    \includegraphics[width=0.48\textwidth]{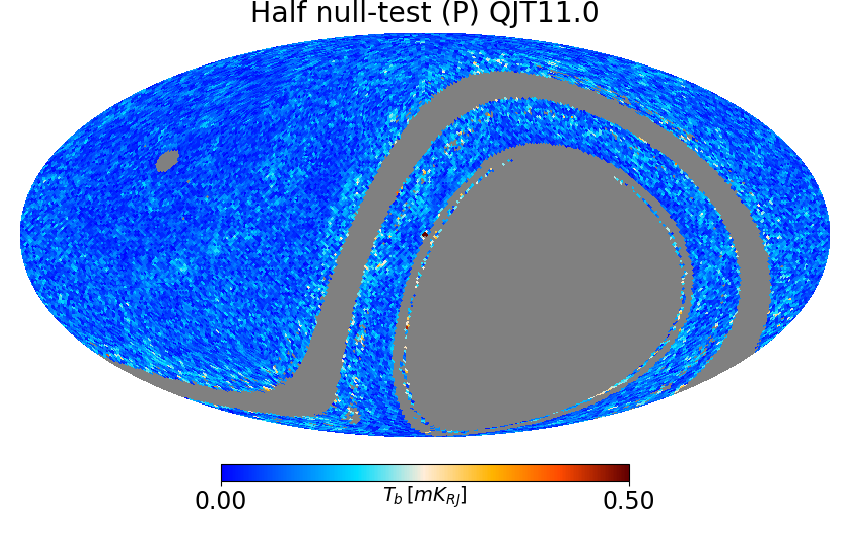}
    \includegraphics[width=0.48\textwidth]{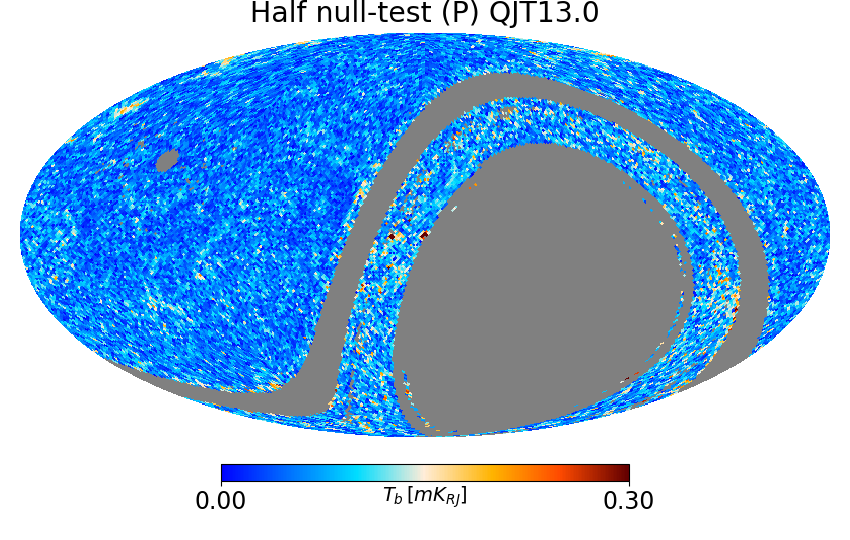}
    \caption{Nulltest of the intensity (first row), Stokes-\textit{Q} (second), Stokes-\textit{U} (third), and polarization amplitude (forth) maps obtained from the \textit{half}-difference nulltest (see \citealp{mfiwidesurvey}), of the 11\,GHz (left) and 13\,GHz (right) QUIJOTE-MFI nominal plus Haze and $\rho$-Ophiuchi raster maps. For comparison purposes, the colour scale is the same as in Fig.\,\ref{figch5:res_int} for intensity, and of Fig.\,\ref{fig:res_p} for polarization. }
    \label{figch5:nulltest}
\end{figure*}

In order to characterize noise structures in the maps we used nulltests. In particular, the  \textit{half}-difference nulltest is the difference between maps obtained from two independent splits selected by date of observation  (see \citealp{mfiwidesurvey}). The \textit{half}-difference maps are expected to show residual noise artifacts or systematics in the data. They are shown in Fig.\,\ref{figch5:nulltest}, for intensity (first row), Stokes \textit{Q} (second), Stokes \textit{U} (third),  polarization amplitude (fourth), at 11 (left) and 13\,GHz (right). It can be observed that residual noise structures are affecting the intensity maps at very large angular scales. However they are very smooth in the Haze region, and are not expected to affect the analysis. In polarization, especially in the \textit{P} maps that are depicted with the same color scale as the polarization residual maps in Fig.\,\ref{fig:res_p}, no significant noise or systematics are observed, therefore the polarization results are expected to be very robust. 

\section{Posterior analysis of the T-T plots} 
\label{secA:ttplot_posterior}

Following the methodology presented in Sect.\,\ref{sec:tt-plots-fsk}, in order to check the goodness of the linear regression of the T-T plots for each angle $\alpha_i$, we compute the posterior distribution of the spectral index parameter $P(\beta)$. We noticed that in low signal-to-noise areas, the wings of the posterior distribution of the spectral index are not reaching zero, and therefore the determination of the spectral index is not appropriate, showing bias towards steep values. In order to be sure that the estimated $\beta$ is unbiased, we have to verify that the posterior of the T-T plots is well defined.

We define the posterior of the slope between the data at frequency $\nu$ (y-axis data, with error $\sigma_y$) and $\nu_0$ (x-axis data, with error $\sigma_x$). It is:
\begin{equation}
    P(m(\beta))= N\cdot e^{-\chi^2/2}
    \label{eq:posterior_betai},
\end{equation}
with $N$ being a normalization factor and $\chi^2$ the chi-square of the linear regression:
\begin{equation}
    \chi^2={\sum_{j}{\frac{(y_j-m(\beta)\cdot x_j-q)^2}{(\sigma_{y_j}^2+m^2(\beta)\cdot \sigma_{x_j}^2)}}},
\end{equation}
where $j$ runs over the pixels enclosed in the area selected for the T-T plots, and $q$ is the best-fit intercept for the given $m$ ($q=<y-mx>$). The slope $m$ is related with the spectral index $\beta$, so we can find the posterior of the spectral index  by converting $m$ into $\beta$  with Eq.\,\ref{eq:beta}. Given that we solve a linear regression for a set of 18 angles $\alpha$, we can compute a posterior distribution $P(\beta)=P(\beta_i)$ for each of them. The final posterior is then the product \mbox{$P_{\rm tot}=\prod_i P(\beta_i)$}, whose maximum should coincide with the $\beta$ in Eq.\,\ref{eq:beta_fin}. \\


We show here the plots of the estimated spectral indices as a function of the projection angle $\alpha$, and the relative posterior distributions. In Fig.\,\ref{fig:fksplot_plawmap} we show the results for \textit{Planck} 30\,GHz-\textit{WMAP} K-band, in Fig.\,\ref{fig:fksplot_qjtwmap} for QUIJOTE 11\,GHz-\textit{WMAP} K-band, and in Fig.\,\ref{fig:fksplot_spasswmap} for the Faraday rotation corrected (Sect.\,\ref{sec:FR}) S-PASS 2.3\,GHz-\textit{WMAP} K-band.

The weighted average of the spectral indices, which is represented as an horizontal black line in the $\beta$ vs $\alpha$ figures, correspond to the final results of this work, which are shown in Fig.\,\ref{fig:betas_ttplot} and quoted in Table\,\ref{tab:betas_ttplots}. Colour corrections are applied independently for the determination of the spectral index at each angle $\alpha$. 

In the plots of the posteriors, each coloured line represents the posterior for a determined projection angle $\alpha$, normalized with its maximum and computed with Eq.\,\ref{eq:posterior_betai}. Here no colour corrections are applied. The red thick line in the plots shows the final posterior normalized with its maximum, obtained as the product of each single posterior distribution. The vertical blue line shows the final spectral index computed with Eq.\,\ref{eq:beta_fin}, with no colour corrections applied. We can notice that there is a very good match between the final posterior distribution and the estimated weighted average spectral index. When we apply colour correction to the spectral index, we obtain the value represented by the vertical black line in the posterior figures, which corresponds to the horizontal black line of the  $\beta$ vs $\alpha$ figures.

We can see from Fig.\,\ref{fig:fksplot_plawmap}-\ref{fig:fksplot_spasswmap} that the posterior distributions of the spectral indices in all the regions and frequencies are closed and approximately Gaussian, and that therefore the spectral indices are well constrained. In addition, we can observe that, when we use the low frequency data such as QUIJOTE 11\,GHz and S-PASS 2.3\,GHz, the posteriors are narrower than that of \textit{Planck} 30\,GHz, thanks to the wider frequency lever with respect to \textit{WMAP} K-band, which leads to a more precise determination of the spectral index at low frequency. 

\begin{figure*}
    \centering
    \includegraphics[width=0.30\textwidth]{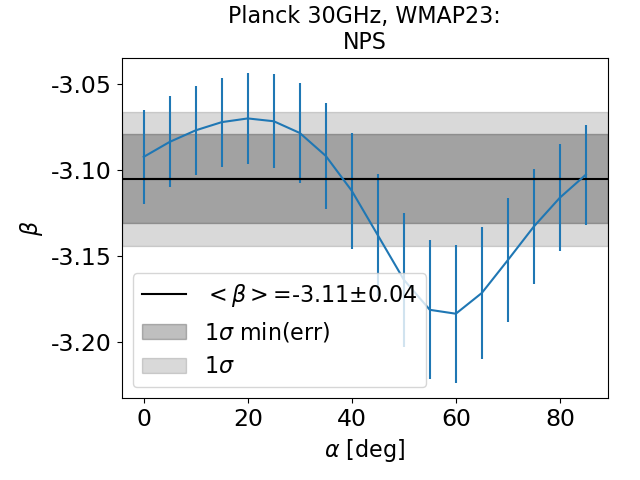}
    \includegraphics[width=0.30\textwidth]{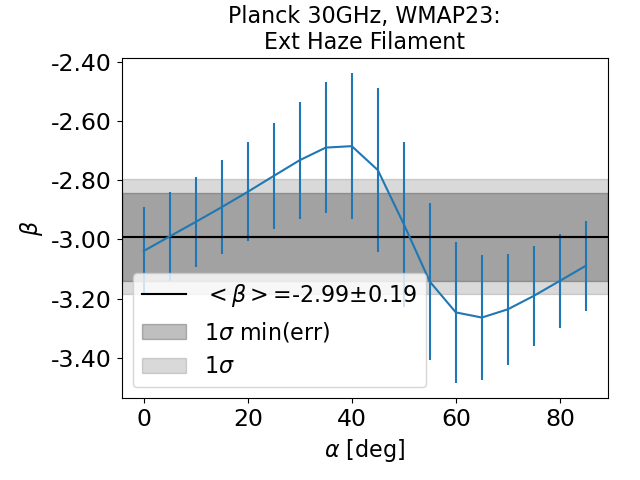}
    \includegraphics[width=0.30\textwidth]{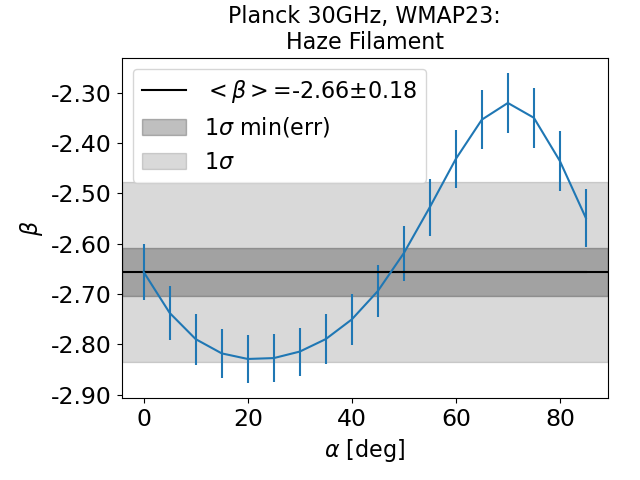}
    
    \includegraphics[width=0.30\textwidth]{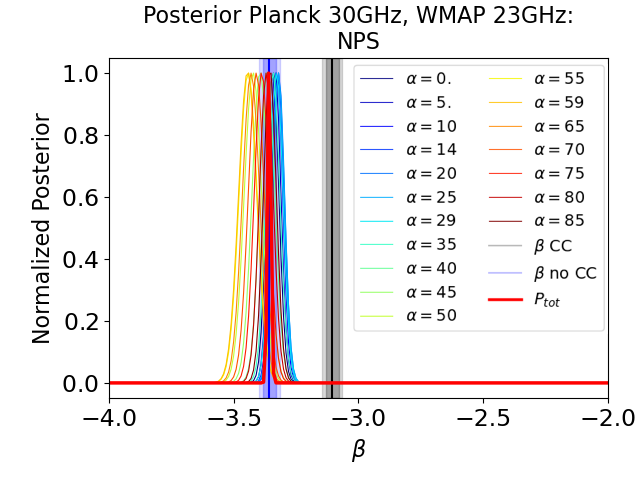}
    \includegraphics[width=0.30\textwidth]{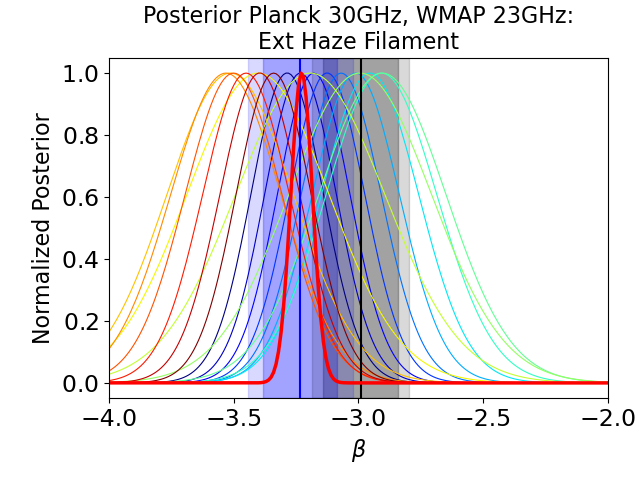}
    \includegraphics[width=0.30\textwidth]{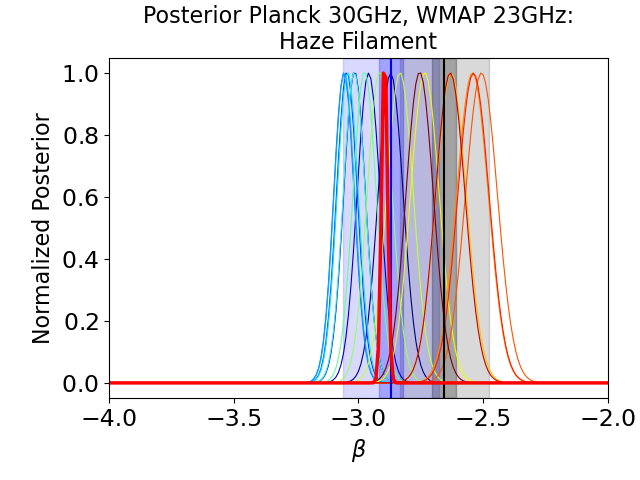}
    \includegraphics[width=0.30\textwidth]{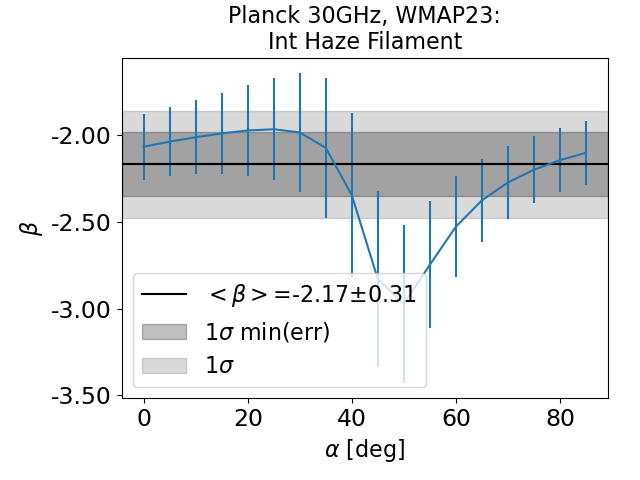}
    \includegraphics[width=0.30\textwidth]{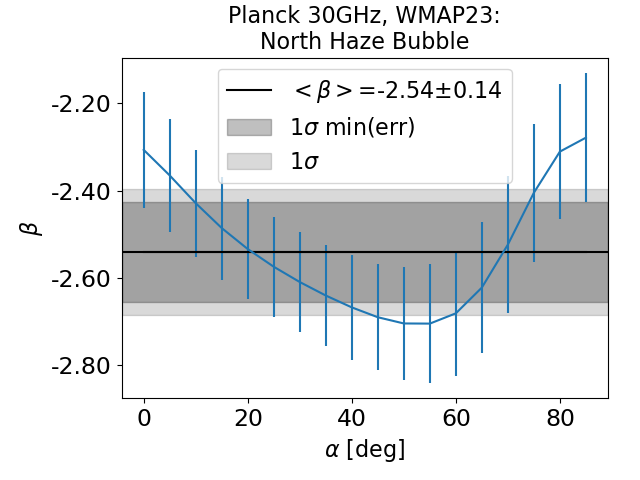}
    \includegraphics[width=0.30\textwidth]{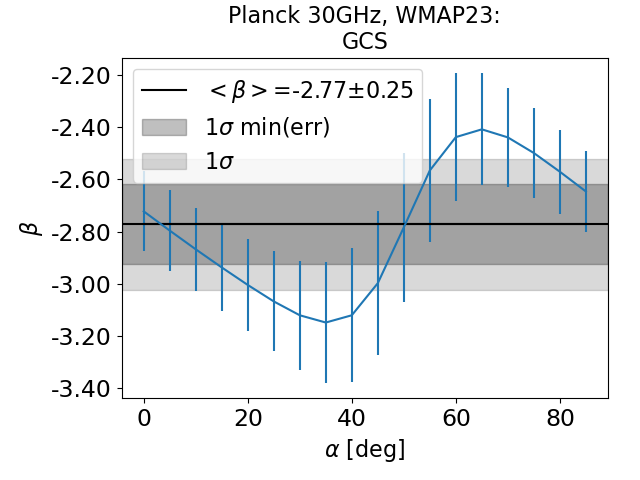}
    
    \includegraphics[width=0.30\textwidth]{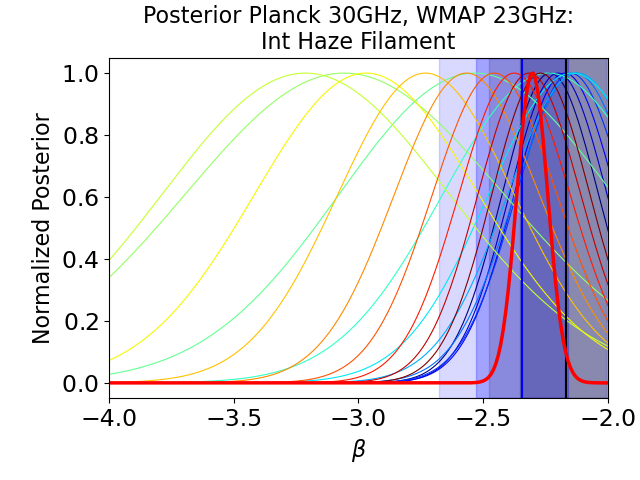}
    \includegraphics[width=0.30\textwidth]{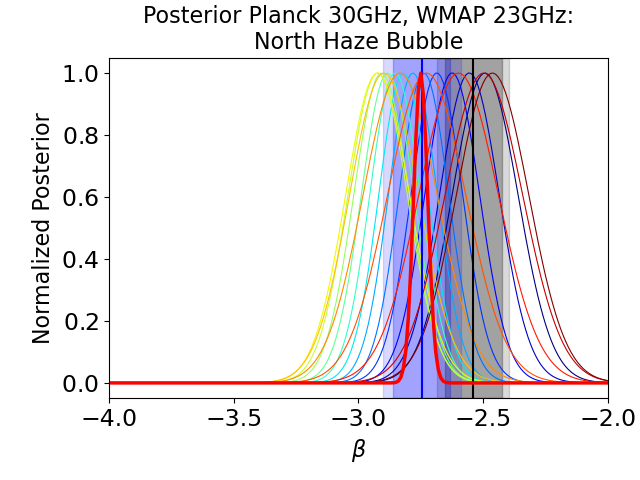}
    \includegraphics[width=0.30\textwidth]{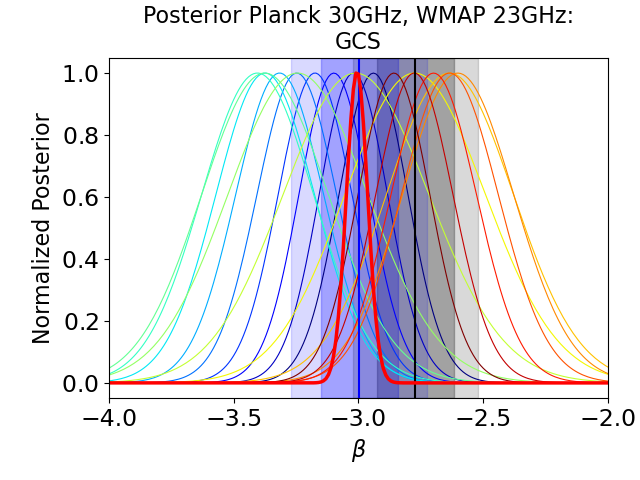}

    \caption{Validation of T-T plots of \textit{Planck} 30\,GHz-\textit{WMAP} K-band for different regions, showing the spectral index $\beta$ as a function of the  projection angle (odd rows) and the posterior distributions for each  projection angle (even rows). The black lines and shaded area show the final estimated spectral index $\beta$  $\pm1\sigma$ uncertainty  with colour corrections applied, while the dark blue lines and shaded area represent the estimated spectral index $\beta$  $\pm1\sigma$ uncertainty before applying colour corrections.}
    \label{fig:fksplot_plawmap}
\end{figure*}

\begin{figure*}
    \centering
    \includegraphics[width=0.30\textwidth]{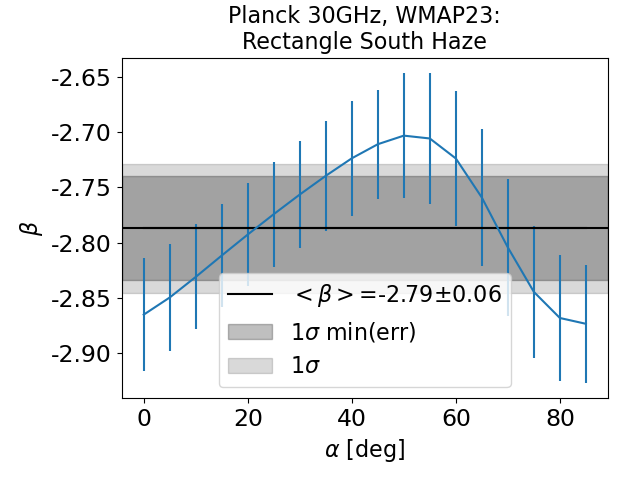}
    \includegraphics[width=0.30\textwidth]{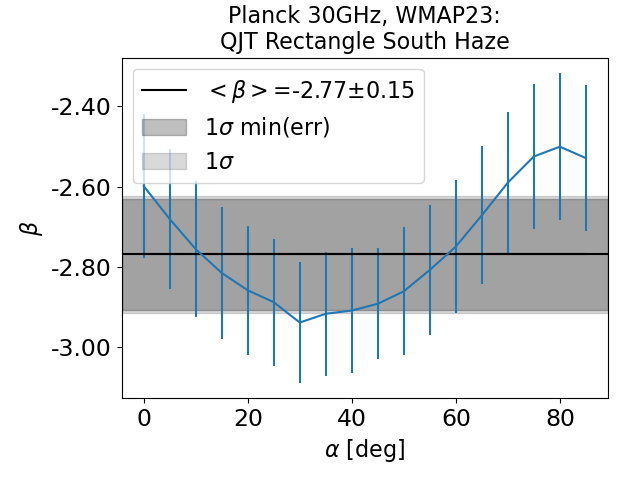}
    \includegraphics[width=0.30\textwidth]{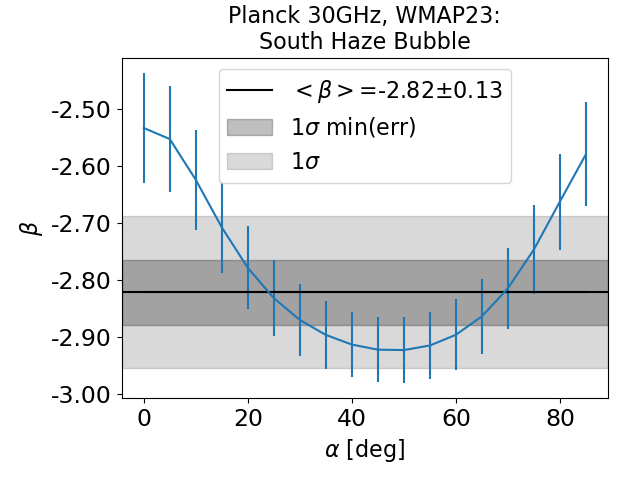}

    \includegraphics[width=0.30\textwidth]{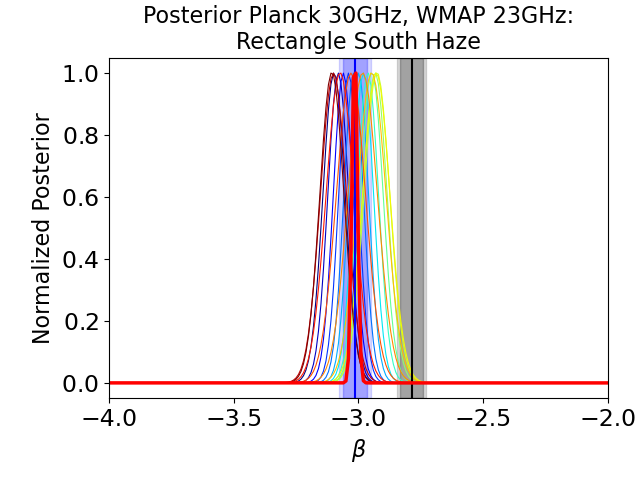}
    \includegraphics[width=0.30\textwidth]{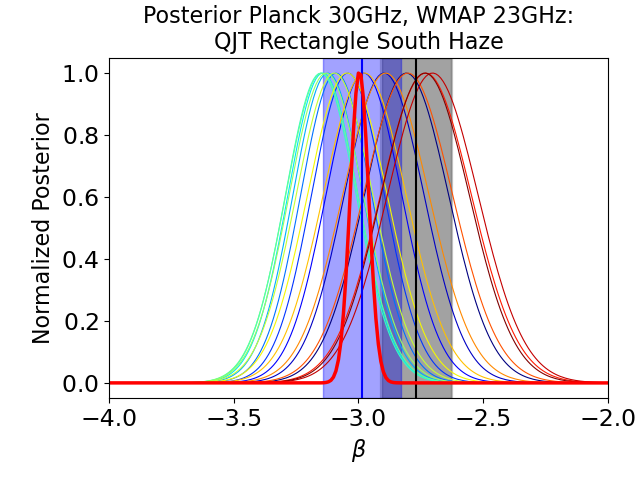}
    \includegraphics[width=0.30\textwidth]{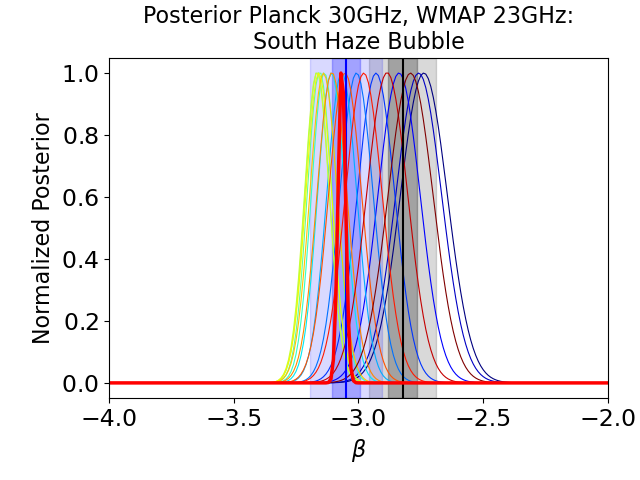}
   
    \includegraphics[width=0.30\textwidth]{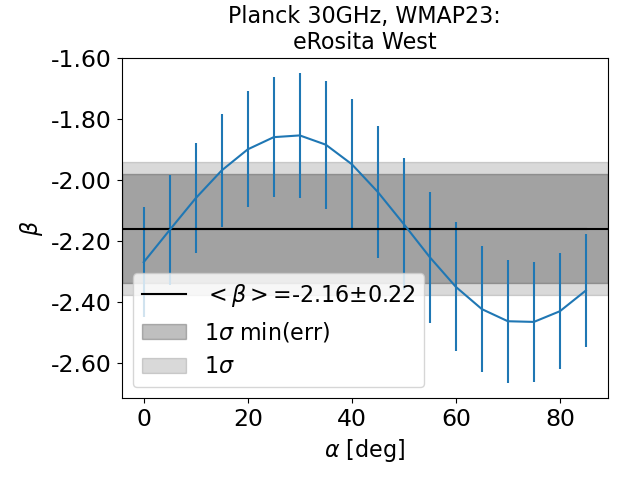}
    \includegraphics[width=0.30\textwidth]{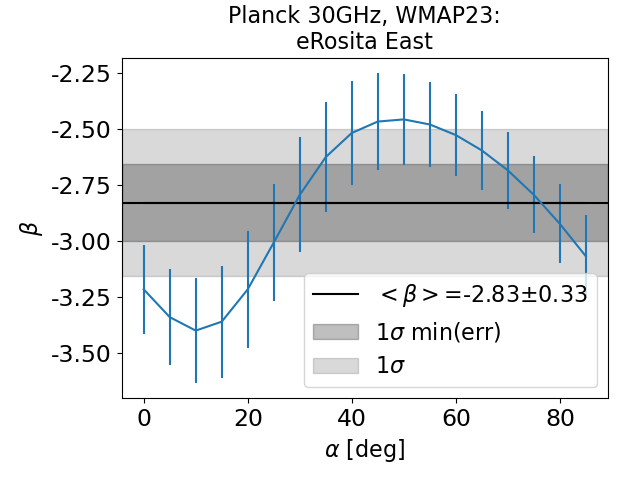}
    \includegraphics[width=0.30\textwidth]{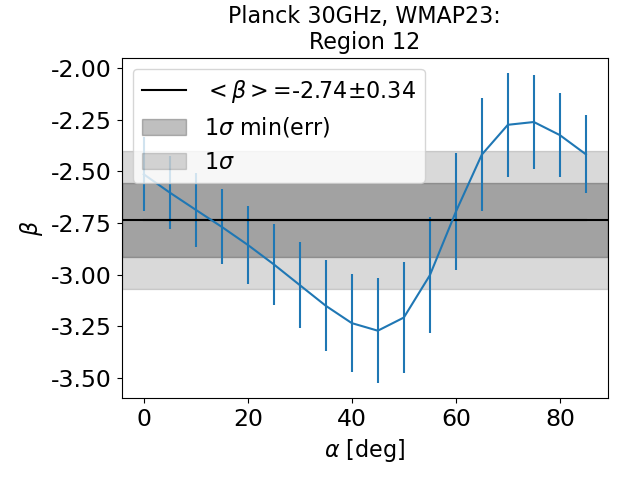}

    \includegraphics[width=0.30\textwidth]{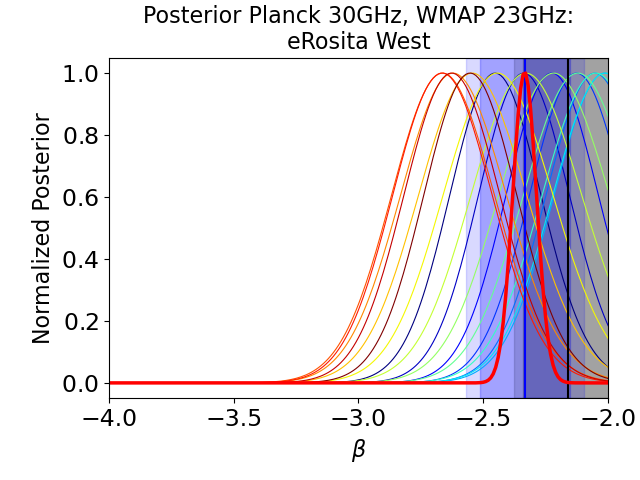}
    \includegraphics[width=0.30\textwidth]{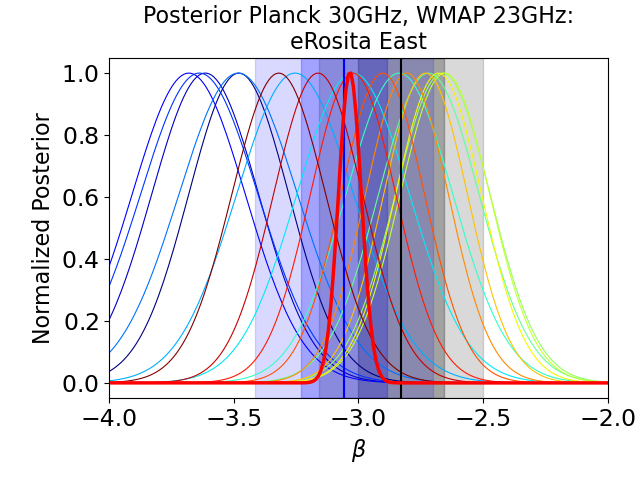}
    \includegraphics[width=0.30\textwidth]{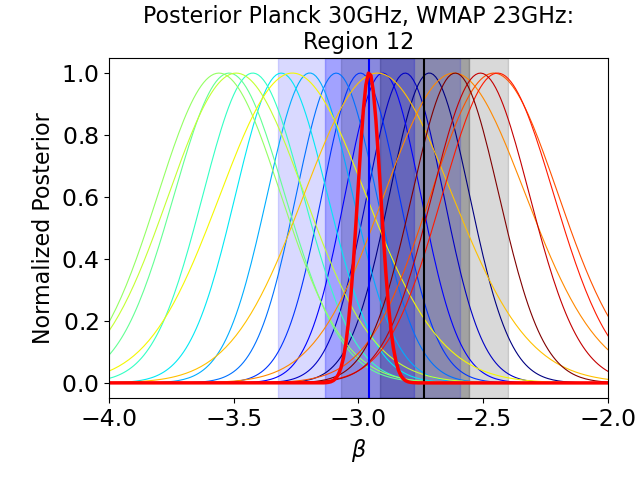}
    
    \contcaption{}
    \label{fig:fksplot_plawmap2}
\end{figure*}

\begin{figure*}
    \centering
    \includegraphics[width=0.30\textwidth]{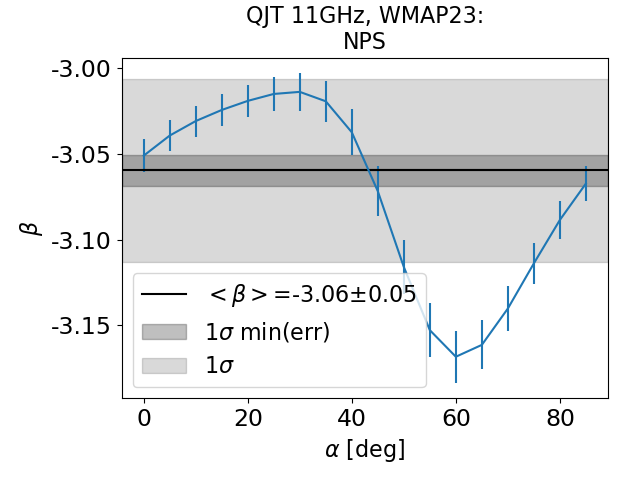}
    \includegraphics[width=0.30\textwidth]{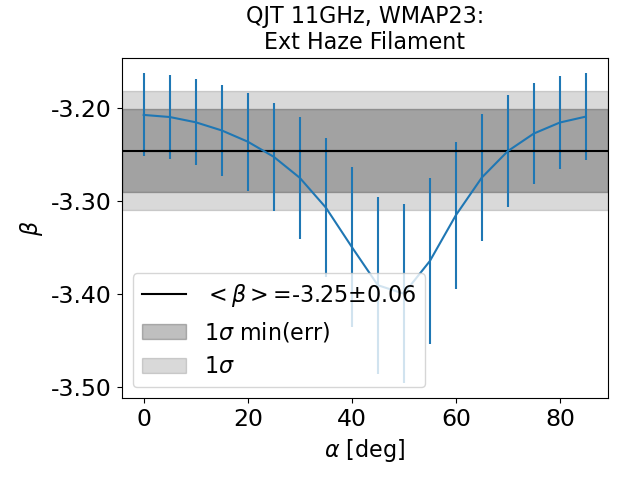}
    \includegraphics[width=0.30\textwidth]{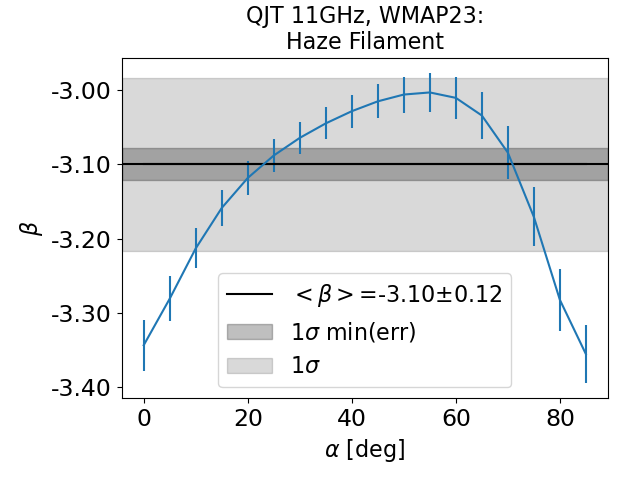}
    
    \includegraphics[width=0.30\textwidth]{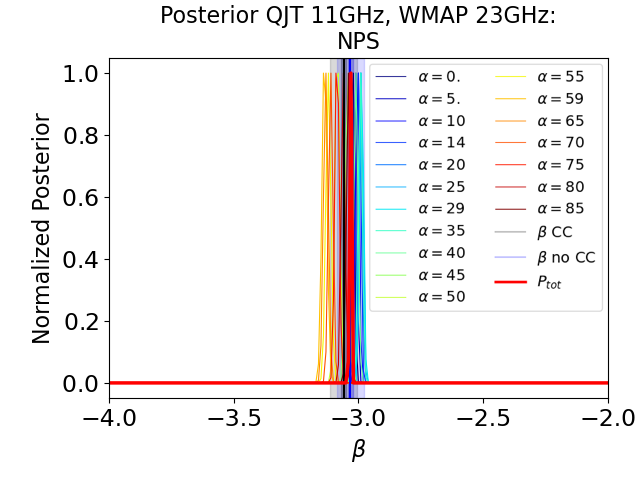}
    \includegraphics[width=0.30\textwidth]{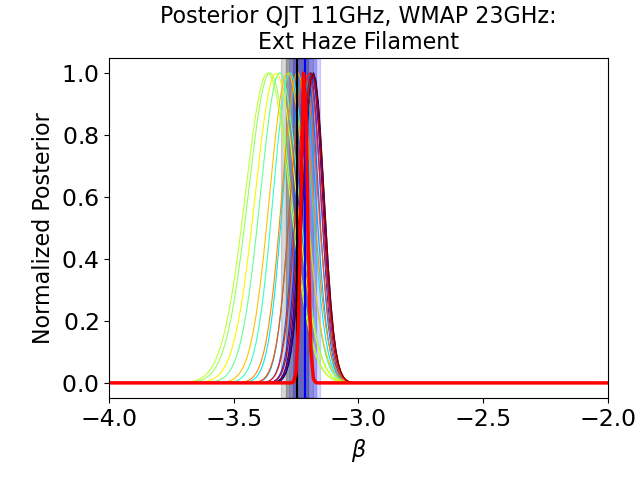}
    \includegraphics[width=0.30\textwidth]{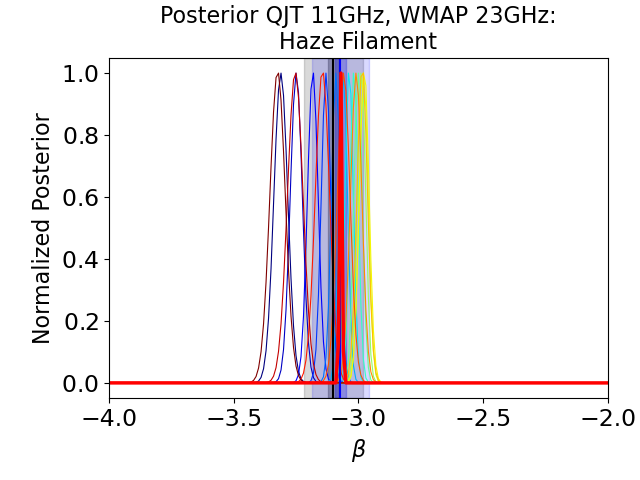}
    \includegraphics[width=0.30\textwidth]{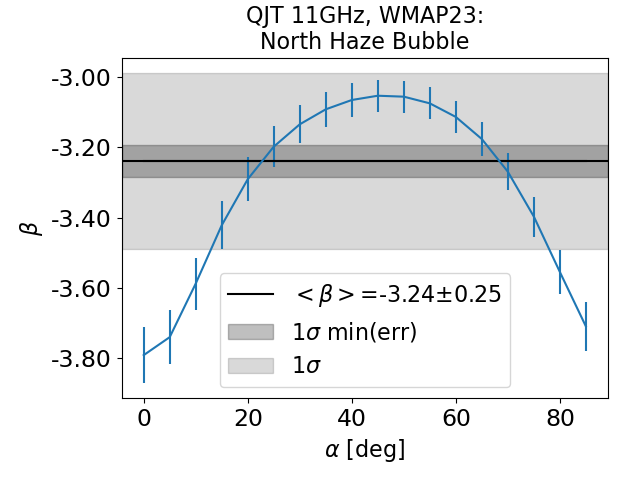}
    \includegraphics[width=0.30\textwidth]{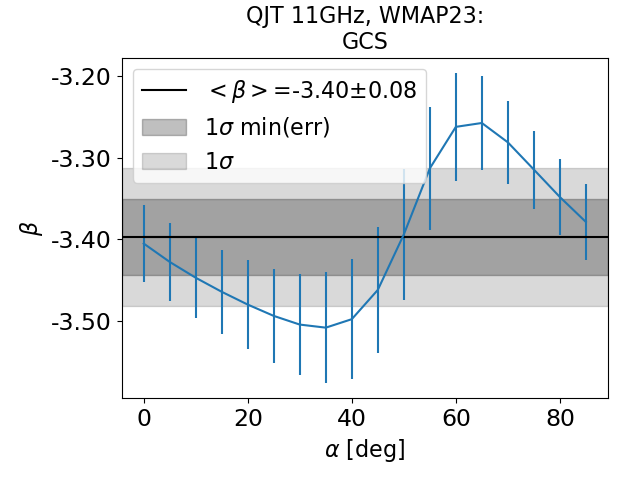}
    \includegraphics[width=0.30\textwidth]{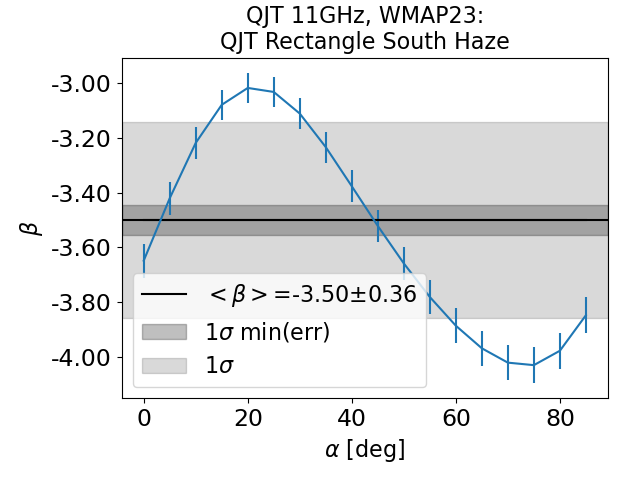}

    \includegraphics[width=0.30\textwidth]{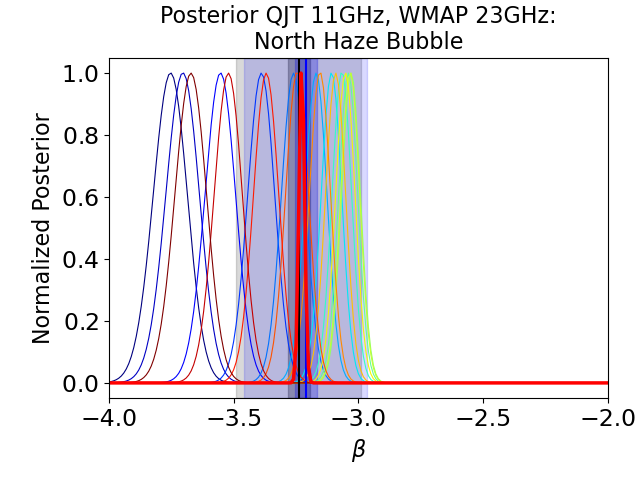}
    \includegraphics[width=0.30\textwidth]{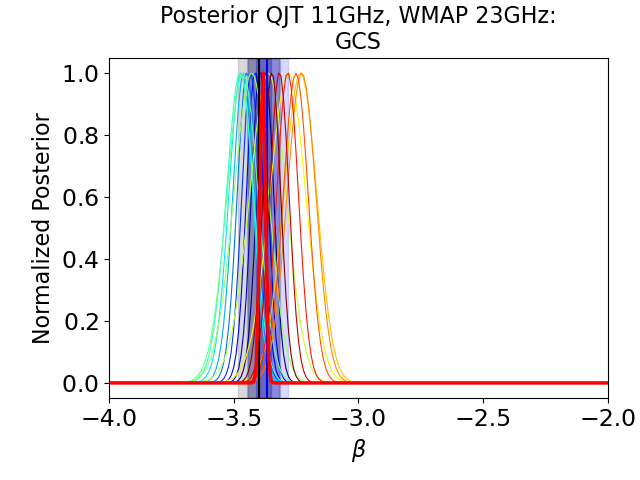}
    \includegraphics[width=0.30\textwidth]{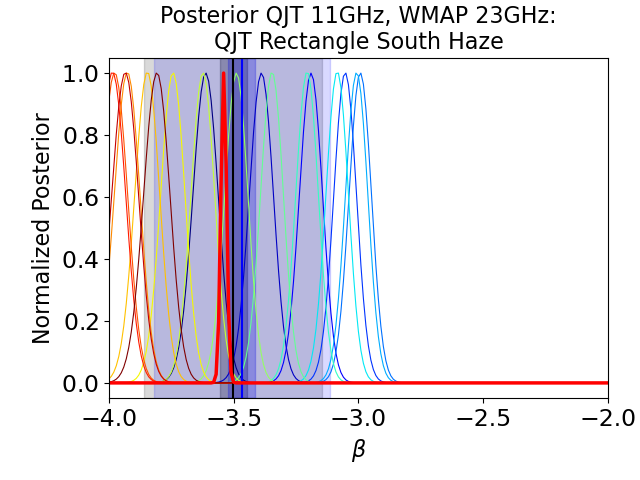}
    
    \includegraphics[width=0.30\textwidth]{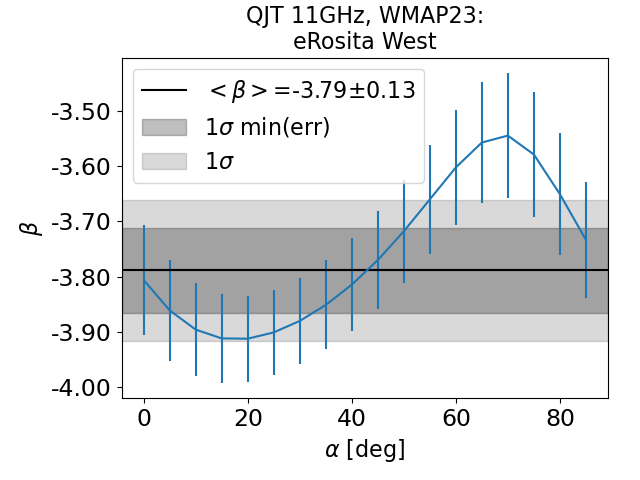}
    \includegraphics[width=0.30\textwidth]{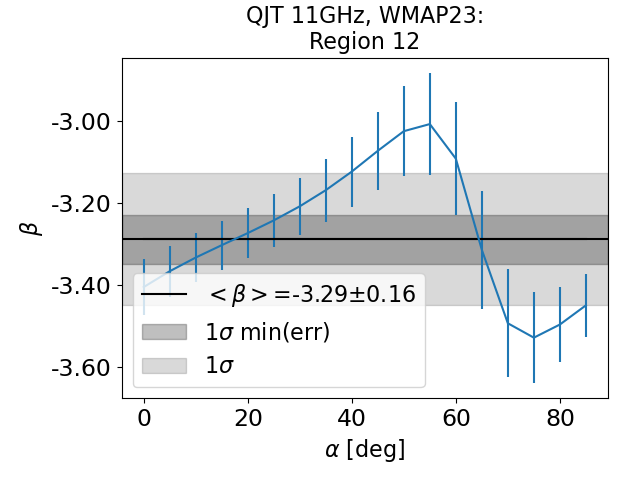}

    \includegraphics[width=0.30\textwidth]{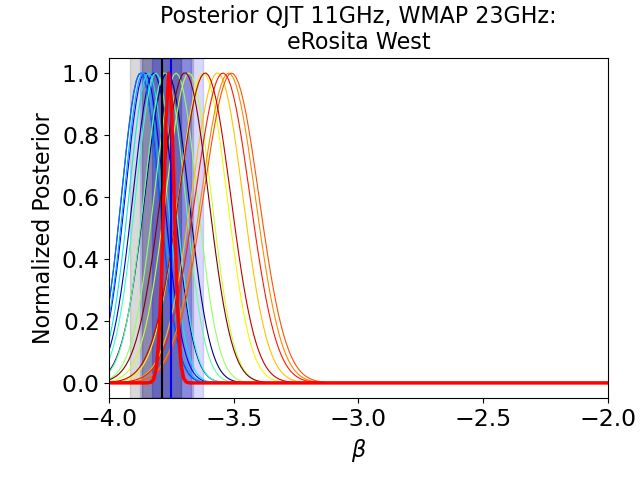}
    \includegraphics[width=0.30\textwidth]{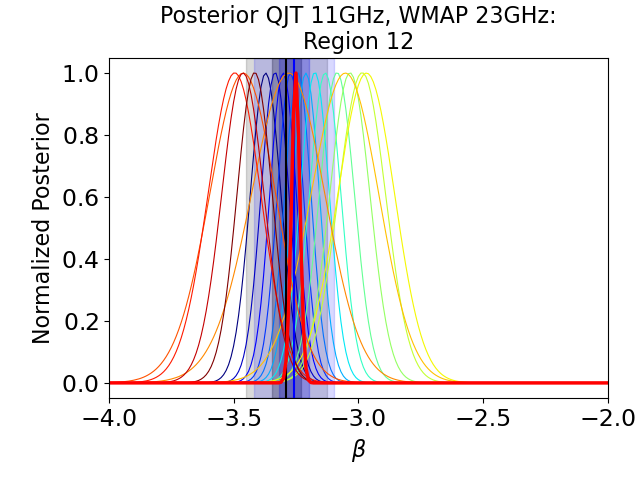}

    \caption{Same as in Fig.\,\ref{fig:fksplot_plawmap} for QUIJOTE 11\,GHz-\textit{WMAP} K-band.}
    \label{fig:fksplot_qjtwmap}
\end{figure*}

\begin{figure*}
    \centering

    \includegraphics[width=0.30\textwidth]{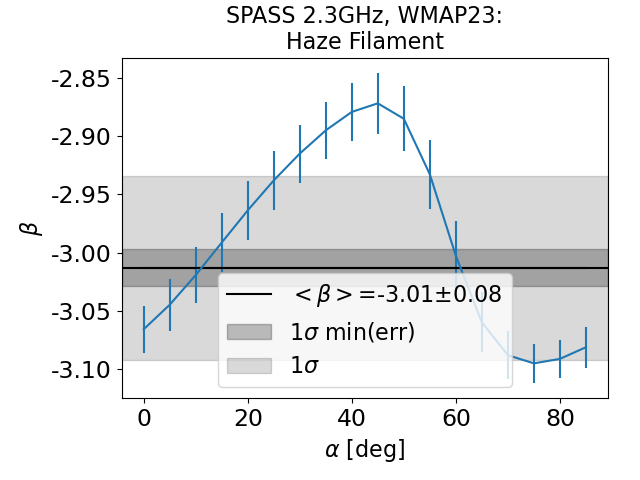}
     \includegraphics[width=0.30\textwidth]{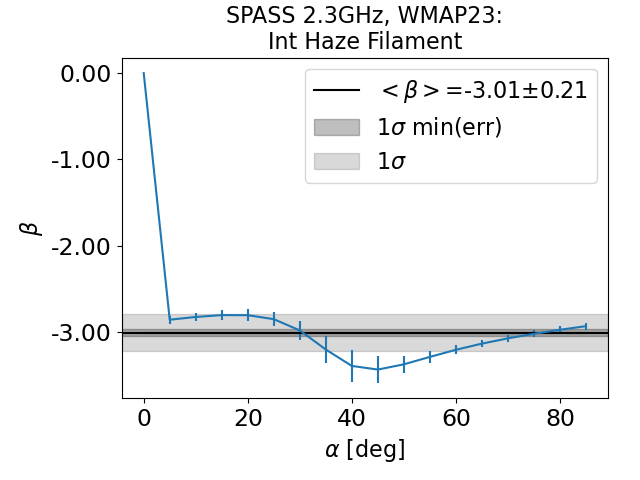}
    \includegraphics[width=0.30\textwidth]{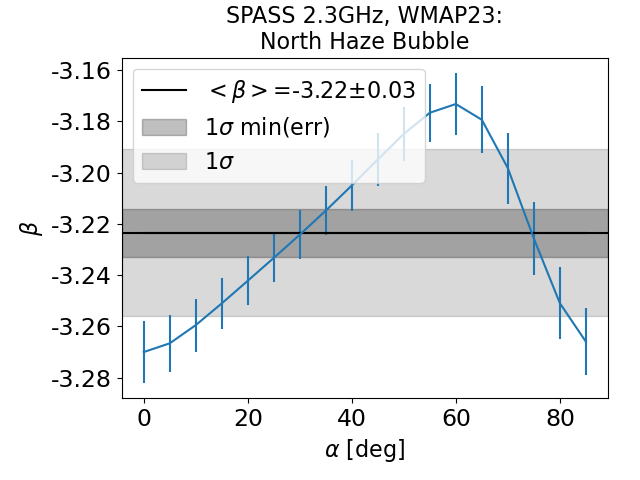}   

    \includegraphics[width=0.30\textwidth]{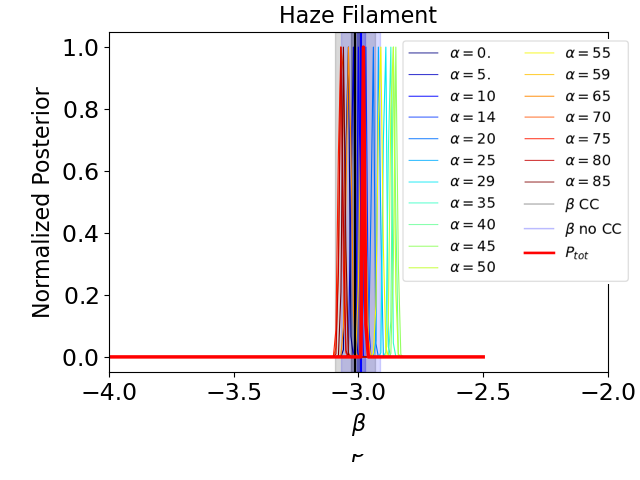}
    \includegraphics[width=0.30\textwidth]{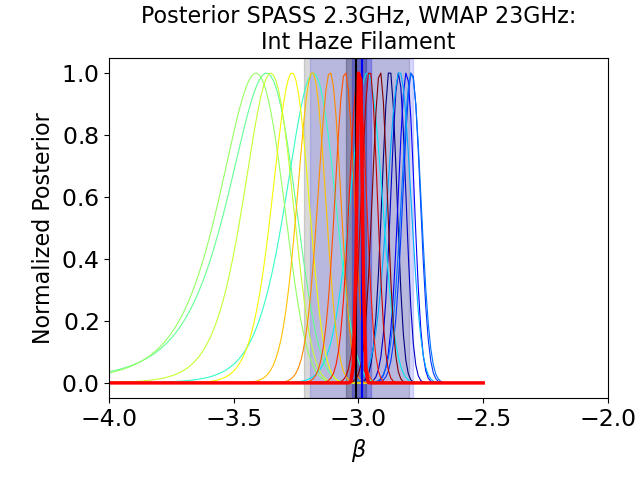}
    \includegraphics[width=0.30\textwidth]{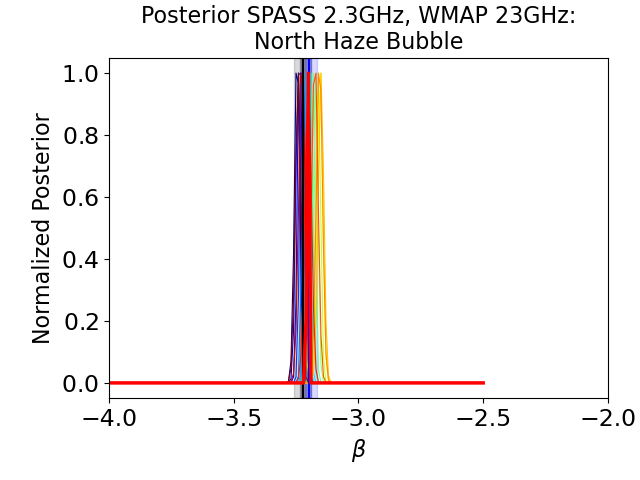}
    \includegraphics[width=0.30\textwidth]{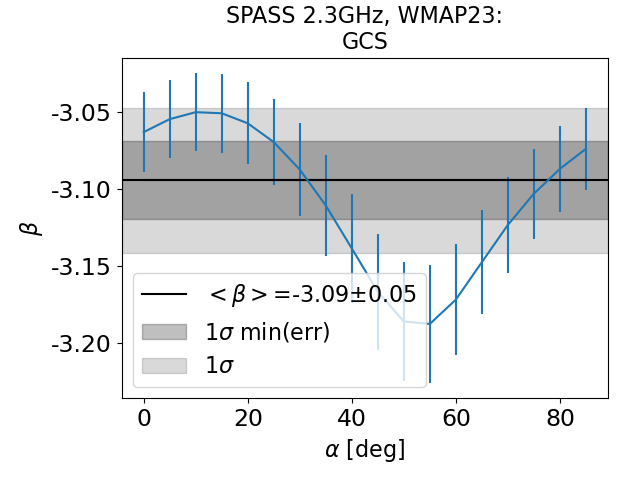}
    \includegraphics[width=0.30\textwidth]{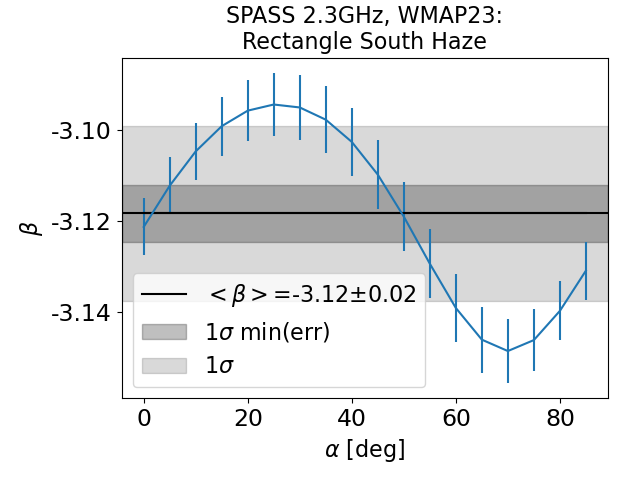}
    \includegraphics[width=0.30\textwidth]{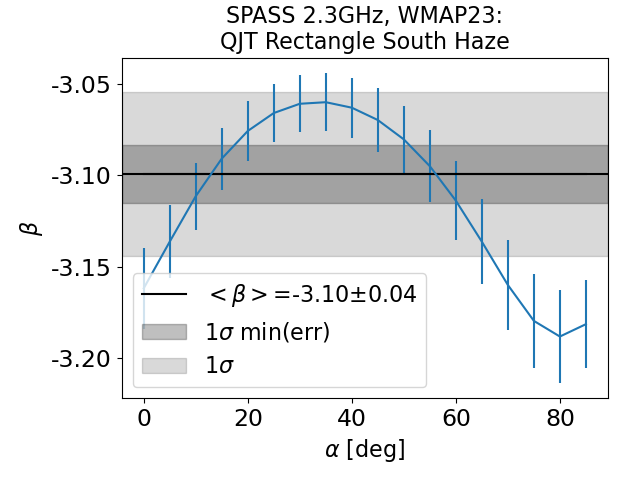}  

    \includegraphics[width=0.30\textwidth]{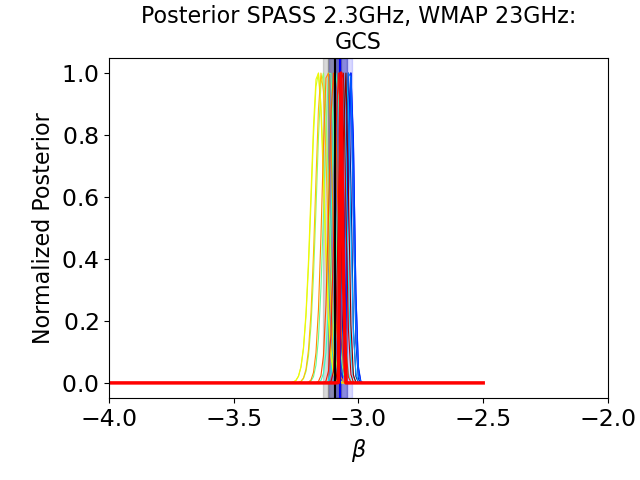}
    \includegraphics[width=0.30\textwidth]{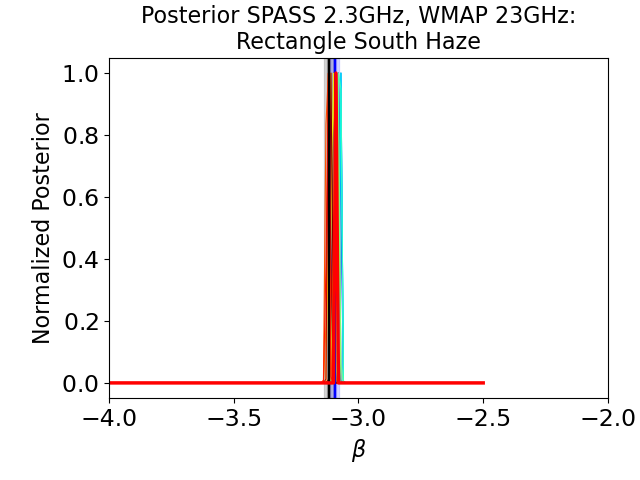}
    \includegraphics[width=0.30\textwidth]{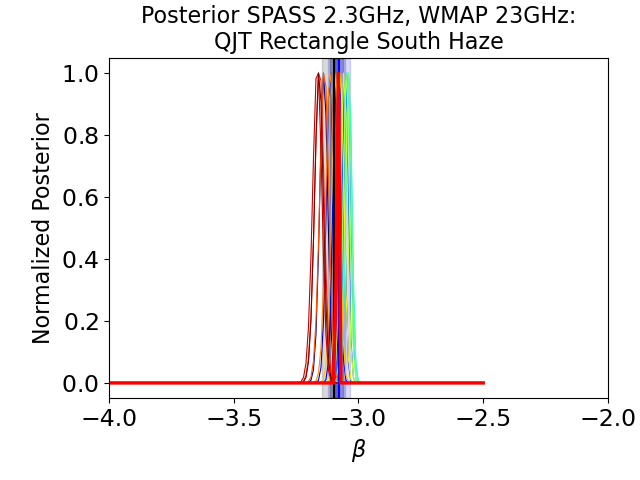}
    
    \includegraphics[width=0.30\textwidth]{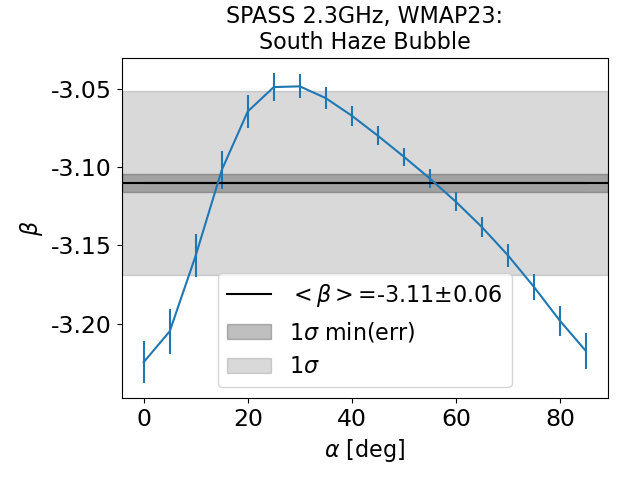}
    \includegraphics[width=0.30\textwidth]{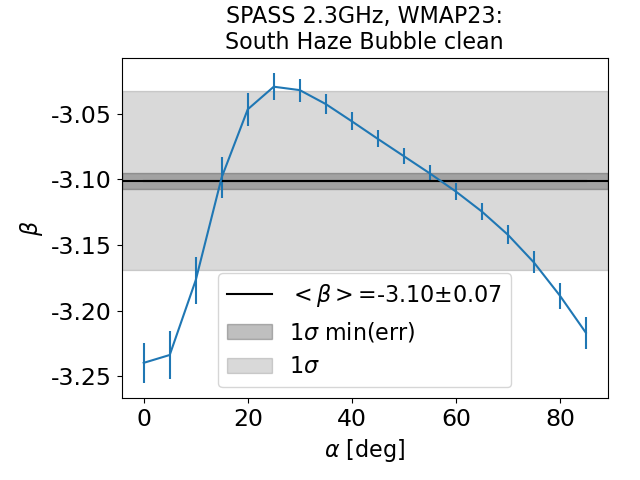}

    \includegraphics[width=0.30\textwidth]{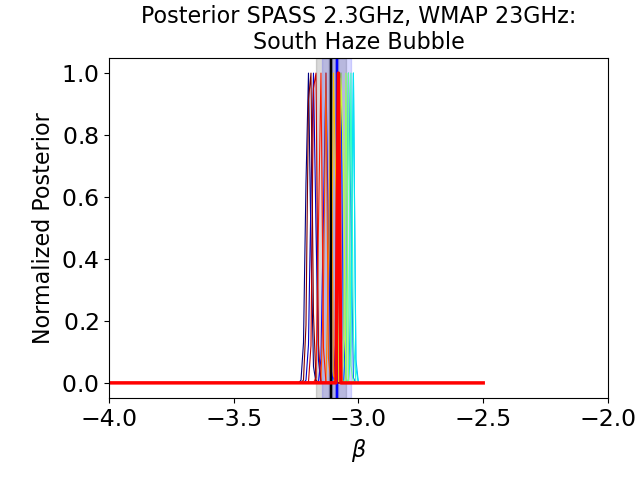}
    \includegraphics[width=0.30\textwidth]{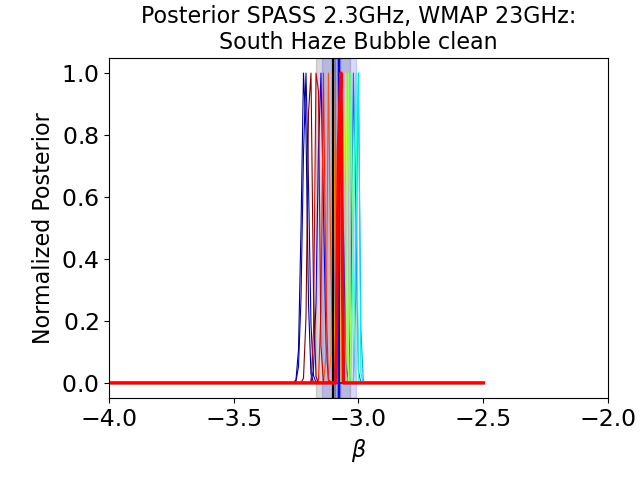}

    \caption{Same as in Fig.\,\ref{fig:fksplot_plawmap} for S-PASS 2.3\,GHz-\textit{WMAP} K-band.}
    \label{fig:fksplot_spasswmap}
\end{figure*}

\begin{figure*}
    \centering
    \includegraphics[width=0.30\textwidth]{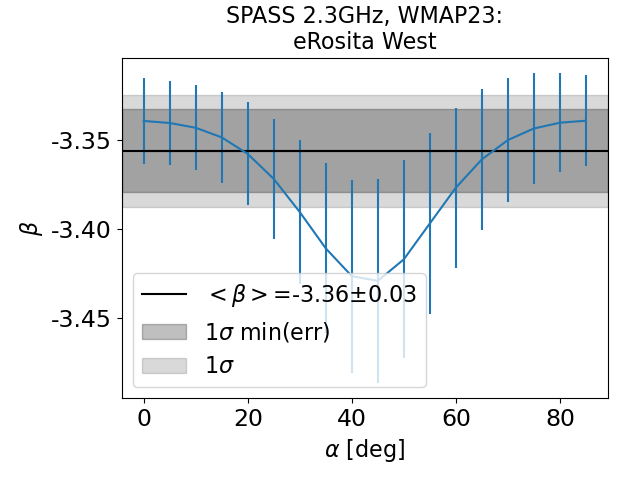}
    \includegraphics[width=0.30\textwidth]{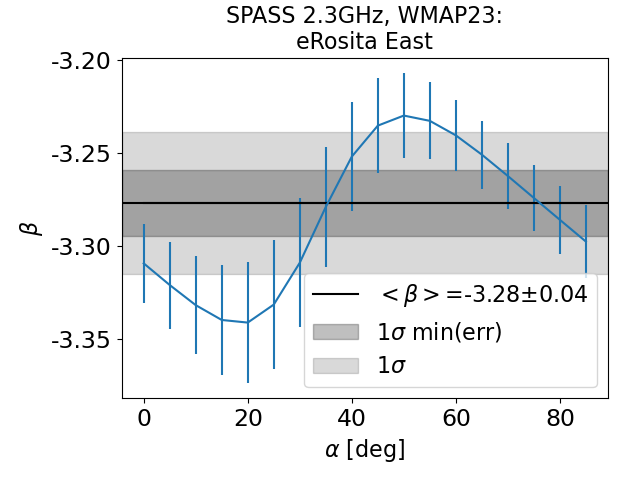}
    \includegraphics[width=0.30\textwidth]{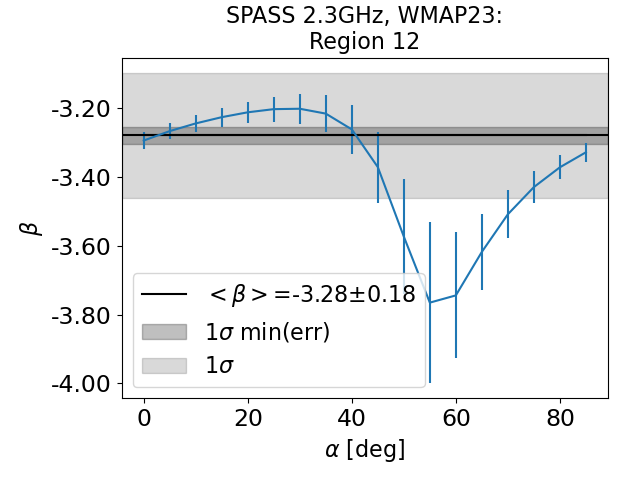}  

    \includegraphics[width=0.30\textwidth]{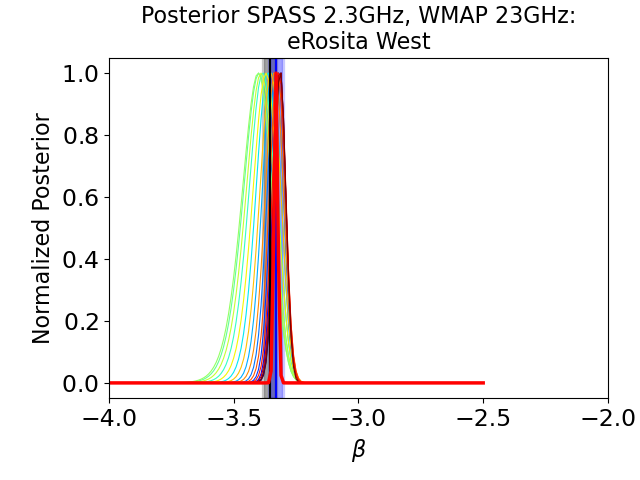}
    \includegraphics[width=0.30\textwidth]{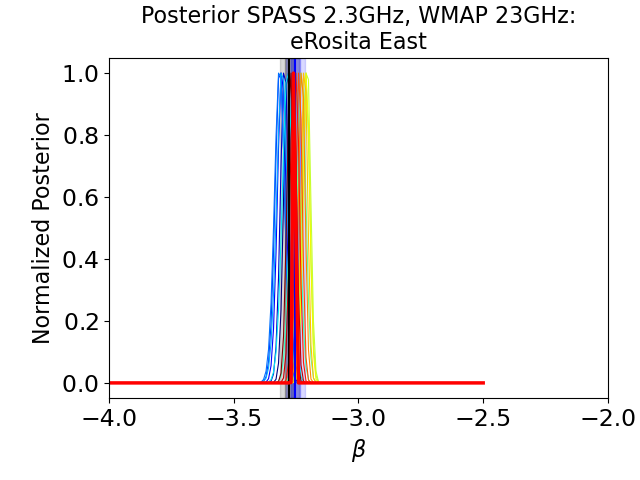}
    \includegraphics[width=0.30\textwidth]{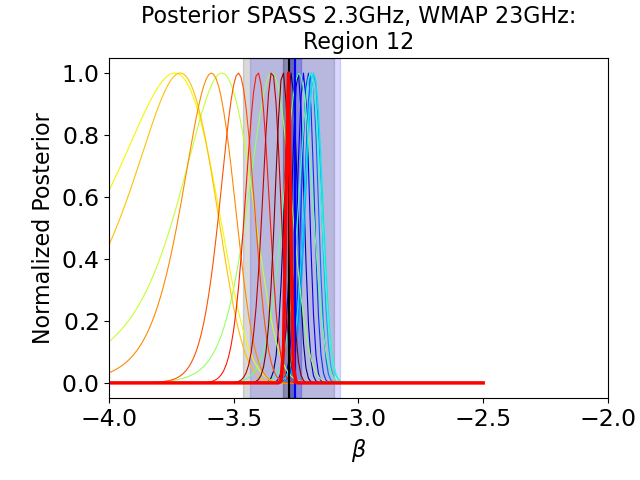}
    
    \contcaption{}
    \label{fig:fksplot_spasswmap2}

\end{figure*}

\bsp	
\label{lastpage}
\end{document}